\pdfoutput=1
\documentclass[10pt,conference]{IEEEtran}
\pdftrailerid{}
\usepackage[T1]{fontenc}
\usepackage{cite}
\usepackage{amsmath,amssymb,amsfonts}
\usepackage{graphicx}
\usepackage{booktabs}
\usepackage{textcomp}
\usepackage{xcolor}
\usepackage{array}
\usepackage{upquote}
\usepackage{listings}
\usepackage{capt-of}  %
\usepackage{tikz}
\usetikzlibrary{positioning,arrows.meta,shapes.geometric,calc}
\usepackage[skins]{tcolorbox}
\usepackage{fontawesome5}
\usepackage{placeins}
\newtcolorbox{rqenv}{enhanced, sharp corners, frame hidden, boxrule=0pt, colback=black!7,
  borderline west={2.5pt}{0pt}{black}, left=7pt, right=4pt, top=2pt, bottom=2pt,
  before skip=4pt, after skip=4pt}
\newcommand{\rqbox}[2]{\begin{rqenv}\faSearch\,\textbf{RQ$_{#1}$.} \textit{#2}\end{rqenv}}
\definecolor{ansback}{RGB}{229,245,238}
\definecolor{ansbar}{RGB}{0,155,85}
\newtcolorbox{ansenv}{enhanced, sharp corners, frame hidden, boxrule=0pt, colback=ansback,
  borderline west={2.5pt}{0pt}{ansbar}, left=7pt, right=4pt, top=2pt, bottom=2pt,
  before skip=4pt, after skip=4pt}
\newcommand{\ansbox}[2]{\begin{ansenv}\faHandPointRight[regular]\,\textbf{Answer to RQ$_{#1}$.} #2\end{ansenv}}
\lstdefinestyle{json}{
  basicstyle=\footnotesize\ttfamily,
  columns=fullflexible,
  keepspaces=true,
  breaklines=true,
  breakindent=1.5em,
  showstringspaces=false,
  frame=none,
  aboveskip=2pt, belowskip=2pt,
  literate={…}{{\ldots}}1
}
\newcommand{\jrole}[1]{\textcolor{blue!50!black}{#1}}
\newcommand{\jpres}[1]{{\setlength{\fboxsep}{0.5pt}\colorbox{green!16}{\textcolor{green!40!black}{#1}}}}
\newcommand{\jext}[1]{{\setlength{\fboxsep}{0.5pt}\colorbox{violet!16}{\textcolor{violet!70!black}{#1}}}}
\newcommand{\jrank}[1]{{\setlength{\fboxsep}{0.5pt}\colorbox{orange!22}{\textcolor{orange!60!black}{#1}}}}
\lstdefinestyle{code}{
  basicstyle=\scriptsize\ttfamily,
  columns=fullflexible,
  keepspaces=true,
  breaklines=true,
  breakindent=1.5em,
  showstringspaces=false,
  frame=none,
  aboveskip=2pt, belowskip=2pt,
  numbers=left,
  numberstyle=\tiny\color{black!60},
  numbersep=4pt,
  xleftmargin=1.2em,
  escapeinside={(*@}{@*)}
}
\definecolor{lstcont}{gray}{0.6}  %
\lstdefinestyle{pyprompt}{
  language=Python,
  deletecomment=[s]{"""}{"""},
  basicstyle=\scriptsize\ttfamily,
  columns=fullflexible,
  keepspaces=true,
  breaklines=true,
  breakindent=1.2em,
  postbreak=\mbox{\textcolor{lstcont}{$\hookrightarrow$}\space},
  showstringspaces=false,
  upquote=true,
  numbers=left,
  numberstyle=\tiny\color{black!45},
  numbersep=7pt,
  xleftmargin=2.2em,
  frame=single,
  framerule=0.4pt,
  rulecolor=\color{black!18},
  backgroundcolor=\color{black!2.5},
  framesep=4pt,
  aboveskip=4pt, belowskip=4pt,
  stringstyle={},
  keywordstyle={},
  commentstyle=\color{black!50}\itshape,
  emph={AGENT_SYSTEM,BASELINE_SYSTEM,TRANSCRIBE_SYSTEM,TRANSCRIBE_USER,SPLIT_SYSTEM,SPLIT_USER,MAIN_FLOW_SYSTEM,MAIN_FLOW_USER,SYNTHESIS_SYSTEM,SYNTHESIS_USER},
  emphstyle=\color{blue!55!black},
  escapeinside={(*@}{@*)},
  literate={…}{{\ldots}}1 {—}{{\rule[0.55ex]{0.6em}{0.35pt}}}1 {–}{{\rule[0.55ex]{0.45em}{0.35pt}}}1 {→}{{$\rightarrow$}}1
}

\newcommand{\phead}[1]{\textcolor{blue!55!black}{#1}}

\usepackage[hidelinks,pdfauthor={},pdftitle={},pdfsubject={},pdfkeywords={},pdfcreator={},pdfproducer={}]{hyperref}

\makeatletter

\providecommand*{\descriptionlabel}[1]{\hspace\labelsep\normalfont\bfseries #1}
\makeatother

\providecommand{\Description}[1]{}

\begin{document}

\title{A Hybrid Approach for Automated Security Pattern Recovery}

\author{\IEEEauthorblockN{Anh Minh Nguyen}
\IEEEauthorblockA{\textit{University College Dublin}\\
Dublin, Ireland}
\and
\IEEEauthorblockN{Tuan Dat Tran}
\IEEEauthorblockA{\textit{University College Dublin}\\
Dublin, Ireland \\
dat.tran@ucdconnect.ie}
\and
\IEEEauthorblockN{Alzubair Hassan}
\IEEEauthorblockA{\textit{University College Dublin}\\
Dublin, Ireland \\
alzubair.hassan@ucd.ie}
\and
\IEEEauthorblockN{Liliana Pasquale}
\IEEEauthorblockA{\textit{University College Dublin}\\
Dublin, Ireland \\
liliana.pasquale@ucd.ie}}

\maketitle

\begin{abstract}

Security-by-design emphasises integrating security into software design and development. Security pattern catalogues operationalise this principle by prescribing reusable architectural solutions to recurring security problems. However, recovering their implementations from source code remains challenging because security patterns are abstract, span multiple components, and admit many valid implementations. We present SecPatRec, a hybrid approach for automated security pattern recovery from source code repositories. We treat a security pattern as a reference architecture and frame its recovery as a deductive task: determining which prescribed roles are present, grounding them in code entities, and recovering their interaction sequences.  SecPatRec pairs a ReAct-style LLM agent equipped with semantic and lexical code search for role detection and grounding with static code property graph analysis for interaction recovery. We curate a human-validated ground-truth dataset of 100 pattern instances covering five security patterns across 93 open-source Java and Python repositories, and introduce a three-level evaluation framework measuring recovery correctness against this ground truth. 
We evaluate SecPatRec using four LLMs, including proprietary and open-source models. Averaged over all models and pattern instances, SecPatRec achieves role-detection accuracy of 0.84, role-grounding similarity of 0.84, and an interaction-recovery score of 0.66.   Compared to a one-shot end-to-end LLM baseline, SecPatRec achieves higher aggregate recovery scores and lower estimated LLM API costs per run, while processing repositories that exceed the baseline’s configured input limit.  The recovered role mappings and statically derived interaction sequences provide structural and behavioural evidence that can support inspection of security-pattern implementations.

\end{abstract}

\section{Introduction}
\label{sec:intro}
Software vulnerabilities can arise from weaknesses in a system's design as well as from implementation mistakes. McGraw estimates that roughly half of software security defects are design flaws~\cite{arce2014avoiding}, not implementation bugs~\cite{mcgraw_software_2006}.  Security-by-design emphasises integrating security into software design and development~\cite{mcgraw_software_2006,yoder_architectural_1997}. The EU Cyber Resilience Act reinforces this goal by establishing cybersecurity requirements for products with digital elements made available on the EU market~\cite{european_parliament_regulation_2024}. Tracing architectural security decisions to their implementation supports inspection of whether the code reflects the intended design~\cite{murphy_software_1995}. However, this inspection is difficult when developers have only a partial view of the security architecture~\cite{thomas_security_2018}, or when architectural documentation is missing, outdated or inconsistent with the code~\cite{ernst_study_2023,levy_understanding_2021}.

Security patterns are one of the primary means of security-by-design: they document reusable architectural solutions to recurring security problems~\cite{scandariato_architecting_2008,schumacher_security_2006,van_den_berghe_reimagined_2022}. A pattern describes a solution through architectural \emph{roles} and the \emph{actions} between them, providing both a blueprint for implementing a security control and a reference for inspecting an existing implementation.  Unlike the Gang of Four design patterns~\cite{gamma_design_1994}, which offer code-level guidelines, security patterns admit multiple valid implementations. A role may be realised through different classes, functions or library calls, depending on the language, framework and application context~\cite{vanhilst_reverse_2007,behrens_what_2018,van_den_berghe_reimagined_2022}, and an implementation may span several components. Consequently, finding individual security mechanisms does not establish how the pattern is implemented. Recovering the roles and their interactions can support architectural inspection and enable larger-scale empirical studies of how security patterns are used in practice, for which evidence remains limited~\cite{yskout_does_2012,yskout_security_2015}.

In this paper, we address \emph{automated security pattern recovery}: given a security pattern and a codebase, determine which prescribed roles are present, ground them in code entities, and recover their interaction sequence. Because a security pattern describes both structure and behaviour, we treat it as a \emph{reference architecture} and frame recovery as a deductive software architecture recovery (DSAR) task~\cite{rukmono_deductive_2024}. The pattern specification guides the search for implementing entities, including code defined within the repository and references to externally defined functionality.

Three challenges motivate our approach. \emph{First, roles are implemented in many ways.} Prior work detects security patterns by architectural analysis and pattern matching~\cite{erlingsson_architecture-centric_2011,alvi_security_2021}; Bunke and Sohr, for example, match connected object process graphs and evaluate on a benchmark of Android applications~\cite{bunke_supporting_2020}. LLM-assisted DSAR adds a different ability: it finds code units by what they mean rather than by their structure~\cite{rukmono_deductive_2024,rukmono_evaluating_2026}. We ask whether this can detect and ground roles in Java and Python repositories too large for a single prompt. \emph{Second, role mappings alone are not enough.} They say which entities play the roles, not how those entities interact, and LLM-assisted DSAR only supports the mapping~\cite{rukmono_deductive_2024,rukmono_evaluating_2026}. Static analysis can supply the call relationships and their order. \emph{Third, each recovery output needs its own measure.} Although architecture-level similarity measures provide useful comparisons of recovered architectures~\cite{garcia_comparative_2013,lutellier_measuring_2018}, evaluating individual patterns requires measurements linked to role presence, code grounding and interaction recovery. Assessing these outputs and comparing recovery configurations requires human-validated annotations at this granularity, which we construct.

In this paper, we propose \textsc{SecPatRec}, a hybrid approach that combines an LLM agent with static code property graph (CPG) analysis~\cite{yamaguchi_modeling_2014}. Guided by the pattern specification, the agent iteratively uses semantic and lexical code search to locate and inspect relevant code, producing role mappings and presence statuses. CPG traversal then uses these mappings to recover interactions between the identified entities. This division combines semantic interpretation of architectural roles with interaction evidence extracted from source code. To evaluate the approach, we construct human-validated ground truth and define three corresponding correctness metrics.

Our contributions are:
\begin{itemize}
  \item \textsc{SecPatRec}, a hybrid approach to automated security pattern recovery, with specifications from the Van Den Berghe catalogue~\cite{van_den_berghe_reimagined_2022} as reference architectures.

  \item \textbf{A human-validated ground-truth dataset} of 100 security pattern instances covering five patterns across 93 open-source repositories: 49 Python and 44 Java projects. Each pattern is represented by ten instances per language. Each instance records role-to-code mappings, role presence statuses and an interaction sequence.

  \item \textbf{A three-level evaluation framework} that measures recovery correctness against the ground truth through role detection accuracy (RDA), mapping localization score (MLS) and interaction recovery correctness (IRC).

  \item \textbf{An empirical evaluation} using four LLMs, including proprietary and open-source models, across five patterns and two languages. We compare \textsc{SecPatRec} with a one-shot end-to-end LLM baseline, examine scalability with repository size, and assess how the code search tools affect correctness, token consumption and estimated LLM API costs.
\end{itemize}

Over the four models and all evaluated pattern instances, \textsc{SecPatRec} achieves a mean role detection accuracy of 0.84, a mean mapping localization score of 0.84 and a mean interaction recovery correctness of 0.66. It achieves higher aggregate recovery scores and lower mean estimated LLM API costs per run than the one-shot baseline, and processes repositories whose packed source exceeds the baseline's configured input limit.

\section{Security Patterns}
\label{sec:background}

Security patterns document reusable solutions to recurring security problems that occur in a particular security context ~\cite{schumacher_security_2003,schumacher_security_2006,fernandez-buglioni_security_2013}. We use the catalogue of Van Den Berghe et al.~\cite{van_den_berghe_reimagined_2022}, published online by KU Leuven's DistriNet group\footnote{\url{https://securitypatterns.distrinet-research.be/patterns/}}, because it describes patterns through a common metamodel\footnote{\url{https://securitypatterns.distrinet-research.be/metamodel/}}. Each solution specifies the participating \emph{roles}, the \emph{actions} they perform, the \emph{data elements} they exchange, and their \emph{behaviour}. Roles represent entities or specialised entity types, such as enforcement points, decision points, storage and cryptographic primitives. Behaviour is described through a walkthrough and a diagram whose numbered interactions convey ordering information. The catalogue also provides coverage mappings to security-risk lists, including the OWASP Top 10 (2021).

\begin{figure}[t]
\centering
\begin{tikzpicture}[
  r/.style={draw, rounded corners=2pt, font=\scriptsize, align=center,
            minimum width=1.5cm, minimum height=0.55cm, inner sep=2pt, fill=blue!8},
  o/.style={r, dashed, fill=white},
  e/.style={-{Stealth[length=1.6mm]}, thick},
  d/.style={e, dashed, gray!80!black},
  s/.style={font=\scriptsize, fill=white, inner sep=1pt},
  u/.style={font=\scriptsize, above=0.5pt, inner sep=1pt},
  tag/.style={font=\scriptsize, text=blue!50!black, inner sep=1pt,
              anchor=south east}
]
\node[o] (sub) at (0,0)    {Subject};
\node[r] (enf) at (2.1,0)  {Enforcer};
\node[r] (vm)  at (4.2,0)  {Verification\\manager};
\node[r] (ps)  at (7.0,2.1) {Password\\store};
\node[r] (ha)  at (7.0,1.05) {Hasher};
\node[r] (pe)  at (7.0,0)   {Pepper store};
\node[r] (en)  at (7.0,-1.05) {Encrypter};
\node[r] (co)  at (7.0,-2.1) {Comparator};
\node[o] (sys) at (2.1,-2.1) {System};
\foreach \n/\i in {enf/1, ps/3, ha/4, pe/5, en/6, co/7}
  \node[tag] at (\n.north east) {$r_\i$};
\node[tag, anchor=south west] at (vm.north west) {$r_2$};  %
\draw[d] (sub) -- node[u] {$b_1$} (enf);
\draw[e] (enf) -- node[u] {$b_2$} (vm);
\draw[e] (vm) -- node[s, pos=0.72] {$b_3$} (ps);
\draw[e] (vm) -- node[s, pos=0.72] {$b_4$} (ha);
\draw[e] (vm) -- node[s, pos=0.72] {$b_5$} (pe);
\draw[e] (vm) -- node[s, pos=0.72] {$b_6$} (en);
\draw[e] (vm) -- node[s, pos=0.72] {$b_7$} (co);
\draw[d] (enf) -- node[s] {$b_8$} (sys);
\draw[d] (sys) to[bend left=35] node[s, pos=0.45] {$b_9$} (sub);
\end{tikzpicture}
\caption{The \textit{Password-based authentication} pattern of the catalogue
\cite{van_den_berghe_reimagined_2022}, reduced to its roles $r_i$ (boxes) and the
actions $b_j$ between them (arrows), in the order of the pattern's
walkthrough.}
\Description{Nine boxes joined by nine arrows labelled b1 to b9. The seven shaded
boxes are the roles r1 to r7 and the two dashed boxes, Subject and System, are
roles that the pattern does not require. The Subject sends its identifier and
password to the Enforcer (b1). The Enforcer passes them to the Verification
manager (b2). The Verification manager then calls, in turn, the Password store for
the stored hash (b3), the Hasher to hash the received password (b4), the Pepper
store for the pepper (b5), the Encrypter to encrypt the hash under the pepper (b6)
and the Comparator to compare the result with the stored value (b7). The Enforcer
forwards the action to the System (b8), and the System may reply to the Subject
(b9).}
\label{fig:pattern-example}
\end{figure}
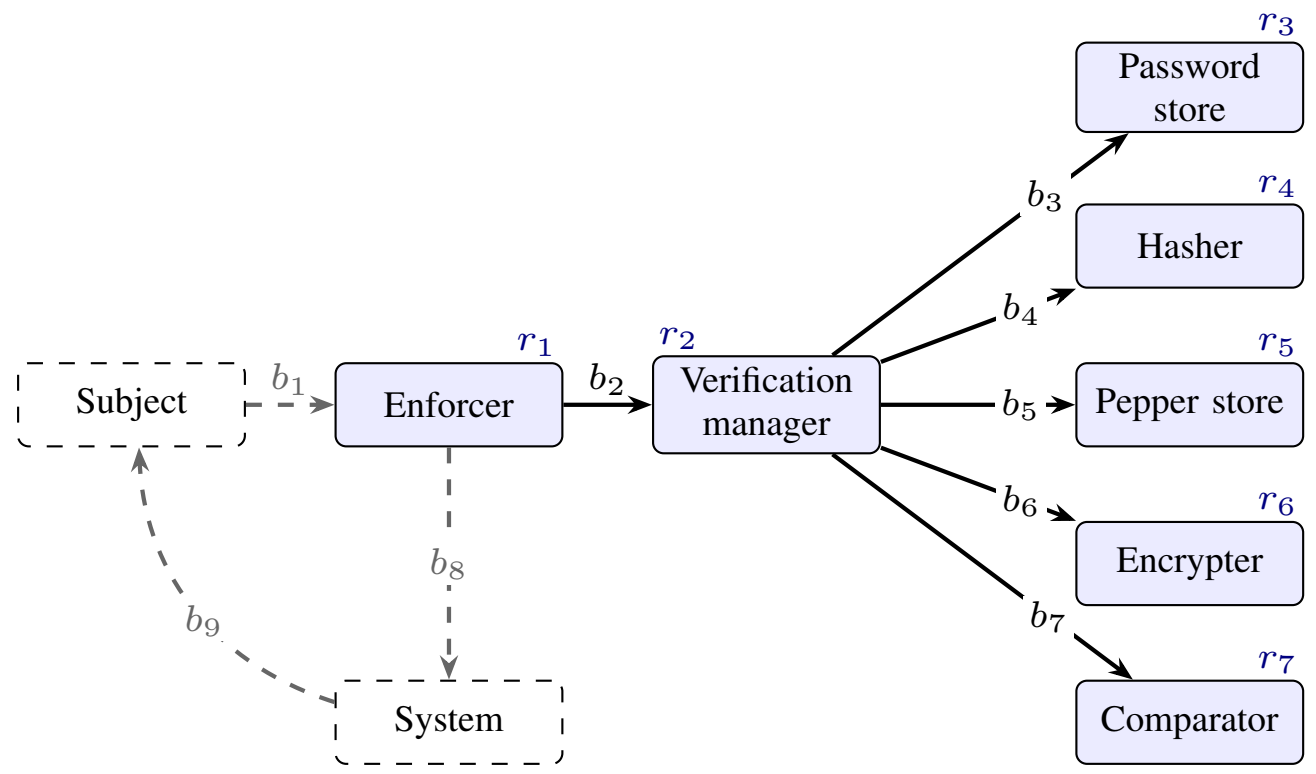

Figure~\ref{fig:pattern-example} presents an adapted view of the successful authentication flow of \textit{Password-based authentication}\footnote{\url{https://securitypatterns.distrinet-research.be/patterns/01_01_002__authentication_pwd/}}. The Subject sends an identifier and password to the Enforcer ($b_1$), which forwards them to the Verification manager ($b_2$). The Verification manager retrieves the stored password-verification value and its salt from the Password store ($b_3$), and requests a hash of the submitted password using that salt from the Hasher ($b_4$). In the peppered variant illustrated by the catalogue, it then retrieves the pepper ($b_5$), uses the Encrypter to encrypt the calculated hash with the pepper as the cryptographic key ($b_6$). The verification manager sends the resulting ciphertext and the stored password-verification value to the Comparator, which compares them with the stored value and returns the verification result ($b_7$). Following successful verification, the Enforcer forwards the request to the System ($b_8$), which may reply to the Subject ($b_9$). Intermediate replies and the failure branch are omitted from the figure.

For recovery, we retain the seven shaded roles and the interactions between them ($b_2$--$b_7$). Subject and System, together with $b_1$, $b_8$ and $b_9$, provide the surrounding context and are outside this recovery specification. Registration and password-reset behaviour described in the catalogue are also outside the selected flow. Details of flow selection are provided in Appendix~\ref{app:adl}. Roles excluded when constructing the reference specification are outside scope; they are distinct from retained roles for which no implementation is identified.

The catalogue describes security responsibilities and interactions without prescribing a unique code structure. Roles do not always correspond to distinct classes, functions or modules, and a library operation may realise several roles. For example, Figure~\ref{fig:code-example} maps the Enforcer, Verification manager and Password store to three functions in a simplified FastAPI login handler, and maps the Comparator to an externally defined \texttt{passlib} password-verification call. Additional responsibilities, such as password hashing, may be encapsulated by the same call. The handler contains no explicit peppering step. Recovering a role therefore requires interpreting the available code and library references in context~\cite{vanhilst_reverse_2007,behrens_what_2018}. Section~\ref{sec:method} formalises the recovery outputs and their presence statuses.

\section{Automated Security Pattern Recovery}
\label{sec:method}
\subsection{Problem Formulation}
\label{sec:refarch}

In this paper, our goal is to recover the architectural components of security patterns from existing codebases. The target catalogue is the Van den Berghe's developed by KU Leuven's Distrinet group \cite{van_den_berghe_reimagined_2022}. We observe that security patterns in the Van den Berghe's catalogue express both the structural properties and the behavioral properties of a security architecture \cite{kazman_software_2021, wirfs-brock_object_2002}. As a result, mining a reference security pattern from a codebase can be approached as a deductive software architecture recovery (DSAR) problem \cite{rukmono_deductive_2024,rukmono_evaluating_2026}, focusing on the tasks of matching architectural roles that the pattern prescribes to specific classes or functions, and to derive the sequence of interactions between the matched roles. We formulate this problem as follow:

A security pattern, or pattern is specified as a tuple $P = (R, B)$, where:
\begin{itemize}
    \item \textbf{Roles ($R$):} A vocabulary $R = \{r_1, \ldots, r_m\}$ of prescribed architectural entities. For example, a password-based authentication pattern might prescribe $R = \{\text{\emph{Enforcer}}, \text{\emph{Verification manager}}, \text{\emph{Hasher}}, \text{\emph{Password store}}\}$.
    \item \textbf{Actions ($B$):} An ordered sequence $B = \langle b_1, \ldots, b_n \rangle$ dictating the behavioral flow between those roles.
\end{itemize}

The pattern itselfs can serves as the reference architecture to be recovered against a codebase $C$. Recovering $P$ from $C$ requires three tasks: (1) detect which roles of $P$ are present in $C$, by identifying the presence status of each role that composes the pattern, (2) map each prescribed role $r \in R$ to a concrete code symbol $e \in \mathcal{E}(C)$ where $\mathcal{E}(C)$ represents the union of all Abstract Syntax Trees (AST) nodes inside the codebase, and (3) recover the sequence of interactions between those mapped symbols. We denote the recovered architecture as a tuple $(M, \mathcal{D}, S)$, where:
\begin{enumerate}
    \item \textbf{Role mappings ($M$):} A set of pairs $m = (r, e)$ with $r \in R$, each mapping a prescribed role to the code symbol $e$ that implements it, where $e$ is either defined in $C$, i.e., $e \in \mathcal{E}(C)$, or only referenced by $C$. 
    \item \textbf{Presence statuses ($\mathcal{D}$):} The set $\mathcal{D} = \{\delta(r) : r \in R\}$ of presence statuses, one for each role. $\delta(r) = \textsc{present}$ when there exists $m \in M$ that maps $r$ to some $e \in \mathcal{E}(C)$. $\delta(r) = \textsc{external}$ when there exists $m \in M$ that maps $r$ to some $e$ whose definition cannot be found in any AST of $C$, such as a library function that $C$ only calls. $\delta(r) = \textsc{missing}$ when there exists no $m \in M$ that maps $r$ to any $e$.
    \item \textbf{An interaction sequence ($S$):} The ordered sequence of actual interactions observed between the mapped symbols in $M$.
\end{enumerate}

Table~\ref{tab:notation} summarizes this formulation.  Figure~\ref{fig:code-example} illustrates the recovery of the pattern from a small login handler that implement the password-based authentication pattern mentioned in Figure~\ref{fig:pattern-example}. The quality of an automated security pattern recovery solution depends on how correctly it can produce the mentioned above outputs. To evaluate these level of correctness, we need to not only define the corresponding set of metrics that align with outputs of the recovery, but also a cureated groundtruth sets of the role mapping $M$, the presence statuses $\mathcal{D}$ and the interaction sequence $S$ that are collected from the original software architecture. This is a chalenging task, as groundtruth architectures in real-world open-source software are scarce, requiring involvement of human expertise to generate them \cite{garcia_comparative_2013,garcia_obtaining_2013}.
\newcommand{\tikzmark}[1]{\tikz[baseline,remember picture]{%
  \node[anchor=base,inner sep=0pt,outer sep=0pt] (#1) {};}}
\newcommand{\delboxes}{\tikz[remember picture,overlay]{%
  \tikzset{
    delpres/.style={draw=green!50!black, draw opacity=0.8, fill=green!40,
                    fill opacity=0.16, rounded corners=1.5pt, thin},
    delext/.style={draw=violet!80!black, draw opacity=0.9, dashed, fill=violet!40,
                   fill opacity=0.16, rounded corners=1.5pt, thin},
    deltag/.style={font=\normalfont\scriptsize, inner sep=0pt}}
  \draw[delpres] ([shift={(-2pt,6.5pt)}]gu-s) rectangle ([shift={(2pt,-2.5pt)}]R |- gu-e);
  \draw[delpres] ([shift={(-2pt,6.5pt)}]au-s) rectangle ([shift={(2pt,-2.5pt)}]R |- au-e);
  \draw[delpres] ([shift={(-2pt,6.5pt)}]lo-s) rectangle ([shift={(2pt,-2.5pt)}]R |- lo-e);
  \draw[delext]  ([shift={(-1.5pt,6pt)}]ex-s) rectangle ([shift={(1.5pt,-2.5pt)}]ex-e);
  \node[deltag, anchor=south east, text=green!40!black]
    at ([shift={(0pt,-1.5pt)}]R |- gu-e) {$\delta(r_3)$ = \textsc{present}};
  \node[deltag, anchor=south east, text=green!40!black]
    at ([shift={(0pt,-1.5pt)}]R |- au-e) {$\delta(r_2)$ = \textsc{present}};
  \node[deltag, anchor=south east, text=green!40!black]
    at ([shift={(0pt,-1.5pt)}]R |- lo-e) {$\delta(r_1)$ = \textsc{present}};
  \node[deltag, anchor=north east, text=violet!70!black]
    at ([shift={(1.5pt,-3pt)}]ex-e) {$\delta(r_7)$ = \textsc{external}};
}}

\begin{figure*}[t]
\centering
\begin{minipage}[t]{0.45\textwidth}
\vspace{0pt}
\begin{lstlisting}[style=code,language=Python]
from fastapi import Depends, FastAPI
from passlib.context import CryptContext
app = FastAPI()
pwd = CryptContext(schemes=["bcrypt"])

(*@\tikzmark{gu-s}@*)def get_user(db, name):
    return db.get(name)(*@\tikzmark{gu-e}@*)

(*@\tikzmark{au-s}@*)def authenticate(db, name, password):
    user = get_user(db, name)
    if user and (*@\tikzmark{ex-s}@*)pwd.verify(password, user.hash)(*@\tikzmark{ex-e}@*):
        return user
    return None(*@\tikzmark{au-e}@*)

(*@\tikzmark{lo-s}@*)@app.post("/login")
def login(form: LoginForm, db=Depends(get_db)):
    user = authenticate(db, form.name, form.password)(*@\tikzmark{R}@*)
    return user(*@\tikzmark{lo-e}\delboxes@*)
\end{lstlisting}
\end{minipage}\hfill
\begin{minipage}[t]{0.53\textwidth}
\vspace{0pt}\centering
\begin{tikzpicture}[
  role/.style={draw, rounded corners=2pt, font=\scriptsize, align=center,
               minimum width=1.95cm, minimum height=0.62cm, fill=blue!8},
  code/.style={draw, font=\ttfamily\scriptsize, align=center,
               minimum width=1.95cm, minimum height=0.62cm, fill=white},
  ext/.style={code, dashed},
  rstep/.style={-{Stealth[length=1.5mm]}, dashed, gray!80!black, thick},
  plays/.style={dotted, thick, blue!60!black},
  call/.style={-{Stealth[length=1.7mm]}, thick, blue!60!black},
  num/.style={circle, fill=orange!85!black, text=white, font=\scriptsize\bfseries,
              inner sep=1pt},
  mtag/.style={font=\scriptsize, text=blue!60!black, fill=white, inner sep=1pt},
  rtag/.style={font=\scriptsize, text=blue!50!black, inner sep=1pt}
]
\node[role] (r1) at (0,2.3) {Enforcer};
\node[role] (r2) at (2.45,2.3) {Verification\\manager};
\node[role] (r3) at (4.9,2.3) {Password\\store};
\node[role] (r4) at (7.35,2.3) {Comparator};
\node[code] (c1) at (0,0) {METHOD\\login};
\node[code] (c2) at (2.45,0) {METHOD\\authenticate};
\node[code] (c3) at (4.9,0) {METHOD\\get\_user};
\node[ext]  (c4) at (7.35,0) {CALL\\pwd.verify};
\foreach \n/\i in {r1/1, r3/3, r4/7}
  \node[rtag, anchor=south east] at (\n.north east) {$r_\i$};
\node[rtag, anchor=south west] at (r2.north west) {$r_2$};
\draw[rstep] (r1) -- (r2);
\draw[rstep] (r2) -- (r3);
\draw[rstep] (r2.north) to[bend left=30] (r4.north);
\foreach \a/\b/\k in {r1/c1/1, r2/c2/2, r3/c3/3, r4/c4/4}
  \draw[plays] (\a) -- node[mtag] {$m_\k$} (\b);
\draw[call] (c1) -- node[num] {$s_1$} (c2);
\draw[call] (c2) -- node[num] {$s_2$} (c3);
\draw[call] (c2.south) to[bend right=28] node[num, pos=0.5] {$s_3$} (c4.south);
\end{tikzpicture}
\end{minipage}
\caption{A login handler (left) and the architecture recovered from it for the
password-based authentication pattern (right): roles $r_i$, role mapping $m_i$
(dotted), pattern actions (dashed) and interaction sequence $s_j$ (solid).
Boxes in the code give $\delta(r_i)$, dashed for \textsc{external}. The handler is simplified from the FastAPI documentation \cite{fastapi}.}
\Description{On the left, a short Python listing: a login route that calls
authenticate, which calls get_user and then the library function pwd.verify. On the
right, four shaded role boxes, Enforcer, Verification manager, Password store and
Comparator, each joined by a dotted line (m1 to m4) to the code symbol below it: login,
authenticate, get_user and the call to pwd.verify, which has a dashed border. Dashed
arrows join the roles in the order of the pattern, and three solid arrows labelled
s1 to s3 between the code symbols show the order of the calls: login to authenticate,
authenticate to get_user, and authenticate to pwd.verify. In the listing, light
solid boxes labelled delta of r1, r2 and r3 equals present mark the three functions,
and a light dashed box labelled delta of r7 equals external marks the pwd.verify
call. The role boxes carry the indices r1, r2, r3 and r7 of Figure 1.}
\label{fig:code-example}
\end{figure*}
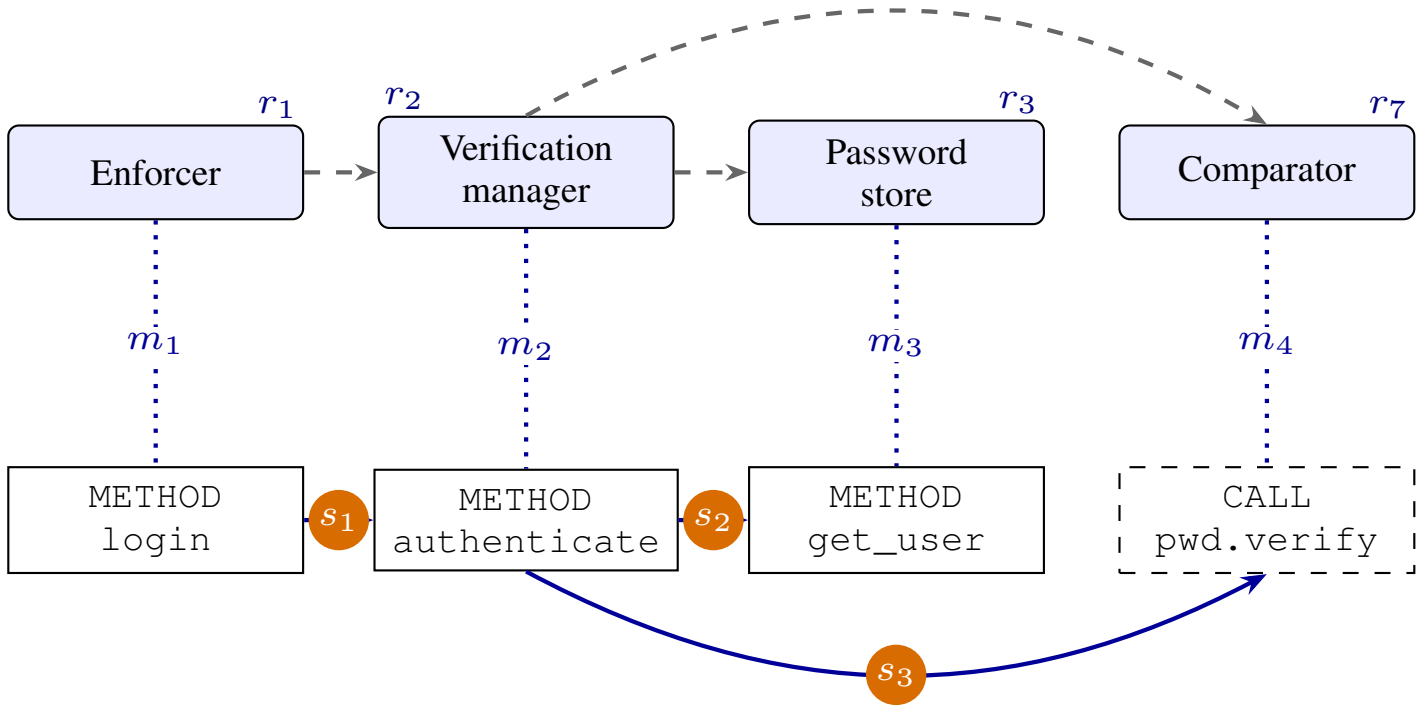

\begin{table}[t]
\caption{Notation.}
\label{tab:notation}
\small
\setlength{\tabcolsep}{4pt}
\begin{tabular}{@{}l>{\raggedright\arraybackslash}p{0.60\columnwidth}@{}}
\toprule
Symbol & Meaning \\
\midrule
\multicolumn{2}{@{}l}{\emph{Pattern specification}}\\
$P = (R, B)$ & pattern specification \\
$R$, $B$ & prescribed roles; their actions, in order \\
\addlinespace
\multicolumn{2}{@{}l}{\emph{Codebase}}\\
$C$, $\mathcal{E}(C)$ & codebase, and its named code symbols extracted from union of all ASTs\\
\addlinespace
\multicolumn{2}{@{}l}{\emph{Recovered Architecture}}\\
$M$, $\mathcal{D}$, $S$ & ground truth: role mapping, presence statuses, interaction sequence \\
\bottomrule
\end{tabular}
\end{table}

\subsection{SecPatRec: Automated Approach for Deductive Security Pattern Recovery}
\label{sec:method-pipeline}
Our goal is to design an automated approach that can recover the architectural components of security patterns from existing codebases, including the role mapping $M$, the presence statuses $\mathcal{D}$ and the interaction sequence $S$, using the pattern specification $P$ as the reference architecture for DSAR. Previous works in DSAR have shown that LLMs can be used for code unit classification effectively \cite{rukmono_deductive_2024,rukmono_evaluating_2026}, which is suitable for the task of generating candidate role mappings $\hat{M}$, and consequently the presence statuses $\hat{\mathcal{D}}$ of the roles. However, for large real-world codebases, chunking all code symbols in $\mathcal{E}(C)$ to feed into LLMs for batched classification tasks could be computationally expensive and time-consuming \cite{xu_large_2025}, and at the same time losing repository-level global contexts as code symbols have been fragmented and processed individually \cite{le-anh_not_2026}. Also, individual code unit classification can not be adopted for the next task of security pattern recovery, which is recovering a sequence of behaviorial interactions between the mapped roles. 

\begin{figure*}[t]
\centering
\resizebox{\textwidth}{!}{%
\begin{tikzpicture}[
  font=\scriptsize,
  comp/.style={draw=teal!70!black, fill=teal!22, thick, rounded corners=2pt,
               align=center, minimum width=3.0cm, minimum height=0.95cm},
  data/.style={draw=gray!70!black, fill=white, thick, align=center,
               trapezium, trapezium left angle=72, trapezium right angle=108,
               inner xsep=0.2cm, minimum width=3.0cm, minimum height=0.95cm},
  tool/.style={draw=gray!60!black, fill=white, rounded corners=1pt, align=center,
               minimum width=1.6cm, minimum height=0.75cm},
  flow/.style={-{Stealth[length=1.8mm]}, thick, gray!55!black},
  off/.style={flow, dashed},
  call/.style={-{Stealth[length=1.8mm]}, thick, blue!60!black, densely dotted},
  note/.style={font=\scriptsize\itshape, text=gray!50!black}
]
\node[data] (refp) at (0,0)    {Reference\\security pattern};
\node[comp] (adlg) at (4.4,0)  {ADL code\\generator};
\node[data] (spec) at (8.8,0)  {Specification\\$P = (R, B)$};
\draw[off] (refp.east) -- (adlg.west);
\draw[off] (adlg.east) -- (spec.west);
\node[note] at (4.4,0.85) {offline, once per pattern};
\node[data, minimum width=1.9cm] (prompt) at (8.8,-1.9) {Prompt};
\draw[flow] (spec.south) -- (prompt.north);
\node[note, anchor=west] at (8.95,-0.95) {inserted into};
\node[data, minimum width=2.2cm] (code) at (0,-3.9) {Codebase $C$};
\node[draw=gray!60!black, rounded corners=2pt, minimum width=3.0cm,
      minimum height=2.1cm] (tools) at (4.4,-3.9) {};
\node[note, anchor=north west] at ([shift={(0.1,-0.04)}]tools.north west)
  {search methods};
\node[tool, minimum width=2.4cm, minimum height=0.65cm] at (4.4,-3.6) {Lexical search};
\node[tool, minimum width=2.4cm, minimum height=0.65cm] at (4.4,-4.4) {Semantic search};
\node[comp] (llm)   at (8.8,-3.9)  {LLM};
\node[data] (map)   at (13.2,-3.9) {Role mapping $\hat{M}$\\Presence status $\hat{\mathcal{D}}$};
\node[data] (final) at (17.6,-3.9) {Recovered architecture\\$(\hat{M}, \hat{\mathcal{D}}, \hat{S})$};
\draw[flow] (prompt.south) -- (llm.north);
\draw[flow] (llm.east) -- (map.west);
\draw[flow] (map.east) -- (final.west);
\draw[call] (llm.west) -- (tools.east);
\draw[call] (tools.west) -- (code.east);
\node[note, text=blue!60!black, below] at (6.7,-3.9) {tool call};
\draw[call] (code.north) -- (0,-2.5) -- (7.7,-2.5) -- ([xshift=-1.1cm]llm.north);
\node[note, text=blue!60!black, above] at (3.85,-2.5) {code context};
\node[comp] (joern) at (4.4,-6.2)  {Joern\\(CPG construction)};
\node[data] (cpg)   at (8.8,-6.2)  {Code property\\graph (CPG)};
\node[comp] (rec)   at (13.2,-6.2) {Interaction recoverer\\(CPG traversal)};
\node[data] (seq)   at (17.6,-6.2) {Interaction sequence\\$\hat{S}$};
\draw[off]  (code.south) |- (joern.west);
\draw[off]  (joern.east) -- (cpg.west);
\draw[flow] (cpg.east) -- (rec.west);
\draw[flow] (map.south) -- (rec.north);
\draw[flow] (rec.east) -- (seq.west);
\draw[flow] (seq.north) -- (final.south);
\node[note] at (6.6,-7.05) {offline, once per codebase};
\node[comp, minimum width=0.6cm, minimum height=0.3cm] at (12.3,0.45) {};
\node[anchor=west] at (12.7,0.45) {Pipeline component};
\node[data, minimum width=0.6cm, minimum height=0.3cm, inner xsep=0.1cm]
  at (12.3,-0.15) {};  %
\node[anchor=west] at (12.7,-0.15) {Data};
\draw[flow] (15.4,0.45) -- (16.0,0.45);
\node[anchor=west] at (16.1,0.45) {Flow};
\draw[off] (15.4,-0.15) -- (16.0,-0.15);
\node[anchor=west] at (16.1,-0.15) {Offline step};
\draw[call] (15.4,-0.75) -- (16.0,-0.75);
\node[anchor=west] at (16.1,-0.75) {Tool call, code context};
\end{tikzpicture}%
}
\caption{Overview of \textsc{SecPatRec}. The specification $P$ is inserted into the
prompt of the LLM, which iteratively searches the codebase $C$ and produces the role mapping
$\hat{M}$ and the presence statuses $\hat{\mathcal{D}}$. The CPG of $C$ is built
offline with Joern, and the interaction recoverer traverses it from the mapped
symbols to produce the interaction sequence $\hat{S}$.}
\Description{A pipeline in three rows. In the top row, a reference security pattern
is turned offline by an ADL code generator into a specification of roles and actions,
which is inserted into a prompt. The prompt is the only input of a ReAct language
model. In the middle row, the language model calls two search methods, lexical search
and semantic search, over the codebase, and the codebase returns the code context to
the language model. The language model produces the role mapping and the presence
statuses. In the bottom row, Joern builds the code property graph of the codebase
offline, and an interaction recoverer traverses it, starting from the role mapping, to
produce the interaction sequence. The role mapping, the presence statuses and the
interaction sequence together form the recovered architecture. Offline steps are
drawn with dashed arrows, and tool calls and code context with dotted arrows.}
\label{fig:secpatrec}
\end{figure*}
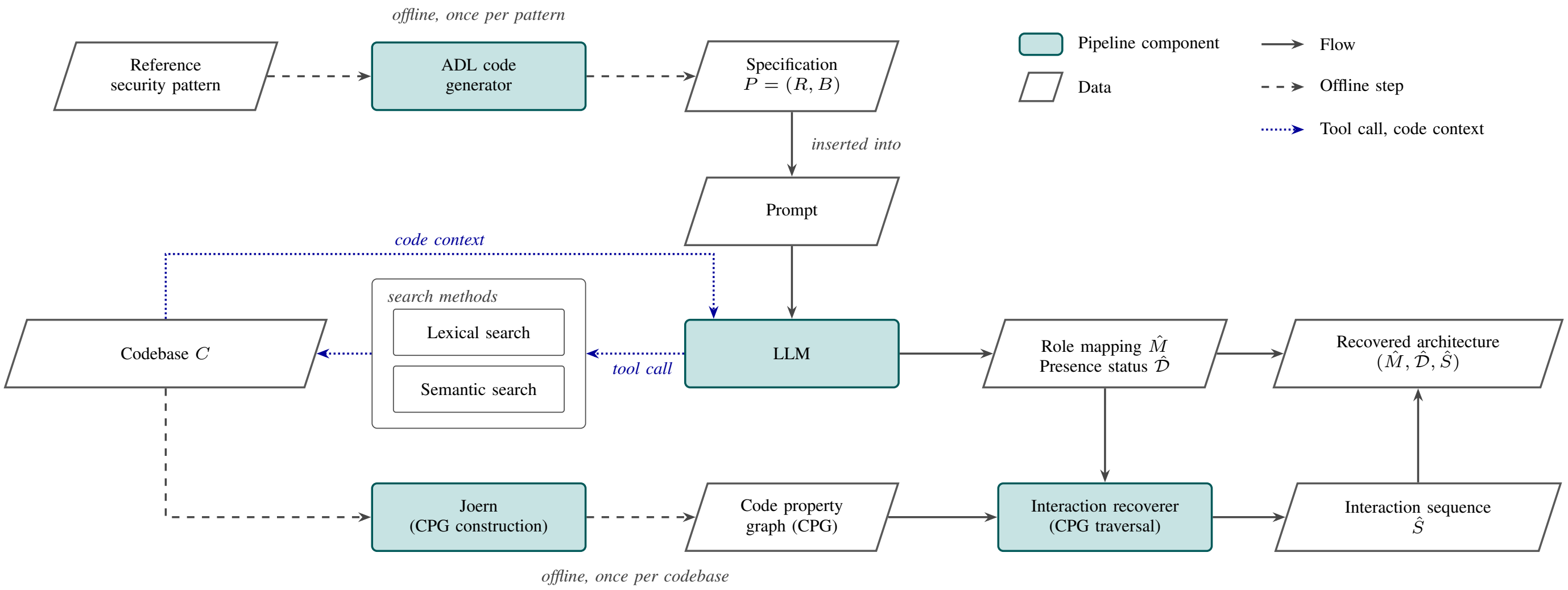

To address the above challenges, we propose a hybrid approach that combines LLMs with code search methods and static code property graph (CPG) analysis \cite{yamaguchi_modeling_2014} to recover the architectural components of Van De Berg's security patterns from existing codebases. LLM that used chain-of-thought reasoning with code search as external tools is suitable for the tasks of security pattern roles' detection and mapping, which requires iterative localization of relevant code contexts from a large source global semantic information\cite{nguyen_sparecodesearch_2025,le-anh_not_2026}, while CPG analysis provides a deterministic way to bridge the mapped roles and their interactions. The overall design of our approach is illustrated in Figure~\ref{fig:secpatrec}. 
\subsubsection{Security Pattern Role Mapping and Presence Detection}
\label{sec:method-roles}

\textbf{Translate Pattern Specification into Architecture Description Language Code} The pattern specification $P$ from the Van De Berg\cite{van_den_berghe_reimagined_2022},is documented as textual descriptions in Markdown format and PNG images of their corresponding UML diagrams. We want to translate these information into a reference architecture description language (ADL) code to guide the LLM chain-of-thought reasoning in the role mapping and presence detection tasks. This translation is done by our developed ADL code generator, which provides Mermaid and JSON code representation of the prescribed roles and their interactions. Details on how we implement this translation are provided in Appendix~\ref{app:adl}. The generated ADL code is then used to construct a prompt for the LLM to perform the role mapping and presence detection tasks. 

\textbf{Iterative Tool-assisted CoT Role Detection and Mapping} We perform role detection and mapping with ReAct~\cite{yao_react_2023}, an iterative reasoning method that lets an LLM interleave chain-of-thought with calls to external tools. At each iteration, the LLM first writes a \emph{Thought} about what it has learned so far from a codebase and what to look for next, then issues one \emph{Action}, i.e., a call to external code search tools, or trigger particular file reads and finally reads the \emph{Observation}, a code context returned by that specific tool call. The iterations continue until the LLM has detected and mapped every role of $P$. The system prompts for this process is inserted with the previously translated JSON ADL code to guide the reasoning. 

The LLM explores the codebase $C$ through two search methods. \emph{Semantic search} takes a description in natural language and returns a code symbols in $C$. We vectorize every symbol with BGE-small~\cite{xiao_c-pack_2024}\footnote{Checkpoint \texttt{BAAI/bge-small-en-v1.5}, \url{https://huggingface.co/BAAI/bge-small-en-v1.5}.}, and then store them in ChromaDB for later retrieval triggers. \emph{Lexical search}, instead, finds a literal string or a regular expression anywhere in the repository and returns the matching lines with their file and line number. The combination of 2 search methods allows the LLM to both semantically and syntatically locate and analyze relevant code elements\cite{nguyen_sparecodesearch_2025,nguyen_vault_2023}. The LLM can also list directories and read files to inspect what code searches return.The final output of this task is a structural set of  role mappings $\hat{M}$ and theoir presence statuses $\hat{\mathcal{D}}$, which are passed on to the next task of recovering the interactions. Outputs of this stage are JSON files that contain the role mapping $\hat{M}$ and the presence statuses $\hat{\mathcal{D}}$ of the roles, formatted as in Figure~\ref{fig:role-mapping-json}a in Appendix~\ref{app:output}.

\subsubsection{Interaction Sequence Recovery}
\label{sec:interaction-sequence-recovery}

\textbf{Constructing the Code Property Graph} Once the role mapping $\hat{M}$ and presence statuses $\hat{\mathcal{D}}$ are obtained, we construct the CPG of the targeted codebase $C$ using Joern\footnote{\url{https://joern.io}}, a Scala-based platform that can model source code for security-relevant tasks \cite{yamaguchi_modeling_2014}. The CPG captures the structural and behavioral relationships between code symbols, allowing us to analyze the interactions between the mapped roles.

\textbf{Traverse the CPG to Recover Interactions between Mapped Roles} We define our own traversal rules to derive the calling relationships between the mapped code symbol entities $\hat{e}$ corresponding to every $\hat{m}$ in $\hat{M}$, and recover the interaction sequence $\hat{S}$ (Figure~\ref{fig:role-mapping-json}b) based on the order of the calls observed in the CPG. Details of how we construct the traversal rules are provided in the Appendix~\ref{app:order}, and the Scala implementation is available in our repository\footnote{\url{https://github.com/anonymousspottedtoad/secpat_interaction_recovery}}. 

\section{Experiment Setup}

\label{sec:setup}

\subsection{Research Questions}
\label{sec:setup-rqs}
To evaluate our automated security pattern recovery approach, we want to understand how well it performs in terms of recovery correctness, costs of computing resources and token consumption, under different LLMs and code search methods.  We also want to valide our assumption that using CPG analysis is more efficient and scalable compared to an end-to-end baseline that use LLM only. We define following research questions (RQs) to guide our evaluation:

\rqbox{1}{Which LLMs has the best \textbf{task-specific performance} for security pattern recovery?}
\rqbox{2}{Which LLMs is the most \textbf{cost-effective} for security pattern recovery tasks when integrated with CPG analysis in our pipeline, in terms of \textbf{token consumption}?}
\rqbox{3}{How does each \textbf{LLM}, when integrated with CPG analysis in our pipeline, can scale across \textbf{diffrent patterns} and \textbf{codebase sizes}, compared to their \textbf{end-to-end LLM baseline} counterparts, in terms of performance and computing resource usage?}
\rqbox{4}{How do \textbf{code search methods} affect the quality of the recovered architecture and the overall computing cost of running security pattern recovery tasks?}

To answer the above questions, we had to first establish a groundtruth dataset of security pattern architectures $M$, $D$, $S$, and at the same time define performance indicators to measure the correctness of our approach in terms of task-specific metrics for security pattern recovery. The details of the groundtruth dataset and the evaluation metrics are described below
\subsection{Groundtruth Dataset of Security Pattern Architectures}
\label{sec:setup-groundtruth}

\subsubsection{Choosing Patterns for Reference and Codebases for Recovery}
The pattern-problem pairs in Van De Berg's catalogue are developed with the aims to cover OWASP Top 10 security risks in web application \footnote{\url{https://securitypatterns.distrinet-research.be/coverage/}}. As a result, we want to use real-world open-source codebase that implements security control for web applications for validation our approach. We use our custom tool \footnote{\url{https://github.com/username/repo}} to automate the cloning of 93 repositories from Github. The choice of repositories is based on the condition that they depend on 2 corresponding web security frameworks: Python’s FastAPI and Java’s Springboot Security. The selection of FastAPI and Spring Boot was driven by their widespread adoption; as of 2025, they are used by 38\% and 65\% of their respective developer communities\cite{jetbrains_state_nodate,jetbrains_state_2025}.  The target patterns used as reference architectures are P1 Password-based authentication, P2 Verifiable-token authentication, P3 Obscure-token access control, P4 Session-based access control and P5 Rate limiting. They all belong the set of security solutions that tackle the security problems related to OWASP's Top 10 2025: A01: Broken Access Control and A07: Authentication Failures \cite{owasp_owasp_nodate}.

Details statistics of the selected codebases and the corresponding security patterns are provided in Table~\ref{tab:dataset-size} and Figure~\ref{fig:boxplotdatasize} in Appendix~\ref{app:annotation-stats}, where code size is measured in KLOC and in tokens of OpenAI's \texttt{o200k\_base} tokenizer\footnote{\url{https://github.com/openai/tiktoken}}. Our source of target codebases for recovery spans 49 repositories in Python and 44 repositories in Java. We preprocess these repositories to collect only .java and .py sources files  with a total of 272,137 methods and 54,785 classes extracted from their ASTs. We also construct the CPGs of these codebases using Joern, consuming a total of 818.5 MB of storage space. Our repositories are diverse in terms of their size, making it suitable for evaluating the scalability and performance of our approach. 

\begin{table}[t]
  \centering
  \caption{Size of the 93 codebases per language.}
  \label{tab:dataset-size}
  \small
  \setlength{\tabcolsep}{3pt}
   \resizebox{\columnwidth}{!}{%
   \begin{tabular}{lrrrrrrr}
     \toprule
      & \#Repos & KLOC & \#Methods & \#Classes & \#Tokens & \#Files & CPG (MB) \\
     \midrule
    Java   & 44 & 3,628.2 & 179,975 & 37,478 & 32.12M & 32,628 & 228.3 \\
    Python & 49 & 2,555.4 &  92,162 & 17,307 & 20.89M & 10,622 & 590.2 \\
     \midrule
    Total  & 93 & 6,183.6 & 272,137 & 54,785 & 53.01M & 43,250 & 818.5 \\
     \bottomrule
   \end{tabular}}
\end{table}
\subsubsection{Constructing Groundtruths}
Obtaining groundtruth for software architecture recovery is inherently difficult, as knowledge of the system architects who designed the architecture is rarely available \cite{garcia_obtaining_2013,garcia_comparative_2013}. As a result, previous works\cite{lutellier_measuring_2018,garcia_obtaining_2013,garcia_comparative_2013} had to rely on \textit{authorative protocol}, in which an automated process is used to generate a set of \textit{preliminary architectures}, and then external human engineers are used to certify them. In this work, we follow a similar approach, but instead use Claude Code to generate a the preliminary role mappings $M$ and presence statuses $\mathcal{D}$, then have human experts verify them. This human-AI collaboration annotation approach is also used in previous software engineering and mining software repositories works\cite{kim_meganno_2024,ahmed_can_2025}. We share our working session of how we use Claude Code to generate preliminary mappings in Appendix~\ref{app:annotation}. We then traverse the CPGs to deterministically recover the preliminary interaction sequence $S$ among provided preliminary $M$, using the same traversal rule as mentioned in \ref{sec:interaction-sequence-recovery}.

To help human annotators verify the preliminary architectures, we provide them with a web-based interface that allows them to inspect the preliminary role mappings and presence statuses, visualizing the candidate interaction sequence recovered, allowing annotators to trace the interactions between the mapped roles and code entities. Annotators are provided with the original Van de Berg's catalogue for reference. They can approve or reject, or modify each preliminary roles mapping and interaction themselves if they think that existing recovery do not reflect the design that the orginal pattern intended. They can also map a single codebase with more than one security patterns, as in real-world open-source projects, developers can apply multiple security solutions to a single codebase \cite{behrens_what_2018}. We have two annotators, both of whom are senior software engineers that has more than 4 years of web application development. Example of how we visualize the candidate groundtruths for human verification is provided in Appendix~\ref{app:annotation}.

After the annotators have finished their verification, we sampled a list of 100 groundtruth architectures, each of which contains a human-verified role-to-code mappings $M$, role presence statuses $\mathcal{D}$ and an interaction sequence $S$ between the mapped roles. In total, the 100 groundtruth architectures contain 500 roles, of which 334 are present, 75 external and 91 missing, and 280 interactions; Table~\ref{tab:ground-truth} gives them per pattern and language. We publish the dataset in an anonymized Hugging Face bucket for reuse and reproducibility\footnote{\url{https://huggingface.co/buckets/AnonymousSpottedToad/securityPatternsRecovery}}. 

\begin{table}[t]
\caption{Groundtruth roles by presence status and interactions per pattern, for Python and Java (10 codebases each).}
\label{tab:ground-truth}
\centering
\small
\setlength{\tabcolsep}{3.2pt}
\begin{tabular}{lrrrrrrrr}
\toprule
& \multicolumn{4}{c}{Python} & \multicolumn{4}{c}{Java} \\
\cmidrule(lr){2-5}\cmidrule(lr){6-9}
Pattern & Pres. & Ext. & Miss. & Int. & Pres. & Ext. & Miss. & Int. \\
\midrule
P1 & 32 & 17 & 21 & 31 & 33 & 14 & 23 & 39 \\
P2 & 31 & 9 & 0 & 24 & 28 & 12 & 0 & 27 \\
P3 & 33 & 0 & 7 & 19 & 30 & 6 & 4 & 28 \\
P4 & 42 & 0 & 18 & 19 & 31 & 11 & 18 & 25 \\
P5 & 40 & 0 & 0 & 34 & 34 & 6 & 0 & 34 \\
\midrule
Total & 178 & 26 & 46 & 127 & 156 & 49 & 45 & 153 \\
\bottomrule
\end{tabular}
\end{table}

\subsection{Evaluation Metrics}
\label{sec:setup-evaluation}
For each task of security pattern recovery, we define task-specific performance indicator to measure the correctness of our approach.
We define one metric for each output of the recovery, comparing the presence statuses $\hat{\mathcal{D}}$ with $\mathcal{D}$, the role mapping $\hat{M}$ with $M$, and the interaction sequence $\hat{S}$ with $S$. We call them role detection accuracy (\textbf{RDA}), mapping localization score (\textbf{MLS}) and interaction recovery correctness (\textbf{IRC}). A metric is computed for one recovery instance, i.e., one pattern recovered from one codebase, and is then averaged over all instances. The implmentation of our evaluation metrics is available in our repository\footnote{\url{https://github.com/anonymousspottedtoad/secpat_evaluation_metrics}}. Appendix~\ref{app:irc-example} shows the three metrics on a small example (Figures~\ref{fig:metrics} and~\ref{fig:l3}).

\subsubsection{Role Detection Accuracy}
Each role of the pattern has one of three statuses: \textsc{present}, \textsc{external} or  \textsc{missing}. Role detection accuracy is the fraction of roles for which the LLM outputs the same presence status $\delta(r)$ compared to the groundtruth $\delta(r)$, so the problem is translated into a multi-label classification problem over the closed label set $R$

\begin{equation}
  \mathrm{RDA} = \frac{\bigl|\{\, r \in R : \hat{\delta}(r) = \delta(r) \,\}\bigr|}{|R|}.
  \label{eq:l1}
\end{equation}

\subsubsection{Mapping Localization Score}
For every role founded by the both the LLM and the groundtruth, i.e $r \in F$ where $F = \{\, r \in R : \delta(r) \neq \textsc{missing} \text{ and } \hat{\delta}(r) \neq \textsc{missing} \,\}$, we want to measure how close the recovered symbol $\hat{e}_r$ that the LLM maps to $r$, compared with the groundtruth symbol $e_r$. We therefore measure how far apart the two symbols are in the static containment tree $T_(C)$ of the codebase, in the same way as fault localization is judged by the distance between the reported and the actual faulty location in the program dependence graph \cite{cleve_locating_2005}: the closer the two symbols, the better the mapping localizes the role in the code. $T_(C)$ is constructed by agrregating all ASTs of source files in the codebase. The distance $d$ is the number of edges required for moving $e_r$ to $\hat{e}_r$ in the same $T_{(C)}$.
\begin{equation}
  \begin{split}
    d(\hat{e}_r, e_r) ={}& \mathrm{depth}(\hat{e}_r) + \mathrm{depth}(e_r) \\
    & - 2\,\mathrm{depth}\bigl(\mathrm{LCA}(\hat{e}_r, e_r)\bigr).
  \end{split}
  \label{eq:l2-distance}
\end{equation}
where LCA is the lowest common ancestor of the two symbols in the containment tree. We normalize this distance by the height $2H$ of the containment tree as no path in a tree of height $H$ is longer than $2H$ edges, then score the tree-induced similarity as $1 - d/2H$. The mapping localization score is the mean score over all roles in $F$, and it is 0 when no role is found by both sides ($|F| = 0$), as then nothing is mapped correctly:
\begin{equation}
  \mathrm{MLS} =
  \begin{cases}
    \dfrac{1}{|F|} \displaystyle\sum_{r \in F} \Bigl( 1 - \dfrac{d(\hat{e}_r, e_r)}{2H} \Bigr) & |F| > 0, \\[2.5ex]
    0 & |F| = 0.
  \end{cases}
  \label{eq:l2}
\end{equation}
This approach to measure structural similarity between two security-relevant code entities using tree-induced similarity has also been used previously in vulnerability detection and root cause analysis literatures \cite{cleve_locating_2005} 

\subsubsection{Interaction Recovery Correctness}

Given two interaction sequences, the groundtruth $S$ and $\hat{S}$ recovered from our CPG analysis methods. We want to measure how close $\hat{S}$ is to $S$, in terms of both the interaction $b$ found in both sequential similairty and the order they are performed. 

During the CPG analysis, there could exist unexpected recovered interactions outside $S$ between any two roles, caused by the mismatch in previous mapping task, or there exist concrete call element between those two roles, but are not in the groundtruth sequences. Let $k = |S \cap \hat{S}|$ be the number of interactions shared by the groundtruth. We calculate the Jaccard index $J = k / |S \cup \hat{S}|$ to measure sequential similarity. We also interest in assessing if the order of the groundtruth interactions is preserved. For every $k$ shared interactions between $S$ and $\hat{S}$,  they form $\frac{1}{2}k(k-1)$ interaction pairs (Figure~\ref{fig:l3}). Let $c$ be the number of interaction pairs that $S$ and $\hat{S}$ put in the same order (concordant), and $q$ the number of pairs they put in opposite orders (discordant). We calculate the unsigned Kendall's tau $\tau = c / \frac{1}{2}k(k-1)$ to measure this agreement in the ordering of the shared interactions. We then multiply $\tau$ with $J$ to calculate the final sequential correctness score IRC between the groundtruth and recovered interaction sequences. The final score is defined as follows: 
\begin{equation}
  \mathrm{IRC} =
  \begin{cases}
    \dfrac{c}{\frac{1}{2}k(k-1)} \cdot \dfrac{k}{|S \cup \hat{S}|} & k > 1, \\[2.5ex]
    \dfrac{k}{|S \cup \hat{S}|} & k \le 1.
  \end{cases}
  \label{eq:l3}
\end{equation}
When $k \le 1$ there is no pair to put in the wrong order, so $\tau$ is 1 and IRC is $J$ alone, which is 0 when $k = 0$.

\subsection{Model Selection}
\label{sec:setup-models}
We choose our set of LLMs to evaluate on based on their benchmarked performace in full-stack web development tasks, since we want to make sure that they have been exposed to modern web security libraries that the groundtruth dataset is built on. Based on the top 20 results from CodeArena's WebDev benchmark\footnote{https://arena.ai/leaderboard/code/webdev}, we select the following 4 LLMs for evaluation: OpenAI's GPT 6.1 Sol, Meta's Muse Spark, Deepseek-v4.0-Flash-0731, and  GLM-5.3.
GPT and Muse are propietary LLMs, so we use their API endpoints to call them. Deepseek and GLM are open-source LLMs, so we host them on our cloud server with 2x NVIDIA's B200 GPUs and 3x B300 GPUs on ModalAI's infrastructure. We use the same prompt template for all LLMs, and the maximum tokens that all 4 LLMs can process are set to 1,000,000 tokens. Each LLM is run once per instance and setting, with a temperature of 0.7.

\section{Evaluation}
\label{sec:evaluation}

\subsection{RQ1: Task-specific Performance}
\label{sec:setup-metrics}
To answer this question, we ran our constructed pipeline on the dataset we built in Section \ref{sec:setup-groundtruth}. We evaluated the performance of each LLM on the three tasks mentioned in Section \ref{sec:setup-evaluation} using their corresponding metrics RDA, MLS and IRC, and find out which LLM performs best on each task. Our overall evaluation results are shown in Figure~\ref{fig:mean-performance-all-models} in Appendix~\ref{app:results}.

\begin{figure}[t]
  \centering
  \includegraphics[width=\columnwidth]{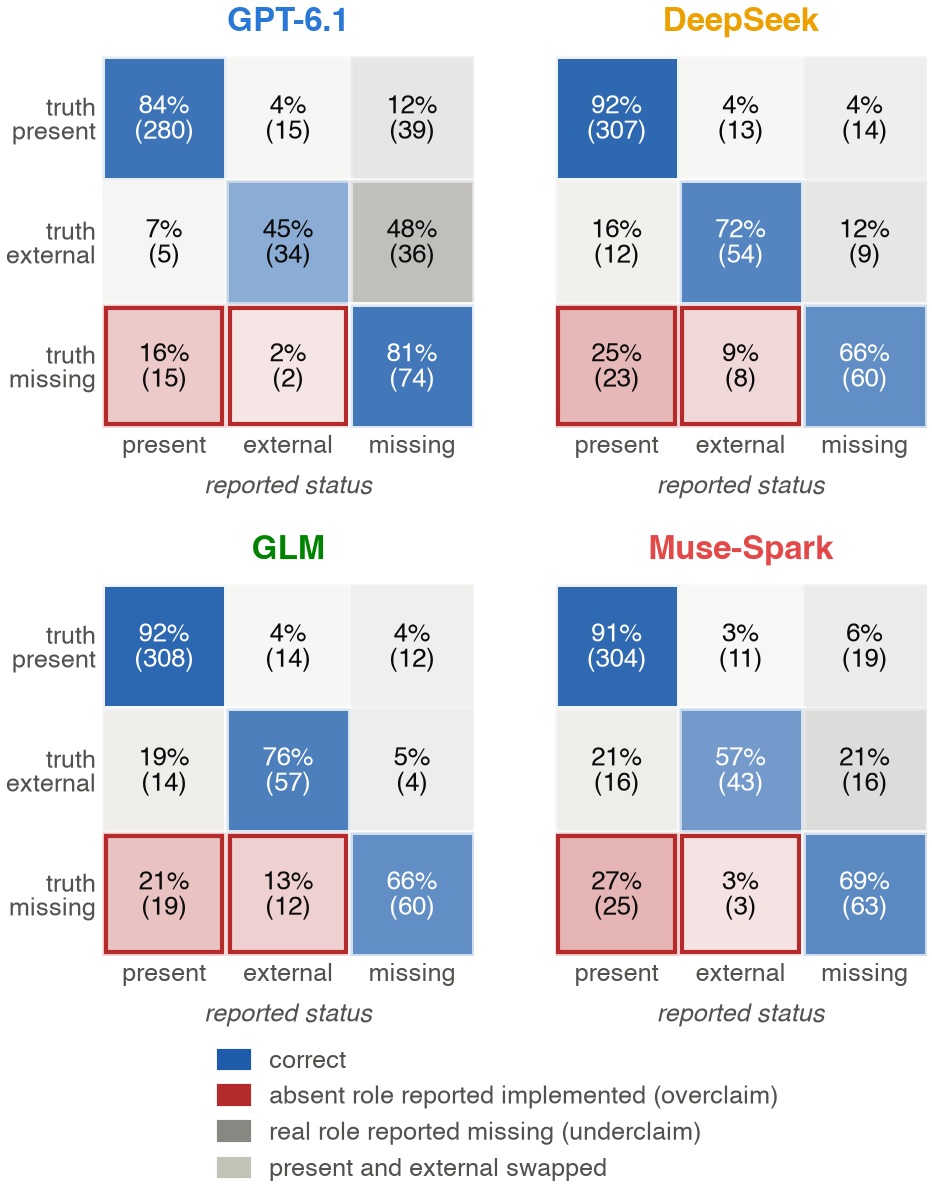}
  \caption{Presence status reported by each LLM against the groundtruth, over the roles of the 100 instances (share of the row, count in brackets). Red frames: a missing role reported as present or external.}
  \label{fig:l1-confusion-matrix}
\end{figure}

Overall, we find that the LLMs perform well on the task of pattern role's presence detection, with an average RDA (over the 100 instances) between 0.79 and 0.86. GLM and DeepSeek are the best performing LLMs on this task, both with an average RDA score of 0.86, while GPT-6.1 has the lowest one (0.79). Looking at the confustion matrix in \ref{fig:l1-confusion-matrix}, we find that GPT-6.1, compared to other LLMs, is more likely to flag a role as missing when it is actually present, or called from an external library. This could be due to the fact that GPT-6.1 is more conservative in its predictions, and prefers to caution when it is not sure about the presence of a security-related component, or when it lacks the knowledge of a specific security library usage that is used in the codebase.

For the task of pattern role's mapping, the LLMs are close to each other, with an average MLS between 0.83 and 0.86. DeepSeek performs the best (0.86), followed by GLM (0.84), Muse-Spark (0.83) and GPT-6.1 (0.83). However, looking specifically at the mappings, we find that GPT-6.1 has more cases when it can perfectly ground the correct code entity. (See Table~\ref{tab:l2-records} in Appendix~\ref{app:results}).

For the task of interaction sequence recovery, we find that GPT 6.1 is the best performing LLM, with an average IRC score of 0.72. It is the best in Java's Springboot case (0.66, against 0.52 to 0.63 for the others), and on par with GLM in Python (0.78 for both). However, we find that all LLMs have a lower performance on this task (average IRC between 0.61 and 0.72) compared to the other two tasks.

\ansbox{1}{GPT-6.1 recovers interactions best (IRC 0.72, against 0.61 to 0.66), and its lead is largest on Java. GLM and DeepSeek detect roles best (RDA 0.86), while GPT-6.1 detects the fewest (0.79) because it is cautious and rarely claims a role that is absent. All four LLMs map roles about equally well (MLS 0.83 to 0.86), but GPT-6.1 maps the most instances exactly (27 of 100).}

\subsection{RQ2: Cost and Efficiency}
\label{sec:setup-performance}
We want to evalutate how many tokens are required for the tasks of pattern role's detection and mappings, with respect to each LLMs. As we are relying on iterative loop of LLM's CoT reasoning to perform the tasks, we want to see how many tokens are required for each iteration, and how many iterations are required for each task. Looking at the results in Figure~\ref{fig:mean-tokens-all-models-pipeline} (Appendix~\ref{app:results}), we find that GPT-6.1 produces the fewest output tokens (1.4k to 1.8k per run), but its higher price makes it the most expensive LLM per run (\$0.04 to \$0.05). GLM uses the fewest new input tokens and is the cheapest, at about \$0.005 per run.

It is also noted that the fees calculated for Deepseek-v4.0-Flash-0731 and GLM are esitimated based on their published pricing information \footnote{\url{https://www.deepseek.ai/pricing}} and \footnote{\url{https://www.glm.ai/pricing}}, but in our experiments, we were hosting them on custom cloud server, and the actual cost of running them is not reflected in the pricing information.

\ansbox{2}{GLM is the most cost-effective LLM: it uses the fewest new input tokens and costs about \$0.005 per run. GPT-6.1 produces the least output (1.4k to 1.8k tokens) but has the highest price per run (\$0.04 to \$0.05). DeepSeek and Muse-Spark produce 4 to 5 times as much output, and Muse-Spark also caches the least.}

\FloatBarrier
\subsection{RQ3: Scalability across Codebase Size and Patterns Complexity}
\label{sec:setup-scalability}

\begin{figure}[t]
  \centering
  \includegraphics[width=\columnwidth]{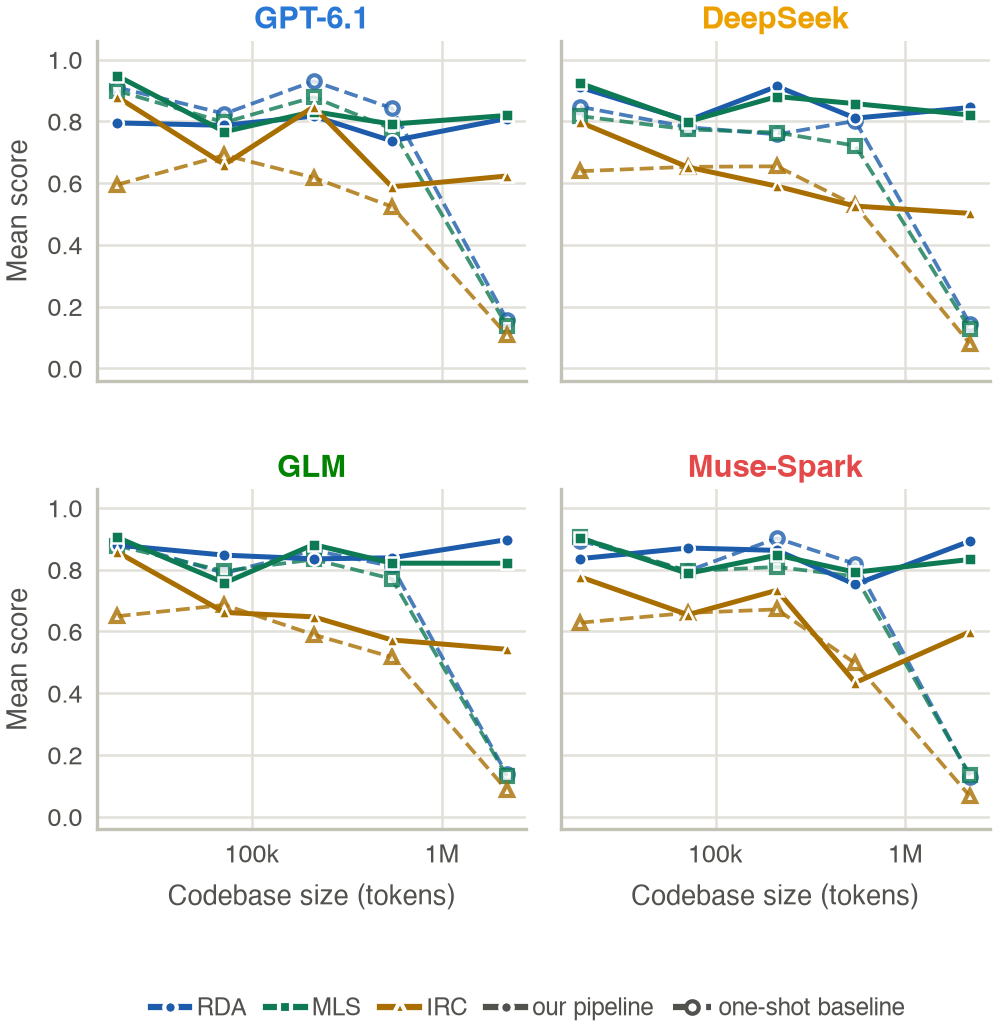}
  \caption{Mean RDA, MLS and IRC against codebase size per LLM: SecPatRec (solid) against the one-shot baseline (dashed), in five size bins of about 20 instances. Instances over the baseline's input limit score 0 for the baseline.}
  \label{fig:performance-scaling-baseline-vs-pipeline}
\end{figure}

For each LLMs, we set up a series of baseline run with a diffrent prompt template (see Appendix~\ref{app:prompts-baseline}) to force them outputting  $\hat{S}$ and $\hat{R}$ in one-shot, without the iterative loop of CoT reasoning, code search tools, or CPG analysis. We then compare the performance of each LLMs on the three tasks with respect to their baseline run counterparts, and find out how the SecPatRec pipeline and baseline differences scale across the five target patterns, and how they can retain their performance when the codebase size increases. 

Looking at Figure~\ref{fig:mean-performance-all-pattern-all-model-baseline-pipeline} in Appendix~\ref{app:results}, we find that SecPatRec outperforms the baseline run in most patterns, partly because the baseline cannot run on the 16 codebases over its input limit, which score 0. On the 84 codebases both can run, the baseline is ahead in 21 of the 60 cases (five patterns, four LLMs and three metrics), mostly on RDA (9 cases), while our pipeline still has a higher overall IRC for every LLM (Table~\ref{tab:baseline-runnable} in Appendix~\ref{app:results}). However, we find that all models perform the worst in P4 Session-based access control, where their IRC (0.37 to 0.58) is the lowest among all five patterns, and is even lower in the case of Java (0.16 to 0.46). This is due to the fact that in Springboot, the Authorization Enforcer is not called by the Authentication Enforcer in the codebase: it is declared as an annotation (e.g., \texttt{@PreAuthorize}), and the Spring framework applies it at run time, after the Authentication Enforcer passes the request on (see Appendix~\ref{app:session-failure}). As no call in the source code links the two enforcers, the CPG analysis cannot find the interactions between them, which leads to a low IRC score. That's why in P4 in Java, the baseline recovers the interactions better than our solution for most LLMs (three of four, e.g., 0.35 against 0.16 for GLM), as the baseline LLMs hallucinate the interaction order themselves, while SecPatRec can only recover the interactions existing inside the static CPG.

Figures~\ref{fig:mean-tokens-baseline-vs-pipeline} and~\ref{fig:cost-vs-size} in Appendix~\ref{app:results} show the token cost side of scalability. The one-shot baseline has to read the whole codebase, so its token cost grows almost in proportion to the codebase size, where a ten times larger codebase costs 7 to 10 times more tokens, compared to only 1.2 to 1.9 times in SecPatRec, because the LLM only need to reads what the code search tools returns. The SecPatRec pipeline is therefore cheaper than the baseline for every codebase above 9k to 36k tokens, depending on the LLM, and at 1M tokens it is 6 (Muse-Spark) to 60 (GPT-6.1) times cheaper.

Figure~\ref{fig:performance-scaling-baseline-vs-pipeline} shows how recovery correctness changes with respect to codebase size. The RDA and MLS of SecPatRec do not decrease as codebases grow, while those number for the baseline drops dramatically, due to the size of 16 codebase exceeding the LLM's token limit. For interaction recovery task, looking at Figure~\ref{fig:irc-by-mls} in Appendix~\ref{app:results}, when every role is mapped to exactly the right symbol from previous steps (MLS $= 1$, 80 instances), the CPG analyzer in SecPatRec recovers the interaction sequence perfectly (IRC $= 1$) on 73 of them, 0.93 on average, against 12\% to 38\% perfect instances in every lower MLS band. The CPG static analysis therefore scales with the codebase size.

\ansbox{3}{The SecPatRec pipeline runs on the 16 of 100 codebases over the baseline's input limit, and on the 84 codebases both can run, it still has a higher IRC than the one-shot baseline for every LLM (+0.02 to +0.12) and a similar or higher MLS ($-0.01$ to $+0.10$; Table~\ref{tab:baseline-runnable}). SecPatRec cost barely grows with codebase size (log-log slope 0.08 to 0.28, against 0.83 to 1.00 for the baseline), so it is cheaper above 9k to 36k tokens and 6 to 60 times cheaper at 1M tokens, while RDA and MLS do not drop as codebases grow.}

\subsection{RQ4: Effects of Code Search Methods}
\label{sec:setup-ablations}

Table~\ref{tab:delta-ablation-codesearch} in Appendix~\ref{app:results} shows how the recovery changes when one search tool is removed from the LLM-based mapping stage. Removing either tool changes RDA and MLS by less than 0.03 for every LLM. To understand why, we counted the tool calls in every run. With both tools available, DeepSeek, GLM and Muse-Spark use keyword search for 80 to 93\% of their searches, while GPT-6.1 uses the two about equally (46 to 52\%). When one tool is removed, its calls move almost one for one to the other tool, and the number of steps per run changes by about one at most. The ReAct loop thus treats keyword and semantic search as interchangeable ways to find candidate code entities, which is why removing either of them, even the one an LLM prefers, does not hurt the recovery.

\ansbox{4}{The ReAct uses keyword and semantic search interchangeably: when one is removed, its calls move to the other, and the run takes about as many steps. Removing either tool therefore changes RDA and MLS by less than 0.03, and changes IRC by $-0.04$ to $+0.09$ depending on the LLM. Removing semantic search saves 165 to 188\,MB of memory without losing accuracy, so keyword search alone is enough when resources are limited.}

\FloatBarrier
\section{Discussion and Threats to Validity}
\label{sec:discussion}

\subsection{Discussion}
\paragraph{Why a hybrid design works}
The RDA and MLS of SecPatRec stay flat as codebases grow (Figure~\ref{fig:performance-scaling-baseline-vs-pipeline}), and when the roles are mapped exactly, the CPG traversal recovers the interaction sequence perfectly in 73 of 80 instances. Most of the IRC error therefore comes from the mapping stage, not from the traversal. This is the main argument for our hybrid design: the LLM handles the part of the task that needs an understanding of the code, while the static analysis reads the interactions from the code itself, without degrading as codebases grow. It also means that improving role mapping is the key to improving our approach.

\paragraph{What the static analysis cannot see}
The CPG only contains the calls written in the source code, so it cannot see the interactions that only appear at run time, such as Spring's \texttt{@PreAuthorize}, Spring's bean injection, or FastAPI's \texttt{Depends}. This is why the one-shot baseline has a higher IRC than SecPatRec on P4 in Java. However, the baseline cannot observe this order in the code either, as it invents the order from its knowledge of Spring, so its higher IRC is not a better recovery. We leave reading the framework's configuration and annotations as future work.

\paragraph{Which LLM to use}
We recommend GLM for low-cost recovery, as it costs about \$0.005 per run and matches the other LLMs on RDA (0.86) and MLS (0.84). GPT-6.1 is the best choice when the interaction sequence or exact mappings matter, with the highest IRC (0.72) and the most exact mappings (27 of 100), but at about 8 to 10 times the cost of GLM. It is also the most conservative LLM: it reports 48\% of the external roles as missing, but it overclaims the least, reporting only 18\% of the missing roles as present or external, against 30\% to 34\% for the others.

\paragraph{Extensibility to Future Security Pattern mining works}
Our groundtruth dataset and evaluation metrics can serve as a benchmark for future security pattern recovery approaches, we publish our dataset at a customize and anonymized Huggingface bucket for public use and recreatibility\footnote{\url{https://huggingface.co/buckets/AnonymousSpottedToad/securityPatternsRecovery}}. The cost of SecPatRec barely grows with codebase size (log-log slope 0.08 to 0.28, against 0.83 to 1.00 for the baseline). Since keyword search alone is enough (RQ4), the vector database and the embedding model can also be dropped. As a result, SecPatRec, a ReACt loop of LLM combining with lexical keyword search and CPG analysis can serve as an instrument to mine how security patterns are implemented at scale. 

\subsection{Threats to Validity}
\paragraph{Construct validity}
The specification $P$ is itself generated by an LLM from the catalogue page and its main flow only, so an error in $P$ would affect both the groundtruth and the recovery; to reduce this threat, we checked the five specifications by hand against their catalogue pages. Finally, scoring the 16 instances over the baseline's input limit as 0 mixes the instances the baseline cannot run with those it runs worse, so we also report the 84 instances it can run, where SecPatRec still has a higher IRC for every LLM (Table~\ref{tab:baseline-runnable}).

\paragraph{Internal validity}
Each LLM was run once per instance and setting, so part of the differences we report may come from run-to-run variability, as the two ablation settings in RQ4 differ from each other as much as from the full pipeline. The ADL generator uses GPT-6 Sol, from the same vendor as GPT-6.1, but its prompts are fixed and its output $P$ is the same for every LLM. Lastly, the 93 repositories and the catalogue are public, so the LLMs may have seen both during training.

\paragraph{External validity}
Our evaluation covers two languages, two frameworks (FastAPI and Spring Security), and five authentication and access control patterns (A01 and A07) from one catalogue, on repositories selected by their dependency on these frameworks. Within this scope, the Java repositories are larger and externalise more roles than the Python ones, so the two languages are not directly comparable. We do not yet investigate how the approach performs on other pattern in the Van De Berg's catalogues, or on other web frameworks such as Django or Express, or on languages for which we have no traversal rules, but will consider that in future works. Our two annotators are experienced web developers rather than the architects of the systems, which is the standard limitation of groundtruths built with an authoritative protocol~\cite{garcia_obtaining_2013}.

\section{Related Work}
\label{sec:relatedwork}

\subsection{Automated Software Architecture Recovery}
Software architecture recovery is a well-studied field \cite{garcia_comparative_2013, ducasse_software_2009}, but existing methods are hard to adapt to security pattern recovery. Traditional static methods cluster code entities with human-specified graph rules \cite{mitchell_automatic_2006,mancoridis_bunch_1999, sartipi_software_2003}, which limits their generalizability to modern codebases across languages and libraries. Learning-based methods learn the representation of code entities \cite{garcia_enhancing_2011} or semantic features of their dependency graphs \cite{zhang_software_2023,bui_treecaps_2021}, which partially solves generalization, but still gives an inductive view of the architecture, not suitable for security patterns, which are prescriptive and enforceable.

Deductive software architecture recovery instead leverages the reasoning of LLMs to recover the architecture of a codebase against a reference architecture \cite{zhao_software_2026, amalfitano_automated_2026, rukmono_deductive_2024,rukmono_evaluating_2026}, which suits the prescriptive nature of security patterns. However, these works only recover the component view of the architecture, not the interaction view, which is crucial for security patterns. Our work is the first deductive approach that recovers both views of security patterns from source code, evaluated on a large dataset of real-world codebases.

\subsection{Evaluating Architecture Recovery Correctness}
To measure the correctness of automated architecture recovery, traditional cluster-based recovery methods rely on the distance between the recovered clusters and those of a groundtruth architecture \cite{garcia_comparative_2013, garcia_obtaining_2013, lutellier_measuring_2018}. Since our method is deductive, mapping the roles of a given pattern to code rather than grouping the code into clusters, these distance-based measures are not suitable for our evaluation. Other works focus on measuring architectural drift, to analyze whether an implementation conforms to its intended design or architecture \cite{de_silva_controlling_2012, rosik_assessing_2011, zamfirov_lessons_2025}. Reflexion models, for example, compare an engineer's high-level model with a model extracted from the source code, and report where the two diverge \cite{murphy_software_1995}. Pruijt et al. compare the accuracy of ten tools that check such conformance \cite{pruijt_accuracy_2017}, and commercial platforms such as Sonar also offer these checks, but they are proprietary and not open-source \cite{curlee_architecture_2026}. However, measuring conformance is also not suitable for our evaluation, as we measure the independent tasks of security pattern recovery, which we have divided into mapping security-relevant code \cite{cleve_locating_2005} and arranging their interactions in sequence \cite{lapata_automatic_2006}.

\subsection{Mining Security Architectural Tactics}
Works on mining security architectural tactics are close to ours, as they share our motivation of mining the architecture-relevant code that enables the security properties of software \cite{mirakhorli_detecting_2016, mujhid_search_2017, santos_catalog_2017}. Mirakhorli and Cleland-Huang detect tactic-related code in open-source systems and monitor it as the system evolves \cite{mirakhorli_detecting_2016}. Mujhid et al. build ArchEngine, a search engine that retrieves and ranks the code snippets implementing a given tactic \cite{mujhid_search_2017}. Santos et al. catalogue 224 security architecture weaknesses, organized by the security tactics that address them \cite{santos_catalog_2017}. Like SecPatRec, these works mine code with lexical code search, using API signatures as keywords. However, they do not attempt to tackle the mapping of the mined code snippets to the architectural components of the security tactics or patterns, which is the core of our recovery task.

\section{Conclusion}
\label{sec:conclusion}
We presented \textsc{SecPatRec}, a hybrid approach to security pattern recovery that combines an LLM agent with code search for role mapping and CPG analysis for interaction recovery. On 100 instances across 93 Java and Python codebases, it achieves a mean RDA of 0.84, MLS of 0.84 and IRC of 0.66 over four LLMs, outperforming a one-shot LLM baseline at a cost that barely grows with codebase size. As the CPG analysis is nearly exact when the roles are mapped exactly, improving role mapping is the key to further improvement. Our dataset, metrics and pipeline can serve as a benchmark and an instrument for mining security patterns at scale.

\clearpage

\bibliographystyle{IEEEtran}
\bibliography{references/recovery}

\clearpage

\appendices
\raggedbottom
\setcounter{dbltopnumber}{3}
\renewcommand{\dbltopfraction}{0.95}
\renewcommand{\dblfloatpagefraction}{0.6}

\section{Extra Technical Details}
\label{app:prompts}

\let\origsection\section
\let\origsubsection\subsection
\let\section\subsection
\let\subsection\subsubsection

\usetikzlibrary{calc}

\lstdefinestyle{adlart}{
  basicstyle=\ttfamily\fontsize{5}{6}\selectfont,
  columns=fullflexible, keepspaces=true, showstringspaces=false,
  breaklines=false, frame=none, aboveskip=0pt, belowskip=0pt,
  escapeinside={(*@}{@*)}}
\newcommand{\adlel}{{\color{gray!60}\ldots}}
\newcommand{\adlfold}[1]{{\setlength{\fboxsep}{0.3pt}\colorbox{orange!20}{#1}}}
\newcommand{\adlfile}[1]{{\sffamily\fontsize{5.5}{6.5}\selectfont\color{gray!45!black}#1}}
\lstdefinestyle{adlprompt}{
  basicstyle=\ttfamily\fontsize{7}{7.9}\selectfont,
  columns=fullflexible, keepspaces=true, showstringspaces=false,
  breaklines=true, breakatwhitespace=true, breakindent=0pt,
  frame=single, framerule=0.4pt,
  rulecolor=\color{gray!50}, backgroundcolor=\color{gray!6},
  xleftmargin=3pt, xrightmargin=3pt, framexleftmargin=3pt, framexrightmargin=3pt,
  aboveskip=0pt, belowskip=2pt,
  escapeinside={(*@}{@*)}}
\newcommand{\adlpart}[1]{{\sffamily\bfseries\color{teal!45!black}#1}}
\makeatletter
\newcommand{\adlfigurehere}{\def\@captype{figure}}
\makeatother
\newsavebox{\adlpage}
\newsavebox{\adlflow}
\newsavebox{\adlspec}

\section{The ADL Code Generator}
\label{app:adl}

\textsc{SecPatRec} reads every security pattern through a single artefact, the
specification $P = (R, B)$. Security patterns, however, are documented in different
ways: each catalogue represents a pattern's structure in its own form of prose and
diagrams \cite{schumacher_security_2006, fernandez-buglioni_security_2013,
van_den_berghe_reimagined_2022}, which the pipeline cannot read directly. An ADL code
generator is therefore needed to bring a pattern into the structure the pipeline
understands. This appendix presents one such case: transforming the patterns of the
KU Leuven catalogue \cite{van_den_berghe_reimagined_2022} into $P$.
Figure~\ref{fig:adl-pipeline} follows \emph{Password-based authentication} (P1),
the pattern of Figure~\ref{fig:pattern-example}, through the generator's six
stages.

\subsection{The KU Leuven Catalogue}

\begin{figure*}[t]
\centering
\colorlet{adlread}{teal!60!black}
\colorlet{adlimg}{orange!80!black}
\begin{tikzpicture}[
  font=\scriptsize,
  shot/.style={anchor=north west, inner sep=0pt},
  tag/.style={font=\tiny, anchor=south west, inner sep=1pt, yshift=1pt},
  hl/.style={draw=adlread, fill=teal!30, fill opacity=0.35, rounded corners=0.8pt,
             thin},
  hlt/.style={font=\sffamily\fontsize{5}{6}\selectfont, text=teal!35!black,
              anchor=south, inner sep=0.3pt, text height=1.5ex, text depth=0.25ex},
]
\def\rs{0.014189}
\node[shot] (rb) at (0,0) {\includegraphics[width=10.5cm]{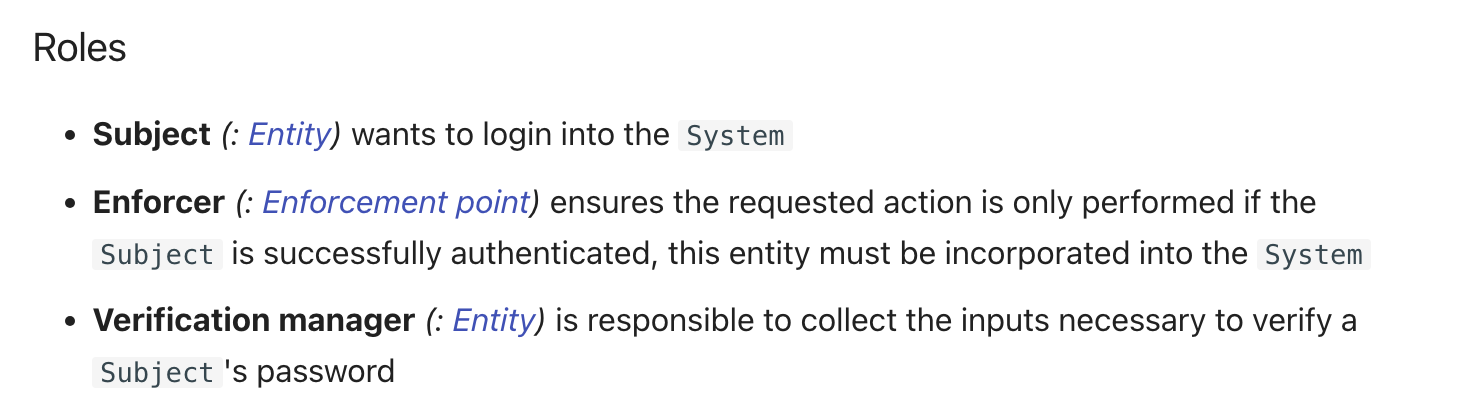}};
\draw[adlread, thick] (rb.north west) rectangle (rb.south east);
\node[tag] at (rb.north west) {\textbf{(a)} Solution, Roles\quad$\Rightarrow$\quad
  \texttt{roles[]}: one entry per role, with its \texttt{name}, \texttt{type} and
  \texttt{description} (stage 5)};
\foreach \x/\y/\t in {44/107/name, 108/173/type, 173/394/description} {
  \draw[hl] ({\x*\rs}, {-(795-740)*\rs}) rectangle ({\y*\rs}, {-(817-740)*\rs});
  \node[hlt] at ({(\x+\y)/2*\rs}, {-(795-740)*\rs}) {\t};
}

\node[shot] (bh) at (0,-3.31) {\includegraphics[width=10.5cm]{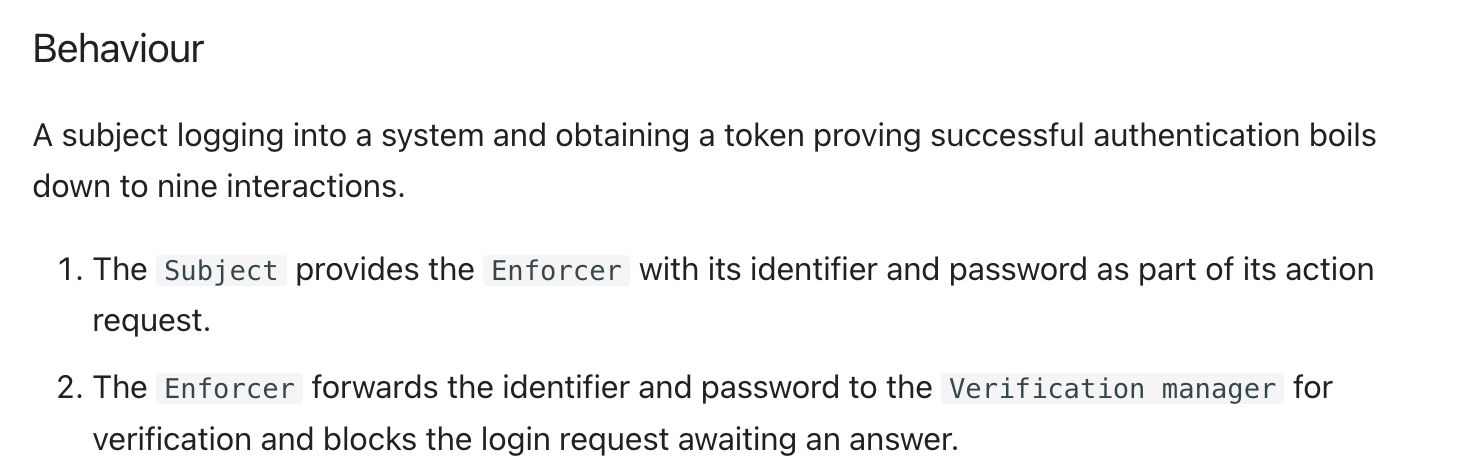}};
\draw[adlread, thick] (bh.north west) rectangle (bh.south east);
\node[tag] at (bh.north west) {\textbf{(b)} Solution, Behaviour\quad$\Rightarrow$\quad
  which flow is the main one (stage 4); what each interaction does (stage 5)};

\node[shot] (dg) at (10.8,0) {\includegraphics[width=7.3cm]{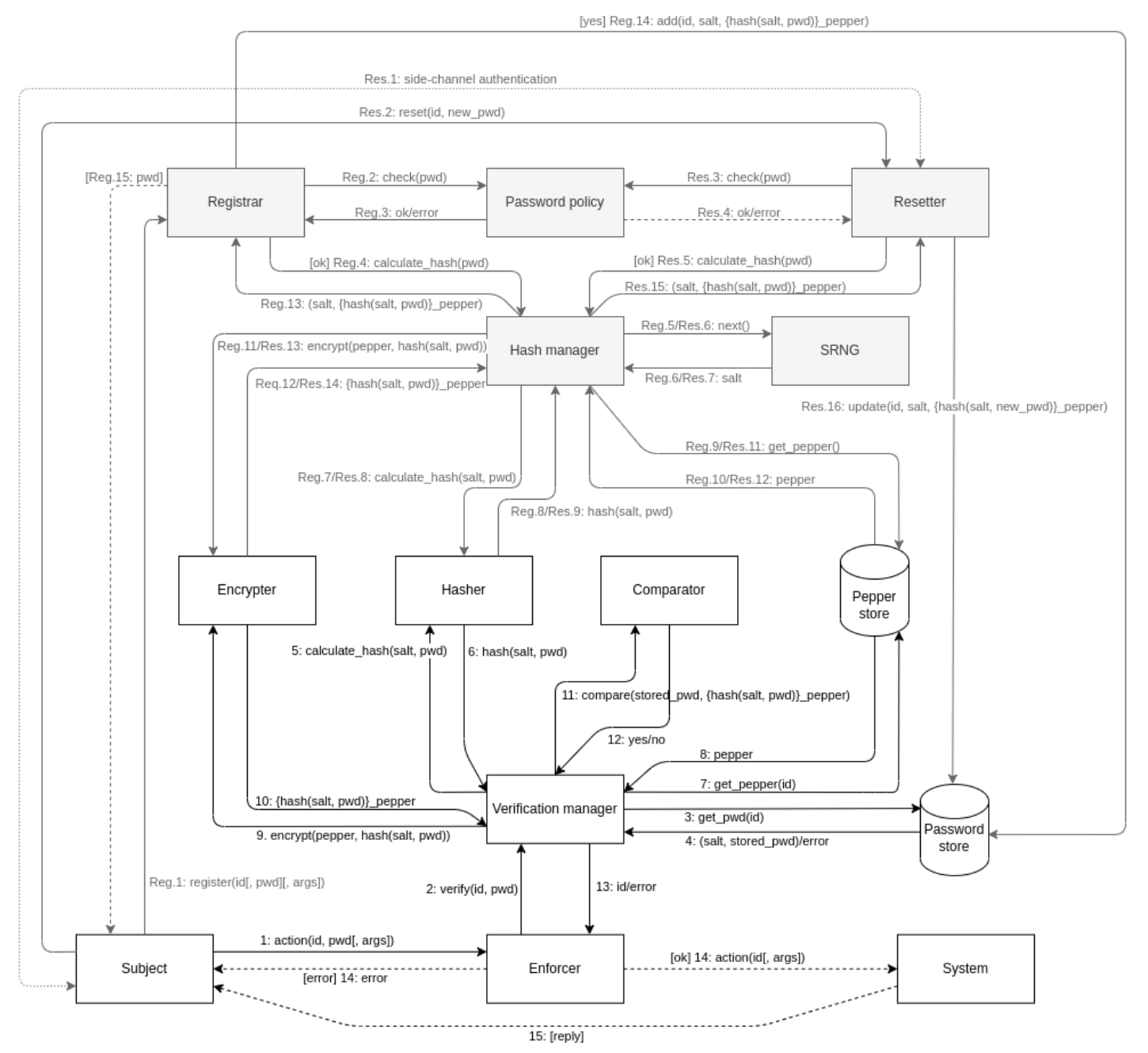}};
\draw[adlimg, thick] (dg.north west) rectangle (dg.south east);
\node[tag] at (dg.north west) {\textbf{(c)} Diagram image\quad$\Rightarrow$\quad
  one Mermaid flowchart (stage 2)};
\end{tikzpicture}
\caption{The parts of the catalogue page of \emph{Password-based authentication}
\cite{van_den_berghe_reimagined_2022} that the generator reads, captured on 24
October 2026: (a)~the roles, with the name, metamodel type and description of the
first one marked; (b)~the numbered walkthrough; and (c)~the diagram image. The page's
name, summary, actions and data elements are read as well. Screenshots
\copyright~imec-DistriNet, CC BY-NC-SA~4.0.}
\Description{Three screenshots of the catalogue page. Panel a is the Roles section,
with the first role's name, type and description boxed and labelled, and a tag
saying it becomes one roles entry each. Panel b is the start of the Behaviour
section's numbered walkthrough, tagged as deciding the main flow and describing
each interaction. Panel c is the pattern's diagram image, tagged as transcribed into
one Mermaid flowchart.}
\label{fig:adl-page}
\end{figure*}
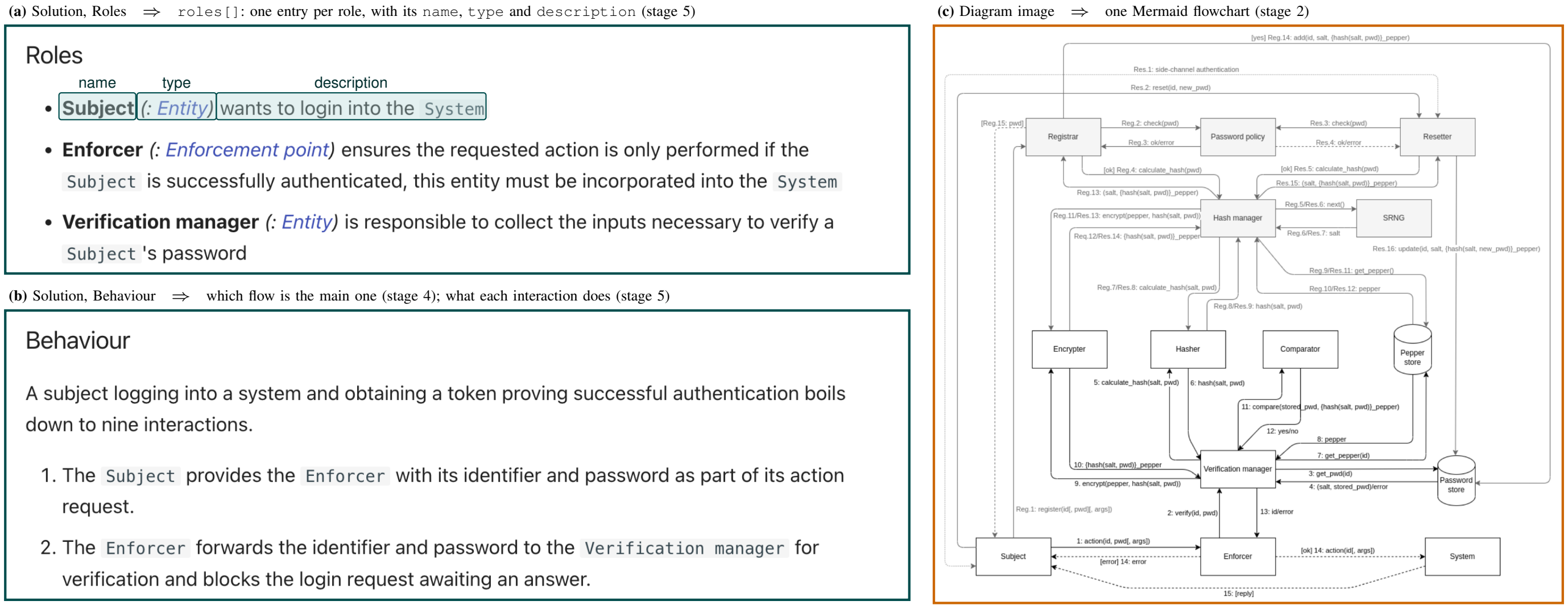

Each pattern of the KU Leuven catalogue is documented on one page, in the same
parts:
\begin{itemize}
  \setlength{\itemsep}{0pt}
  \item \emph{Name, Summary}: what the pattern is and what it achieves.
  \item \emph{Parent, Problem}: the general pattern it refines and the problem it
        solves.
  \item \emph{Roles}: the participants, each with a metamodel type.
  \item \emph{Actions, Data elements}: the calls and data the roles exchange.
  \item \emph{Behaviour}: the interactions between roles, as numbered steps.
  \item \emph{Diagram}: the same interactions, drawn as boxes and numbered arrows.
  \item \emph{Considerations, Implementations, References}: design advice,
        example libraries and literature.
\end{itemize}
Our main concern is the pattern's structure: the Name, Summary, Roles, Actions,
Data elements, Behaviour and the diagram (Figure~\ref{fig:adl-page}). The other
parts guide a designer but define no role or interaction, so the generator keeps
them only as context.

A pattern's diagram usually holds several flows: beside the main flow, which
carries the core idea of the pattern, it draws lifecycle flows such as
registration and password reset (P1 has all three). Our concern is the main flow,
so the generator includes a component that selects it (\emph{Select}, below). The
other flows are not lost: without this component, the generator keeps every flow.

\begin{lrbox}{\adlpage}\begin{minipage}[t]{3.29cm}
\begin{lstlisting}[style=adlart]
# Password-based authentication
(*@\adlel@*)
## Solution
### Roles
- Subject (: [Entity](...))
  wants to login into the
  `System`
- Enforcer (: [Enforcement
  point](...)) ensures (*@\adlel@*)
(*@\adlel@*)
### Actions
### Data elements
### Behaviour
1. The `Subject` provides the
   `Enforcer` with its (*@\adlel@*)
(*@\adlel@*)
\end{lstlisting}
\end{minipage}\end{lrbox}
\begin{lrbox}{\adlflow}\begin{minipage}[t]{5.39cm}
\begin{lstlisting}[style=adlart]
flowchart LR
Subject["Subject"]
Enforcer["Enforcer"]
VerificationManager["Verification manager"]
(*@\adlel@*)
Subject -->|"1: action(id, pwd[, args])"| Enforcer
(*@\adlel@*)
Enforcer -->|"[ok] 14: action(id[, args])"| System
Enforcer -->|"[error] 14: error"| Subject
System -->|"15: [reply]"| Subject
(*@\adlel@*)
HashManager -->|"Reg.5/Res.6: next()"| SRNG
Subject -->|"Res.2: reset(id, new_pwd)"| Resetter
(*@\adlel@*)
\end{lstlisting}
\end{minipage}\end{lrbox}
\begin{lrbox}{\adlspec}\begin{minipage}[t]{5.19cm}
\begin{lstlisting}[style=adlart]
{ "name": "Password-based authentication",
  "summary": "A subject proves they own (*@\adlel@*)",
  "roles": [ (*@\adlel@*),
    { "name": "Password store",
      "type": "Storage", "required": true,
      "description": "keeps track of (*@\adlel@*)" },
    (*@\adlel@*) ],
  "behaviour": [ (*@\adlel@*),
    { "step": "3", "kind": "action",
      "from": "Verification manager",
      "to": "Password store",
      "element": "get_pwd(id)",
      "description": "The `Verification manager`
        requests the stored hash value and
        salt from the `Password store`.
        (*@\adlfold{The \textasciigrave Verification manager\textasciigrave{} receives}@*)
        (*@\adlfold{the requested hash value (\adlel)}@*)" },
    (*@\adlel@*) ] }
\end{lstlisting}
\end{minipage}\end{lrbox}
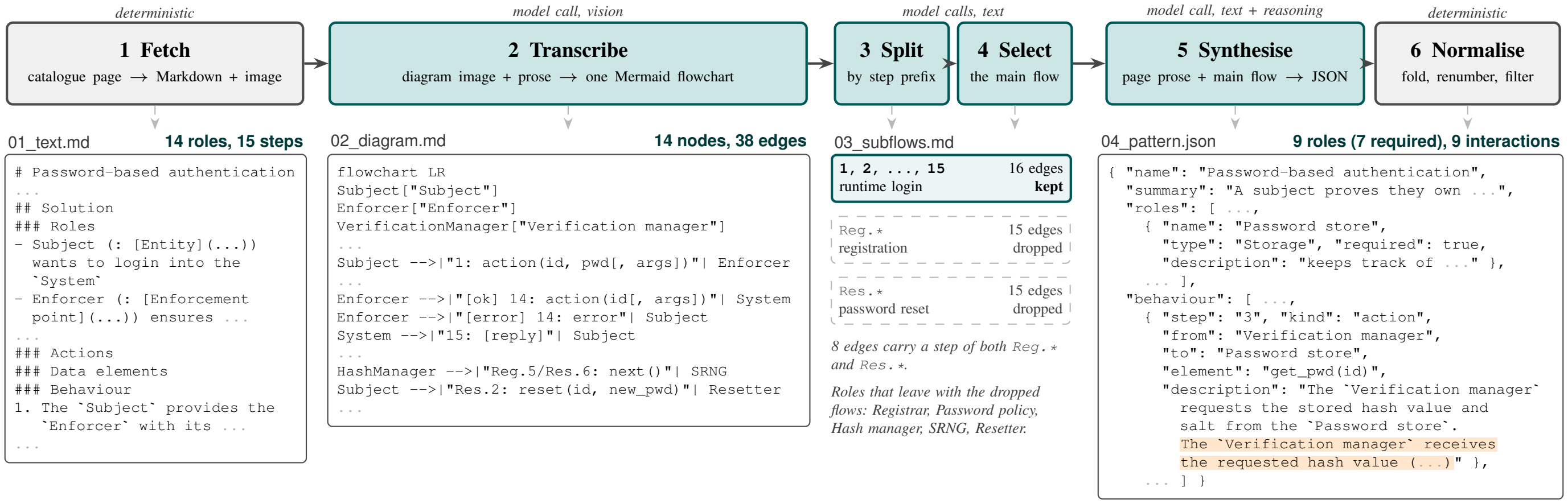
\begin{figure*}[t]
\centering
\begin{tikzpicture}[
  font=\scriptsize,
  stage/.style={draw, thick, rounded corners=2pt, align=center, inner sep=2pt,
                minimum height=0.95cm, text width=#1},
  det/.style={stage=#1, draw=gray!60!black, fill=gray!12},
  llm/.style={stage=#1, draw=teal!70!black, fill=teal!20},
  kind/.style={font=\tiny\itshape, text=gray!45!black, anchor=south, inner sep=1pt},
  art/.style={draw=gray!60!black, fill=white, rounded corners=1.5pt,
              anchor=north west, inner sep=3pt},
  fname/.style={anchor=south west, inner sep=1pt},
  cnt/.style={font=\sffamily\bfseries\fontsize{5.5}{6.5}\selectfont,
              text=teal!45!black, anchor=south east, inner sep=1pt},
  flow/.style={-{Stealth[length=1.7mm]}, thick, gray!55!black},
  writes/.style={-{Stealth[length=1.4mm]}, gray!55, densely dashed},
  chip/.style={draw, rounded corners=1.5pt, align=left, anchor=north west,
               text width=2.6cm, inner sep=2.5pt, font=\tiny},
  kept/.style={chip, draw=teal!70!black, thick, fill=teal!8},
  gone/.style={chip, draw=gray!55, dashed, text=gray!55!black},
  note/.style={font=\tiny\itshape, text=gray!50!black, align=left,
               anchor=north west, text width=2.75cm, inner sep=0pt},
]
\node[det=3.3cm]  (s1) at (1.75,0)   {\textbf{1\enspace Fetch}\strut\\
  {\tiny catalogue page $\rightarrow$ Markdown + image\strut}};
\node[llm=5.4cm]  (s2) at (6.55,0)   {\textbf{2\enspace Transcribe}\strut\\
  {\tiny diagram image + prose $\rightarrow$ one Mermaid flowchart\strut}};
\node[llm=1.18cm] (s3) at (10.31,0)  {\textbf{3\enspace Split}\strut\\
  {\tiny by step prefix\strut}};
\node[llm=1.18cm] (s4) at (11.74,0)  {\textbf{4\enspace Select}\strut\\
  {\tiny the main flow\strut}};
\node[llm=2.85cm] (s5) at (14.3,0)   {\textbf{5\enspace Synthesise}\strut\\
  {\tiny page prose + main flow $\rightarrow$ JSON\strut}};
\node[det=2.0cm]  (s6) at (17.0,0)   {\textbf{6\enspace Normalise}\strut\\
  {\tiny fold, renumber, filter\strut}};
\node[kind] at (s1.north) {deterministic};
\node[kind] at (s2.north) {model call, vision};
\node[kind] at ($(s3.north)!0.5!(s4.north)$) {model calls, text};
\node[kind] at (s5.north) {model call, text + reasoning};
\node[kind] at (s6.north) {deterministic};
\foreach \a/\b in {s1/s2, s2/s3, s3/s4, s4/s5, s5/s6}
  \draw[flow] (\a.east) -- (\b.west);

\node[art] (a1) at (0,-1.05) {\usebox{\adlpage}};
\node[fname] at (a1.north west) {\adlfile{01\_text.md}};
\node[cnt] at (a1.north east) {14 roles, 15 steps};
\node[art] (a2) at (3.75,-1.05) {\usebox{\adlflow}};
\node[fname] at (a2.north west) {\adlfile{02\_diagram.md}};
\node[cnt] at (a2.north east) {14 nodes, 38 edges};
\node[fname] at (9.6,-1.05) {\adlfile{03\_subflows.md}};
\node[kept] (c1) at (9.6,-1.05)
  {\textbf{\texttt{1, 2, \dots, 15}}\hfill 16 edges\\ runtime login\hfill\textbf{kept}};
\node[gone] (c2) at ([yshift=-1.5mm]c1.south west)
  {\texttt{Reg.*}\hfill 15 edges\\ registration\hfill dropped};
\node[gone] (c3) at ([yshift=-1.5mm]c2.south west)
  {\texttt{Res.*}\hfill 15 edges\\ password reset\hfill dropped};
\node[note] at ([yshift=-1.8mm]c3.south west)
  {8 edges carry a step of both \texttt{Reg.*} and \texttt{Res.*}.\\[3pt]
   Roles that leave with the dropped flows: Registrar, Password policy,
   Hash manager, SRNG, Resetter.};
\node[art] (a4) at (12.7,-1.05) {\usebox{\adlspec}};
\node[fname] at (a4.north west) {\adlfile{04\_pattern.json}};
\node[cnt] at (a4.north east) {9 roles (7 required), 9 interactions};

\foreach \s in {s1, s2, s3, s4, s5, s6}
  \draw[writes] (\s.south) -- (\s.south |- 0,-0.77);
\end{tikzpicture}
\caption{The ADL code generator on \emph{Password-based authentication} (P1). Top:
the six stages, grey where deterministic and tinted where a model call. Bottom: the
artefact each stage writes, as verbatim excerpts (prose re-wrapped, runs elided).
Stage~5 reads the page prose as well as the main flow; in its output, a response
survives only as a sentence appended to the request it answers (shaded).}
\Description{Six stage boxes in a row, Fetch, Transcribe, Split, Select, Synthesise
and Normalise, with Transcribe, Split, Select and Synthesise tinted as model calls.
Below them, the artefacts each writes for the password-based authentication
pattern: the page as Markdown with 14 roles and 15 behaviour steps; a Mermaid
flowchart with 14 nodes and 38 edges whose labels keep their step
prefixes such as 1, Reg.5/Res.6 and Res.2; three sub-flows, of which the runtime
login flow numbered 1 to 15 is kept and the registration and reset flows are
dropped together with five roles; and the final JSON specification with 9 roles,
7 of them required, and 9 interactions, one of which shows a response sentence
appended to its request.}
\label{fig:adl-pipeline}
\end{figure*}

\subsection{Stages}

The generator turns these parts into $P$ in six stages
(Figure~\ref{fig:adl-pipeline}). Fetching the page and normalising the result are
deterministic; the four stages between them are model calls, each with a fixed
prompt (Section~\ref{app:adl-prompts}), on
one pinned model, GPT-6 Sol, independent of the models under study
(Section~\ref{sec:setup-models}). Each stage writes its result to disk, so an
error in $P$ can be traced to the stage that introduced it.

\textbf{Fetch.} An HTML parser renders the catalogue page as Markdown, keeping its
\emph{Roles}, \emph{Actions}, \emph{Data elements} and \emph{Behaviour} sections
apart, strips the site's navigation and references, and downloads the diagram
image. For P1, the page names fourteen roles and narrates fifteen numbered steps.

\textbf{Transcribe.} A vision call transcribes the diagram into one
Mermaid\footnote{\url{https://mermaid.js.org}} flowchart: one node per box and one
edge per arrow, each label kept verbatim with
its step prefix. The prompt has the model read each direction off the arrowhead and
emit exactly one edge per step number. P1's flowchart has 14 nodes and 38 edges.

\textbf{Split and select.} A text-only call groups the edges by step-prefix family
(\texttt{1..N}, \texttt{Reg.*}, \texttt{Res.*}) into one sequence diagram each, and
a second call keeps only the pattern's primary runtime flow. For P1, this keeps the
login flow (steps 1--15) and drops registration and reset, together with five roles
no login invokes: the Registrar, Password policy, Hash manager, SRNG and Resetter.

\textbf{Synthesise.} A text-only call with reasoning emits $P$ from the page prose
and the main flow: each role with its name, metamodel type, \texttt{required} flag
and description, and each interaction with its \texttt{from} and \texttt{to} roles,
call \texttt{element} and description. Its central rule is that \emph{an entry is an
interaction, not an arrow}: a response is folded into the request it answers, as a
sentence of its description, so P1's 16 arrows become the nine interactions
$b_1, \dots, b_9$ of Figure~\ref{fig:pattern-example} (Figure~\ref{fig:adl-fold}).
A returning edge that answers no request, such as the System's reply to the
Subject, keeps its own entry of kind \texttt{data}.

\begin{center}
\begin{minipage}{\columnwidth}
\centering
\begin{tikzpicture}[
  font=\scriptsize,
  hd/.style={font=\tiny, anchor=south west, rotate=40, inner sep=1pt},
  life/.style={gray!40, thin},
  band/.style={draw=teal!45, fill=teal!9, rounded corners=1.5pt, thin},
  req/.style={-{Stealth[length=1.5mm]}, thick, teal!55!black},
  rsp/.style={-{Stealth[length=1.2mm]}, densely dashed, gray!65!black},
  bj/.style={font=\tiny, fill=white, inner sep=0.5pt, text=teal!40!black},
  sn/.style={font=\tiny, text=gray!55!black, anchor=east, inner sep=1pt},
]
\def\dx{0.8}
\def\dy{0.27}
\foreach \f/\t/\a/\z in {0/1/1/15, 1/2/2/13, 2/3/3/4, 2/4/5/6, 2/5/7/8,
                         2/6/9/10, 2/7/11/12}
  \fill[band] ({\f*\dx-0.07}, {-\a*\dy+0.11}) rectangle ({\t*\dx+0.07}, {-\z*\dy-0.11});
\foreach \i/\n in {0/Subject, 1/Enforcer, 2/Verification manager,
                   3/Password store, 4/Hasher, 5/Pepper store, 6/Encrypter,
                   7/Comparator, 8/System} {
  \node[hd] at (\i*\dx-0.05, 0.05) {\n};
  \draw[life] (\i*\dx, 0) -- (\i*\dx, {-16*\dy-0.15});
}
\foreach \r/\s/\f/\t/\k/\b in {
    1/1/0/1/1/1,   2/2/1/2/1/2,   3/3/2/3/1/3,   4/4/3/2/0/,
    5/5/2/4/1/4,   6/6/4/2/0/,    7/7/2/5/1/5,   8/8/5/2/0/,
    9/9/2/6/1/6,  10/10/6/2/0/,  11/11/2/7/1/7,  12/12/7/2/0/,
   13/13/2/1/0/, 14/{[ok] 14}/1/8/1/8, 15/{[error] 14}/1/0/0/,
   16/15/8/0/1/9} {
  \node[sn] at (-0.12, {-\r*\dy}) {\s};
  \ifnum\k=1
    \draw[req] (\f*\dx, {-\r*\dy}) -- node[bj] {$b_{\b}$} (\t*\dx, {-\r*\dy});
  \else
    \draw[rsp] (\f*\dx, {-\r*\dy}) -- (\t*\dx, {-\r*\dy});
  \fi
}
\node[font=\tiny\itshape, text=teal!40!black, anchor=west, inner sep=1pt]
  at (8*\dx+0.05, {-16*\dy}) {kind \texttt{data}};
\end{tikzpicture}
\adlfigurehere
\caption{Folding P1's main flow. Its 16 arrows become the nine interactions
$b_1, \dots, b_9$ of Figure~\ref{fig:pattern-example}: each response (dashed) is
folded into the request it answers, and a band spans the interaction from one to
the other. The bands nest like a call stack. The System's reply opens no band,
since the Subject never called the System, and stays as $b_9$.}
\Description{A sequence diagram with nine lifelines, Subject, Enforcer,
Verification manager, Password store, Hasher, Pepper store, Encrypter, Comparator
and System, and sixteen numbered arrows. Nine solid arrows are labelled b1 to b9.
Seven dashed arrows are responses. Shaded bands group each request with the
response folded into it: five short bands pair the Verification manager's calls
with their replies, a taller band between the Enforcer and the Verification manager
pairs step 2 with step 13, and the tallest, between the Subject and the Enforcer,
pairs step 1 with the error returned in step 14. The final reply from the System to
the Subject is solid, labelled b9 and marked kind data.}
\label{fig:adl-fold}
\end{minipage}
\end{center}

\textbf{Normalise.} Deterministic passes enforce the prompt's rules on the returned
JSON: they fold any response the model left as its own entry, renumber the steps
$1, \dots, N$, drop roles that no interaction uses, and mark the initiating party
and the protected System as not required.

The four artefacts of every pattern are released with our dataset, so each
specification can be checked against the catalogue page it was compiled from.

\subsection{Prompts}
\label{app:adl-prompts}

Figures~\ref{fig:prompt-transcription}--\ref{fig:prompt-synthesis} reproduce the
prompts of the four model calls verbatim, as they appear in the generator's source. Each
pairs a system prompt, fixed across patterns, with the template of its per-run
user message, whose placeholders are filled from the catalogue page (\texttt{name},
\texttt{text}) and from the previous stage's output (\texttt{mermaid},
\texttt{catalogue}).

\begin{figure*}[p]
\begin{lstlisting}[style=adlprompt]
(*@\adlpart{System prompt}@*)
You are a diagram transcription expert. You are given a single diagram image from the DistriNet (KU Leuven) security-pattern catalogue together with the pattern's description text. The image is the authoritative depiction of the pattern.

These images are box-and-arrow collaboration diagrams (roles drawn as boxes, connected by numbered message arrows) (*@\textemdash@*) NOT lifeline sequence diagrams. Transcribe the WHOLE diagram into ONE Mermaid `flowchart` (graph) that reproduces the image completely (*@\textemdash@*) every element, nothing added, nothing left out. Do NOT split it into separate diagrams and do NOT use `sequenceDiagram`.

Represent the image as a graph:
- One NODE per role / participant / box, with a valid single-token id and a bracketed label holding the real (possibly multi-word) name: `Enforcer["Enforcer"]`, `PasswordStore["Password store"]`.
- One directed EDGE per arrow, labelled with the arrow's EXACT message text INCLUDING its full step prefix/number (`1.`, `Reg.2`, `Res.4`, (*@\ldots@*)): `Subject -->|"1. action(id, pwd[])"| Enforcer`.

Follow this METHOD so that no edge is missed and every direction is correct:
1. First, list every box (role) in the image. Create one node for each.
2. Then walk the arrows IN STEP-NUMBER ORDER (1, 2, 3 (*@\ldots@*) then `Reg.1`, `Reg.2` (*@\ldots@*) then `Res.1`, `Res.2` (*@\ldots@*)). Every step number visible in the image MUST become exactly one edge (*@\textemdash@*) if you see labels up to 15, you must output 15 numbered edges, with none skipped and none duplicated.
3. For EACH arrow determine its direction by the ARROWHEAD: the edge goes FROM the box at the arrow's tail TO the box the arrowhead points at. Write `Source -->|"label"| Target` in exactly that direction. A reply/return arrow points back the other way (*@\textemdash@*) keep it as its own edge in its own direction. Do not guess direction from left/right position; read the arrowhead. When unsure, cross-check the numbered step against the PATTERN DESCRIPTION, which states who sends each message to whom.
4. Include return/response arrows, guards/conditions (e.g. `[ok]`, `[error]`), and any notes (add a note as its own node or fold it into the relevant edge label).
5. Before finishing, re-scan the image and confirm every arrow you can see has a matching edge and that each edge's endpoints and direction match the arrowhead.

Preserve label wording exactly as shown; do NOT summarise, merge two arrows into one, rename, or invent anything. If some text is partly illegible, transcribe your best reading.

Mermaid rules: start with `flowchart LR` (or `flowchart TD`). Node ids must be single tokens with NO spaces (*@\textemdash@*) always put the readable name in a bracketed `["..."]` label. ALWAYS wrap edge labels in double quotes because they contain digits, dots, parentheses, and commas: `A -->|"..."| B`.

Output ONLY the Mermaid source, starting with `flowchart`. No prose, no code fences.

(*@\adlpart{User message (template)}@*)
Pattern: {name}.

--- PATTERN DESCRIPTION ---
{text}
--- END PATTERN DESCRIPTION ---

Transcribe the ENTIRE attached diagram into one complete Mermaid `flowchart` now: one node per role/box, one directed edge per arrow, every edge label (with its step prefix/number) quoted and exact. Capture every node, every arrow, and every label (*@\textemdash@*) do not miss any detail.
\end{lstlisting}
\caption{The prompt of stage~2, Transcribe (vision call): diagram image + page prose $\rightarrow$ one Mermaid flowchart. The diagram is attached to the user message as an image.}
\label{fig:prompt-transcription}
\end{figure*}

\begin{figure*}[p]
\begin{lstlisting}[style=adlprompt]
(*@\adlpart{System prompt}@*)
You are given the full Mermaid `flowchart` transcription of a security pattern's diagram (roles as nodes, numbered messages as directed edges) plus the pattern's description text. This is a TEXT task (*@\textemdash@*) work only from the flowchart; there is no image.

Every edge label starts with a step PREFIX (*@\textemdash@*) a number or tag such as `1.`, `2.`, `Reg.2`, or `Res.4`. Edges that share a prefix family (`1..N`, `Reg.*`, `Res.*`, (*@\ldots@*)) form ONE sub-flow.

A family is decided by the prefix ALONE, never by topic. Plainly numbered edges `1`, `2`, `3`, (*@\ldots@*) `N` are ONE single family however many they are and however many different roles they touch (*@\textemdash@*) they are the consecutive steps of one walkthrough, so NEVER cut a plain numeric chain into thematic groups ("verification", "enforcement", "update", (*@\ldots@*)). A new family starts only at a distinct non-numeric tag (`Reg.*`, `Res.*`, `A.*`, (*@\ldots@*)) or a restart of the numbering. Branch conditions and repeats do not start a family either: `[allow] 9: (*@\ldots@*)` and `[block] 9: Error` are both step 9 of the numeric family. An edge with NO step prefix at all (e.g. a dotted `Policy changes` annotation) belongs to no family (*@\textemdash@*) leave it out entirely.

Do the following:
1. Group the flowchart's edges into sub-flows by their prefix family.
2. For each sub-flow emit a Mermaid `sequenceDiagram` containing exactly that family's edges, ordered by step number. Convert each directed edge `A -->|"n: msg"| B` into a message `A->>B: n: msg`, preserving the flowchart's direction EXACTLY (source (*@$\rightarrow$@*) target). Use `-->>` when the label is clearly a reply/return/response, otherwise `->>`. Keep EVERY edge of the family (*@\textemdash@*) none may be dropped, merged, or duplicated (*@\textemdash@*) with its label verbatim.
3. Give each sub-flow a short descriptive NAME grounded in the description (e.g. "Runtime authentication", "Subject registration", "Password reset") (*@\textemdash@*) never a generic "Flow 1".

Participant ids must be single Mermaid-valid tokens with NO spaces; for a multi-word node name use an alias (*@\textemdash@*) `participant PasswordStore as Password store` (*@\textemdash@*) and reference the id in every message. Do NOT rename roles or invent edges. If the flowchart has only one prefix family, return a single sub-flow.

Output ONLY a JSON object of exactly this shape (*@\textemdash@*) no prose, no code fences:
{"subflows": [{"name": "<human name>", "prefix": "<edge prefix>", "mermaid": "<Mermaid source>"}]}

(*@\adlpart{User message (template)}@*)
--- PATTERN DESCRIPTION ---
{text}
--- END PATTERN DESCRIPTION ---

--- FULL DIAGRAM (Mermaid flowchart) ---
{mermaid}
--- END FULL DIAGRAM ---

Split the flowchart into sub-flows by edge-prefix family, name each from the description, and return the JSON now. Every PREFIXED edge in the flowchart must appear in exactly one sub-flow, with its label and direction unchanged.
\end{lstlisting}
\caption{The prompt of stage~3, Split (text-only): flowchart $\rightarrow$ one sequence diagram per step-prefix family.}
\label{fig:prompt-subflows}
\end{figure*}

\begin{figure*}[p]
\begin{lstlisting}[style=adlprompt]
(*@\adlpart{System prompt}@*)
You are a security-pattern analyst. You are given a pattern's description text and a numbered list of candidate sub-flows extracted from its diagram. Exactly ONE of them is the MAIN flow (*@\textemdash@*) the core interaction the pattern is fundamentally about, i.e. the primary runtime behaviour laid out in the description's main 'Solution' / 'Behaviour' narrative (the numbered steps). The others are auxiliary or lifecycle flows (e.g. registration, password change / reset, key rotation, provisioning, setup).

Choose the single sub-flow that best corresponds to that main behaviour. Judge by the sub-flow names and edge prefixes against the description (*@\textemdash@*) the main flow is the one whose messages match the description's principal numbered walkthrough.

Output ONLY JSON, no prose, no code fences: {"index": <zero-based index of the main sub-flow>}

(*@\adlpart{User message (template)}@*)
--- PATTERN DESCRIPTION ---
{text}
--- END PATTERN DESCRIPTION ---

Candidate sub-flows (index. name (*@\textemdash@*) edge prefix):
{catalogue}

Which one is the MAIN flow described by the text? Return {"index": n}.
\end{lstlisting}
\caption{The prompt of stage~4, Select (text-only): candidate sub-flows $\rightarrow$ the index of the main flow.}
\label{fig:prompt-mainflow}
\end{figure*}

\begin{figure*}[p]
\begin{lstlisting}[style=adlprompt]
(*@\adlpart{System prompt}@*)
You are a security-pattern analyst. From a DistriNet security pattern's cleaned text (`Roles`, `Actions`, `Data elements`, `Behaviour`) and the Mermaid diagram of its MAIN flow, produce the pattern's canonical JSON. Copy names and wording from the text; never invent anything.

An entry in `behaviour` is an INTERACTION, not an arrow. The request edge `A (*@$\rightarrow$@*) B` and the response edge `B (*@$\rightarrow$@*) A` answering it share ONE entry (*@\textemdash@*) the request's, in the requester's direction (*@\textemdash@*) with the response's sentence appended to its `description`, so the entry states what was asked and what came back. The response gets no entry and no field of its own. Pair them like a call stack: a response answers the most recent unanswered request between those two roles, so it may be several steps later. Example (*@\textemdash@*) arrows `1 Subject(*@$\rightarrow$@*)Enforcer: action(token)`, `2 Enforcer(*@$\rightarrow$@*)Verifier: verify(token)`, `3 Verifier(*@$\rightarrow$@*)Store: get_principal(token)`, `4 Store(*@$\rightarrow$@*)Verifier: principal/error`, `5 Verifier(*@$\rightarrow$@*)Enforcer: principal/error`, `6 Enforcer(*@$\rightarrow$@*)System: action(principal)` (*@$\rightarrow$@*) four entries (1, 2, 3, 6), with 4 folded into 3 and 5 into 2. This includes the flow's closing `reply`/`error` back to the initiator. Only a returning edge whose target never made the request keeps its own entry (`kind: "data"`).

Output ONLY a JSON object:
- "name": the pattern's name; "summary": its `Summary` paragraph (else "").
- "roles": one entry `{"name", "type", "required", "description"}` per role (*@\textemdash@*) `type` is the metamodel type in parentheses ("Entity", "Storage", (*@\ldots@*)). Include ONLY roles used as some interaction's `from` or `to`: drop the `Roles` section's auxiliary-flow roles (registration, reset, provisioning) the main flow never touches, and conversely add a main-flow participant the section omits (often the protected `System`), describing it from the text. "required" is false for the `Subject` and the `System` (*@\textemdash@*) the pattern's environment rather than its machinery (*@\textemdash@*) and true for every other role.
- "behaviour": the main flow's interactions, each `{"step", "from", "to", "kind", "element", "description"}`. `step` numbers the entries sequentially from "1" (*@\textemdash@*) not the diagram's edge numbers, whose sequence the merging breaks. `from`/`to` are the requester and responder, read off the request arrow's direction and named exactly as in `roles`. `kind` is "action" (a call) or "data" (an unfoldable returning edge). `element` is the request arrow's CONCRETE label as instantiated in the diagram (*@\textemdash@*) `hash(salt, pwd)`, not the `Actions`/`Data elements` section's placeholder `hash(x, y)`; "" if nothing matches. Every numbered `Behaviour` step is accounted for: a request as an entry, a response inside the description of the entry it answers.

(*@\adlpart{User message (template)}@*)
--- PATTERN TEXT ---
{text}
--- END PATTERN TEXT ---

--- MAIN FLOW (Mermaid) ---
{mermaid}
--- END MAIN FLOW ---

Return the pattern JSON now (*@\textemdash@*) name, summary, roles[], behaviour[] (*@\textemdash@*) merging each request with the response that answers it into a single entry, numbered sequentially.
\end{lstlisting}
\caption{The prompt of stage~5, Synthesise (text-only, with reasoning): page prose + main flow $\rightarrow$ the specification $P$.}
\label{fig:prompt-synthesis}
\end{figure*}

\let\section\origsection
\let\subsection\origsubsection
\FloatBarrier

\subsection{Prompts for Role Detection and Mapping}
\label{app:prompts-agent}

The agent uses the same system prompt for every pattern, codebase and LLM
(Figures~\ref{fig:prompt-agent} and~\ref{fig:prompt-agent-cont}, with the line
numbers that the captions refer to). The specification $P$ is not part of it: $P$
is sent as the first user message, after a short note that explains its fields
(\texttt{name}, \texttt{summary}, \texttt{roles[]} and \texttt{behaviour[]}), and
it is the only part of the prompt that changes from one run to the next.
Section~\ref{sec:method-roles} describes the ReAct loop and the four tools that
the LLM uses to explore the codebase; a fifth tool, \texttt{submit\_mapping},
returns the final role mapping and ends the run.

\begin{lstlisting}[style=pyprompt]
AGENT_SYSTEM = """\
You are a security pattern analyst. Given a security pattern and a codebase, decide \
which of the pattern's roles the codebase implements, and pin each implemented role to \
the exact code that plays it.

(*@\phead{\#\# The loop}@*)

Work in ReAct steps — Thought (reason about what you know and what to do next),
Action (one tool call), Observation (its result) — until every role is accounted
for, then call `submit_mapping`, which ends the run.

(*@\phead{\#\# How to find the roles}@*)

`behaviour[]` is ONE execution path, not a checklist. A codebase usually holds
several flows sharing the same vocabulary; only one is the pattern's, and every
role must sit on it.

- Find the entry point first — the `to` of step 1 — by its `description`: the
  code a request lands on is not automatically it. Where a framework injects a
  guard (dependency, filter, middleware, decorator, interceptor), the guard is the
  role that checks and the handler it guards processes afterwards. This is the
  only role you may find by searching.
- Then walk `behaviour[]` in order: for each step `X → Y`, read X's code and find
  the call that performs the step; what it reaches is Y. Follow non-role helpers
  through — the role is whatever ends up doing the work.
- Every later role must arrive as the callee of a role already grounded; name the
  caller and line in your Thought ("reached from Enforcer's login at
  routes.py:90"). If you cannot, keep tracing. A name-search hit you never saw
  called from the path is not the answer — re-read the previous role instead.
- Off the path means `missing`: work done only where the path never reaches
  (registration, setup, another endpoint) does not play the role — say in
  `notes` where it lives. DO NOT invent entities.
- Cite the declaration of the callable that does the work. For a store or
  registry, cite the handle or accessor the calling role uses at the call site —
  not a model or class declared elsewhere, not a table or field.
- Cite a library call only for the role the path explicitly calls it for: if the
  repo calls `bcrypt.checkpw`, it is the Verifier, not also the Hasher or
  Comparator it may run internally. A role you never see the repo perform is
  `missing`.

Tools (arguments and limits are in each one's schema):

- `list_dir` — recursive tree of a directory.
- `search` — semantic search of the indexed code, returning the source of every
  strong match.
- `grep` — exact text search over the whole repo.
- `read_file` — a real file by path, with line numbers.

\end{lstlisting}
\vspace{-2pt}
\captionof{figure}{The system prompt of the agent (lines 1--48). Four of its five
  parts are shown here, in order: the task (lines 2--4), the ReAct loop (6--10),
  how to find the roles (12--39) and the tools (41--47). The fifth part, the
  expected output, continues in Figure~\ref{fig:prompt-agent-cont}. The
  specification $P$ is not part of the system prompt: it is sent as the first user
  message.}
\label{fig:prompt-agent}
\vspace{6pt}

\begin{lstlisting}[style=pyprompt,firstnumber=49]
(*@\phead{\#\# Expected output \rule[0.55ex]{0.6em}{0.35pt} submit\_mapping}@*)

`role_mappings` holds exactly one row per role in `roles[]`, missing ones
included; each row was reached from an earlier one.

```json
{
  "role_mappings": [
    {
      "patternRole": "Enforcer",
      "status": "present",
      "entity": "login",
      "file": "src/api/routes.py",
      "lines": "88",
      "notes": "Entry point; calls authenticate_user at routes.py:90."
    },
    {
      "patternRole": "Verification manager",
      "status": "present",
      "entity": "authenticate_user",
      "file": "src/auth/service.py",
      "lines": "42",
      "notes": "Reached from Enforcer's login at routes.py:90; drives the check."
    },
    {
      "patternRole": "Encrypter",
      "status": "missing",
      "entity": "",
      "file": "",
      "lines": "",
      "notes": "Nothing on the traced path encrypts; the only such code is off-path."
    }
  ]
}
```

- `patternRole` — the role's `name`, spelled exactly as in the pattern JSON.
- `status` — `"present"` (fully implemented by in-repo code), `"external"` (a
  library, framework, or service the repo merely imports, calls, or configures),
  or `"missing"` (no implementation on the traced path).
- `entity` — the bare name the path reaches, as declared or called:
  `"authenticate_user"`, never `"AuthManager.authenticate_user"`. Prefer the
  function or method over its class.
- `file`, `lines` — the repo-relative path and the *single* definition or
  assignment line (`"42"`, never a body span); for an external symbol, the
  import or call site.
- `notes` — which role and line reached this one, and what the code does. For
  `"missing"`, where you looked and where any off-path implementation lives.

`entity`, `file` and `lines` are empty strings only when the role is `"missing"`.
"""
\end{lstlisting}
\vspace{-2pt}
\captionof{figure}{The system prompt of the agent, continued (lines 49--99): the
  expected output. An example call of \texttt{submit\_mapping} shows the format of
  the role mapping, and the list below it explains each field, including which
  code symbol a role is mapped to and which line of its file is given.}
\label{fig:prompt-agent-cont}
\vspace{6pt}

\textbf{Prompts of the ablation settings.} The two settings of RQ4
(Section~\ref{sec:setup-ablations}), each without one of the two search tools, use
the same system prompt with that tool taken out of it. Without keyword search,
line~46, which lists \texttt{grep}, is removed. Without semantic search,
lines~44--45, which list \texttt{search}, are removed, and two phrases that mention
searching are reworded so that they do not point to the missing tool: ``by
searching'' (line~22) becomes ``by looking it up directly'', and ``name-search
hit'' (line~28) becomes ``name-match hit''. The rest of the prompt is unchanged.

\subsection{The One-shot Baseline}
\label{app:prompts-baseline}

The one-shot baseline of Section~\ref{sec:setup-scalability} sends the LLM the
system prompt of Figure~\ref{fig:prompt-baseline} and one user message. The user
message holds the specification $P$ and the whole repository, each between two
tags: \texttt{<pattern>}$P$\texttt{</pattern>}, followed by
\texttt{<repository>}\ldots\texttt{</repository>}, which lists the directory tree
of the repository and then every source file, each marked with its path. The
baseline has no tools: the LLM gives its whole answer, the role mapping and the
interaction order, in one reply, and the same request is sent to every LLM. A
request longer than 900K tokens, counted with the \texttt{o200k\_base} tokenizer,
is not sent; these are the 16 instances over the baseline's input limit.

\begin{lstlisting}[style=pyprompt]
BASELINE_SYSTEM = """You are a security-pattern recovery analyst.

You are given one security pattern as JSON and the full source of one repository.
For every role in the pattern, decide whether the repository implements it and
ground it to the exact code, then give the order those roles interact in when the
repository runs.

Answer with one JSON object of exactly this shape, and nothing else:

{
  "role_mappings": [
    {
      "patternRole": "the pattern's role name, verbatim",
      "status": "present | external | missing",
      "entity": "the code entity's bare declared name",
      "file": "repo-relative path",
      "lines": "the entity's declaration line",
      "notes": "why this code represents the role"
    }
  ],
  "interaction_order": [
    {
      "from": "role that invokes",
      "to": "role it invokes",
      "file": "repo-relative file holding the call site",
      "line": 0,
      "notes": "the call, in a few words"
    }
  ]
}

"role_mappings" holds one entry per role in the pattern, none left out:

  status   "present" for in-repo code; "external" for a library,
           framework or service the repo only calls or configures; "missing"
           if not found.
  entity   The bare name as declared, with no enclosing class or module
           qualifier - "authenticate_user", or "hashpw" for a library symbol.
           Logic with no symbol of its own is cited by the innermost named
           entity containing it. Empty only for a role not found.
  file     Where the entity is declared; for a library symbol, where the repo
           imports or calls it. Empty if the role is missing.
  lines    The entity's single declaration line, e.g. "12". Empty if missing.

"interaction_order" lists the interactions in execution order, one entry each,
its "line" the call site's line number or 0 where that cannot be pinpointed.
It is [] if the order cannot be told."""
\end{lstlisting}
\vspace{-2pt}
\captionof{figure}{The system prompt of the one-shot baseline.}
\label{fig:prompt-baseline}
\vspace{6pt}

\FloatBarrier

\subsection{Computing the Metrics on an Example}
\label{app:irc-example}
Figures~\ref{fig:metrics} and~\ref{fig:l3} show how RDA, MLS and IRC (Section~\ref{sec:setup-evaluation}) are computed on a small example.

\newcommand{\lrow}[8]{%
  \pgfmathsetmacro{\yy}{-0.5*#1-0.85}%
  \node[rtag] at (0,\yy) {#2};%
  \node[#3] at (3.55,\yy) {#4};%
  \node[#5] at (4.8,\yy) {#6};%
  \ifnum#7=1 \node[ok] at (5.85,\yy) {\checkmark};\else \node[bad] at (5.85,\yy) {$\times$};\fi
  \ifnum#8=1 \fill[gray!55!black] (6.4,\yy) circle (1.6pt);\fi
  \draw[gray!25] (0,\yy-0.25) -- (6.6,\yy-0.25);}
\newcommand{\elbow}[3][]{\draw[#1] (#2.south) -- ++(0,-0.225) -| (#3.north);}
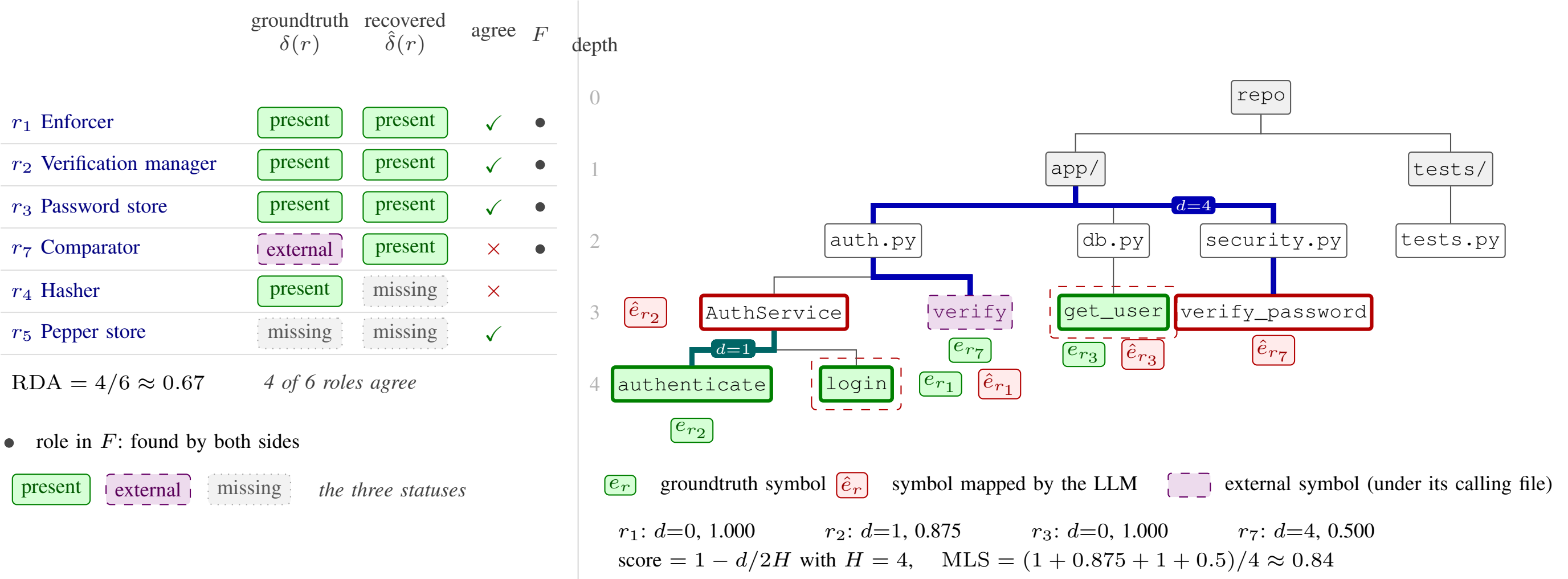
\begin{figure*}[t]
\centering
\resizebox{\textwidth}{!}{%
\begin{tikzpicture}[
  font=\scriptsize,
  chip/.style={rounded corners=1.5pt, minimum width=1.0cm, minimum height=0.36cm, inner xsep=3pt, inner ysep=1pt},
  spres/.style={chip, draw=green!50!black, fill=green!16, text=green!35!black},
  sext/.style={chip, draw=violet!80!black, dashed, fill=violet!14, text=violet!70!black},
  smiss/.style={chip, draw=gray!70, dotted, fill=gray!10, text=gray!70!black},
  rtag/.style={text=blue!50!black, anchor=west},
  hdr/.style={align=center, text=gray!45!black},
  ok/.style={text=green!45!black},
  bad/.style={text=red!70!black},
  tn/.style={draw=gray!70!black, fill=white, rounded corners=1pt, inner sep=1.6pt,
             font=\ttfamily\scriptsize, minimum height=0.4cm},
  dn/.style={tn, fill=gray!12},
  xn/.style={tn, dashed, draw=violet!80!black, fill=violet!14, text=violet!70!black},
  gn/.style={tn, draw=green!50!black, very thick, fill=green!16},
  ln/.style={tn, draw=red!70!black, very thick},
  edge/.style={gray!65!black, thin},
  pA/.style={teal!80!black, line width=1.7pt, opacity=0.85},
  pB/.style={blue!70!black, line width=1.7pt, opacity=0.85},
  dchip/.style={rounded corners=1.5pt, inner sep=1.3pt, text=white, font=\tiny\bfseries},
  gtag/.style={draw=green!50!black, fill=green!16, text=green!35!black, rounded corners=1.5pt, inner sep=1.3pt},
  ltag/.style={draw=red!70!black, fill=red!8, text=red!60!black, rounded corners=1.5pt, inner sep=1.3pt},
  note/.style={font=\scriptsize\itshape, text=gray!50!black}
]
\draw[gray!30] (6.85,0.1) -- (6.85,-6.9);

\node[hdr] at (3.55,-0.3) {groundtruth\\$\delta(r)$};
\node[hdr] at (4.8,-0.3) {recovered\\$\hat{\delta}(r)$};
\node[hdr] at (5.85,-0.3) {agree};
\node[hdr] at (6.4,-0.3) {$F$};
\lrow{1}{$r_1$ Enforcer}{spres}{present}{spres}{present}{1}{1}
\lrow{2}{$r_2$ Verification manager}{spres}{present}{spres}{present}{1}{1}
\lrow{3}{$r_3$ Password store}{spres}{present}{spres}{present}{1}{1}
\lrow{4}{$r_7$ Comparator}{sext}{external}{spres}{present}{0}{1}
\lrow{5}{$r_4$ Hasher}{spres}{present}{smiss}{missing}{0}{0}
\lrow{6}{$r_5$ Pepper store}{smiss}{missing}{smiss}{missing}{1}{0}
\node[anchor=west] at (0,-4.45) {$\mathrm{RDA} = 4/6 \approx 0.67$};
\node[note, anchor=west] at (3.0,-4.45) {4 of 6 roles agree};
\fill[gray!55!black] (0.1,-5.15) circle (1.6pt);
\node[anchor=west, align=left] at (0.3,-5.15) {role in $F$: found by both sides};
\node[spres, minimum width=0.85cm] at (0.6,-5.7) {present};
\node[sext, minimum width=0.85cm] at (1.75,-5.7) {external};
\node[smiss, minimum width=0.85cm] at (2.95,-5.7) {missing};
\node[note, anchor=west] at (3.65,-5.7) {the three statuses};
\node[align=center] at (3.3,-6.95) {\footnotesize (a) Role detection accuracy (RDA)};

\begin{scope}[xshift=7.2cm]
\foreach \d in {0,...,4} \node[text=gray!60] at (-0.15,-1.05-0.85*\d) {\d};
\node[hdr] at (-0.15,-0.45) {depth};
\node[dn] (root) at (7.75,-1.05)  {repo};
\node[dn] (app)  at (5.55,-1.9)  {app/};
\node[dn] (tests) at (10.0,-1.9) {tests/};
\node[tn] (auth) at (3.15,-2.75)  {auth.py};
\node[tn] (db)   at (6.0,-2.75)   {db.py};
\node[tn] (sec)  at (7.9,-2.75)   {security.py};
\node[tn] (tpy)  at (10.0,-2.75)  {tests.py};
\node[ln] (as) at (1.975,-3.6) {AuthService};
\node[xn] (vf) at (4.3,-3.6)   {verify};
\node[gn] (gu) at (6.0,-3.6)   {get\_user};
\node[ln] (vp) at (7.9,-3.6)   {verify\_password};
\node[gn] (au) at (1.0,-4.45)  {authenticate};
\node[gn] (lo) at (2.95,-4.45) {login};
\foreach \n in {gu, lo}
  \draw[red!70!black, dashed, rounded corners=1.5pt] ([shift={(-2.4pt,-2.4pt)}]\n.south west) rectangle ([shift={(2.4pt,2.4pt)}]\n.north east);
\foreach \p/\c in {root/app, root/tests, app/auth, app/db, app/sec, tests/tpy, auth/as, auth/vf, as/au, as/lo, db/gu, sec/vp}
  \elbow[edge]{\p}{\c}
\elbow[pA]{as}{au}
\elbow[pB]{auth}{vf}
\elbow[pB]{app}{auth}
\elbow[pB]{app}{sec}
\elbow[pB]{sec}{vp}
\node[dchip, fill=teal!80!black] at (1.49,-4.03) {$d{=}1$};
\node[dchip, fill=blue!70!black] at (6.95,-2.33) {$d{=}4$};
\node[gtag] at (3.95,-4.45) {$e_{r_1}$};
\node[ltag] at (4.65,-4.45) {$\hat{e}_{r_1}$};
\node[gtag] at (1.0,-5.0)   {$e_{r_2}$};
\node[ltag] at (0.45,-3.6)  {$\hat{e}_{r_2}$};
\node[gtag] at (5.65,-4.1)  {$e_{r_3}$};
\node[ltag] at (6.35,-4.1)  {$\hat{e}_{r_3}$};
\node[gtag] at (4.3,-4.05)  {$e_{r_7}$};
\node[ltag] at (7.9,-4.05)  {$\hat{e}_{r_7}$};
\node[gtag] at (0.15,-5.65) {$e_r$};
\node[anchor=west] at (0.5,-5.65) {groundtruth symbol};
\node[ltag] at (2.9,-5.65) {$\hat{e}_r$};
\node[anchor=west] at (3.25,-5.65) {symbol mapped by the LLM};
\node[xn, minimum width=0.5cm, minimum height=0.28cm] at (6.9,-5.65) {};
\node[anchor=west] at (7.2,-5.65) {external symbol (under its calling file)};
\node[anchor=west] at (0.0,-6.2)  {$r_1$: $d{=}0$, 1.000};
\node[anchor=west] at (2.45,-6.2) {$r_2$: $d{=}1$, 0.875};
\node[anchor=west] at (4.9,-6.2)  {$r_3$: $d{=}0$, 1.000};
\node[anchor=west] at (7.35,-6.2) {$r_7$: $d{=}4$, 0.500};
\node[anchor=west] at (0.0,-6.55) {score $= 1 - d/2H$ with $H = 4$, \quad $\mathrm{MLS} = (1 + 0.875 + 1 + 0.5)/4 \approx 0.84$};
\node[align=center] at (5.2,-6.95) {\footnotesize (b) Mapping localization score (MLS)};
\end{scope}
\end{tikzpicture}}
\caption{RDA and MLS on a small example. (a)~RDA: 4 of the 6 roles have the same presence status in the groundtruth and in the recovery; the 4 roles in $F$ are found by both. (b)~MLS: for each role in $F$, the distance $d$ in the containment tree between the groundtruth symbol $e_r$ (green) and the mapped symbol $\hat{e}_r$ (red), scored $1 - d/2H$.}
\Description{Two panels side by side. Panel a is a table of six roles of the password-based authentication pattern, numbered as in Figure 1, with the groundtruth presence status and the recovered presence status of each role as a coloured label: present in solid green, external in dashed violet, missing in grey. Four of the six rows agree, so RDA is 4 over 6. The roles r1, r2, r3 and r7 are marked as found by both sides. Panel b is the containment tree $E(C)$ of a codebase, from the repository through directories and files to a class and its methods. For each of the four roles, a green tag marks the groundtruth symbol and a red tag marks the symbol the language model mapped. Role r1 is mapped to login, exactly, at distance 0. Role r2 is the method authenticate in the groundtruth and its class AuthService in the recovery, at distance 1. Role r3 is get user, exactly. Role r7 is an external verify call placed under the file auth.py in the groundtruth and a function verify password in another file in the recovery, at distance 4. The scores are 1, 0.875, 1 and 0.5, and MLS is about 0.84.}
\label{fig:metrics}
\end{figure*}

\usetikzlibrary{intersections}
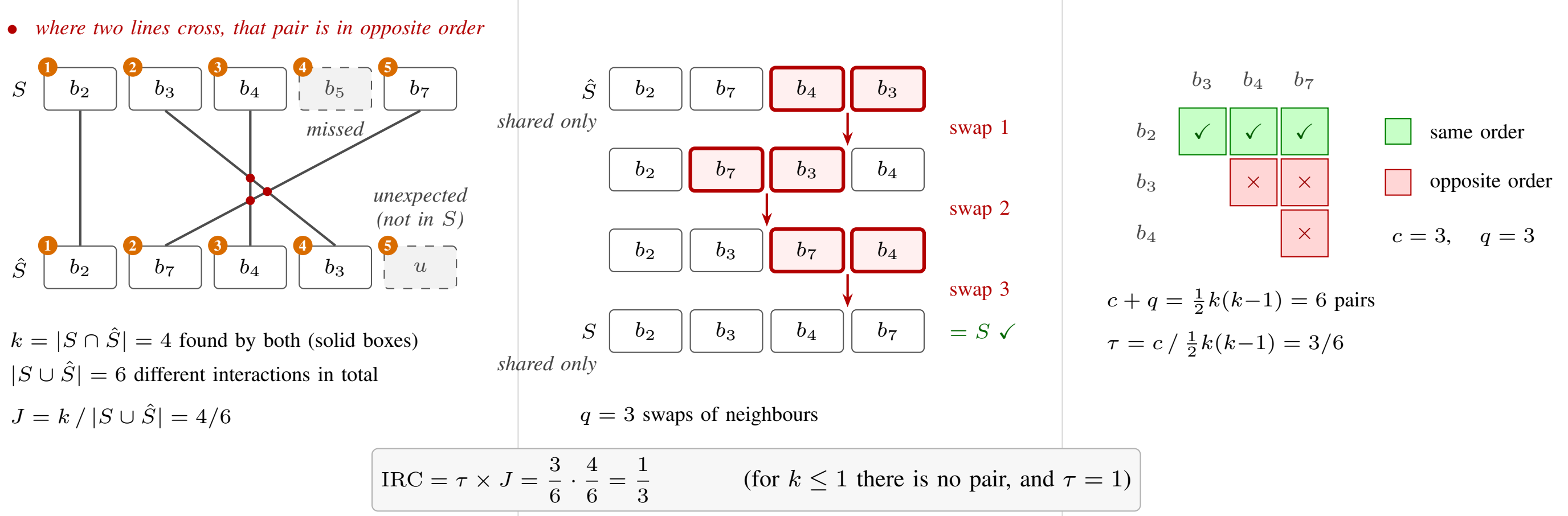
\begin{figure*}[t]
\centering
\resizebox{\textwidth}{!}{%
\begin{tikzpicture}[
  font=\scriptsize,
  ib/.style={draw=gray!70!black, fill=white, rounded corners=1.5pt, minimum width=0.85cm,
             minimum height=0.5cm, inner sep=1pt},
  ibx/.style={ib, dashed, fill=gray!10, text=gray!55!black},
  ibs/.style={ib, draw=red!70!black, very thick, fill=red!6},
  num/.style={circle, fill=orange!85!black, text=white, font=\tiny\bfseries, inner sep=0.8pt},
  cok/.style={draw=green!50!black, fill=green!22, minimum size=0.55cm, inner sep=0pt, text=green!35!black},
  cbad/.style={draw=red!70!black, fill=red!16, minimum size=0.55cm, inner sep=0pt, text=red!60!black},
  note/.style={font=\scriptsize\itshape, text=gray!50!black},
  hdr/.style={align=center, text=gray!45!black}
]
\draw[gray!30] (6.1,0.1) -- (6.1,-6.5);
\draw[gray!30] (12.5,0.1) -- (12.5,-6.5);

\fill[red!70!black] (0.15,-0.3) circle (1.5pt);
\node[note, anchor=west, text=red!70!black] at (0.3,-0.3) {where two lines cross, that pair is in opposite order};
\node[anchor=east] at (0.45,-1.0) {$S$};
\node[anchor=east] at (0.45,-3.1) {$\hat{S}$};
\node[ib]  (s1) at (0.95,-1.0) {$b_2$};
\node[ib]  (s2) at (1.95,-1.0) {$b_3$};
\node[ib]  (s3) at (2.95,-1.0) {$b_4$};
\node[ibx] (s4) at (3.95,-1.0) {$b_5$};
\node[ib]  (s5) at (4.95,-1.0) {$b_7$};
\node[ib]  (r1) at (0.95,-3.1) {$b_2$};
\node[ib]  (r2) at (1.95,-3.1) {$b_7$};
\node[ib]  (r3) at (2.95,-3.1) {$b_4$};
\node[ib]  (r4) at (3.95,-3.1) {$b_3$};
\node[ibx] (r5) at (4.95,-3.1) {$u$};
\foreach \b/\n in {s1/1,s2/2,s3/3,s4/4,s5/5,r1/1,r2/2,r3/3,r4/4,r5/5}
  \node[num] at ([shift={(0.05,0)}]\b.north west) {\n};
\path[name path=l2, draw=gray!60!black, thick] (s1.south) -- (r1.north);
\path[name path=l3, draw=gray!60!black, thick] (s2.south) -- (r4.north);
\path[name path=l4, draw=gray!60!black, thick] (s3.south) -- (r3.north);
\path[name path=l7, draw=gray!60!black, thick] (s5.south) -- (r2.north);
\fill[name intersections={of=l3 and l4}, red!70!black] (intersection-1) circle (1.5pt);
\fill[name intersections={of=l3 and l7}, red!70!black] (intersection-1) circle (1.5pt);
\fill[name intersections={of=l4 and l7}, red!70!black] (intersection-1) circle (1.5pt);
\node[note, anchor=north] at (s4.south) {missed};
\node[note, anchor=south, align=center] at ([yshift=2pt]r5.north) {unexpected\\(not in $S$)};
\node[anchor=west] at (0,-3.95) {$k = |S \cap \hat{S}| = 4$ found by both (solid boxes)};
\node[anchor=west] at (0,-4.35) {$|S \cup \hat{S}| = 6$ different interactions in total};
\node[anchor=west] at (0,-4.85) {$J = k\,/\,|S \cup \hat{S}| = 4/6$};
\node[align=center] at (3.0,-6.25) {\footnotesize (a) Jaccard index $J$: what both sequences found};

\begin{scope}[xshift=6.6cm]
\node[anchor=east] at (0.55,-1.0)  {$\hat{S}$};
\node[anchor=east] at (0.55,-3.85) {$S$};
\node[note, anchor=east, align=right] at (0.55,-1.4) {shared only};
\node[note, anchor=east, align=right] at (0.55,-4.25) {shared only};
\node[ib]  at (1.0,-1.0)  {$b_2$}; \node[ib]  at (1.95,-1.0) {$b_7$};
\node[ibs] (a3) at (2.9,-1.0)  {$b_4$}; \node[ibs] (a4) at (3.85,-1.0) {$b_3$};
\node[ib]  at (1.0,-1.95) {$b_2$}; \node[ibs] (b2) at (1.95,-1.95) {$b_7$};
\node[ibs] (b3) at (2.9,-1.95) {$b_3$}; \node[ib]  at (3.85,-1.95) {$b_4$};
\node[ib]  at (1.0,-2.9)  {$b_2$}; \node[ib]  at (1.95,-2.9) {$b_3$};
\node[ibs] (c3) at (2.9,-2.9)  {$b_7$}; \node[ibs] (c4) at (3.85,-2.9) {$b_4$};
\node[ib]  at (1.0,-3.85) {$b_2$}; \node[ib]  at (1.95,-3.85) {$b_3$};
\node[ib]  at (2.9,-3.85) {$b_4$}; \node[ib]  at (3.85,-3.85) {$b_7$};
\draw[-{Stealth[length=1.6mm]}, red!70!black, thick] (3.375,-1.28) -- (3.375,-1.67);
\draw[-{Stealth[length=1.6mm]}, red!70!black, thick] (2.425,-2.23) -- (2.425,-2.62);
\draw[-{Stealth[length=1.6mm]}, red!70!black, thick] (3.375,-3.18) -- (3.375,-3.57);
\node[anchor=west, text=red!70!black] at (4.45,-1.475) {swap 1};
\node[anchor=west, text=red!70!black] at (4.45,-2.425) {swap 2};
\node[anchor=west, text=red!70!black] at (4.45,-3.375) {swap 3};
\node[anchor=west, text=green!40!black] at (4.45,-3.85) {$=S$ \checkmark};
\node[anchor=west] at (0.1,-4.85) {$q = 3$ swaps of neighbours};
\node[align=center] at (2.9,-6.25) {\footnotesize (b) Swaps: $q$ neighbour swaps turn $\hat{S}$ into $S$};
\end{scope}

\begin{scope}[xshift=12.9cm]
\node[hdr] at (0.0,-0.95) {};
\foreach \x/\t in {1.25/b_3, 1.85/b_4, 2.45/b_7}
  \node[text=gray!55!black] at (\x,-0.9) {$\t$};
\foreach \y/\t in {-1.5/b_2, -2.1/b_3, -2.7/b_4}
  \node[text=gray!55!black, anchor=east] at (0.85,\y) {$\t$};
\node[cok] at (1.25,-1.5) {\checkmark}; \node[cok] at (1.85,-1.5) {\checkmark}; \node[cok] at (2.45,-1.5) {\checkmark};
\node[cbad] at (1.85,-2.1) {$\times$};   \node[cbad] at (2.45,-2.1) {$\times$};
\node[cbad] at (2.45,-2.7) {$\times$};
\node[cok, minimum size=0.3cm] at (3.55,-1.5) {};
\node[anchor=west] at (3.8,-1.5) {same order};
\node[cbad, minimum size=0.3cm] at (3.55,-2.1) {};
\node[anchor=west] at (3.8,-2.1) {opposite order};
\node[anchor=west] at (3.35,-2.75) {$c = 3$, \quad $q = 3$};
\node[anchor=west] at (0.0,-3.5)  {$c + q = \tfrac{1}{2}k(k{-}1) = 6$ pairs};
\node[anchor=west] at (0.0,-4.0)  {$\tau = c\,/\,\tfrac{1}{2}k(k{-}1) = 3/6$};
\node[align=center] at (2.4,-6.25) {\footnotesize (c) Pairs: $c$, $q$ and $\tau$};
\end{scope}

\node[draw=gray!60, rounded corners=2pt, inner sep=3pt, fill=gray!6] at (8.9,-5.6)
  {$\mathrm{IRC} = \tau \times J = \dfrac{3}{6} \cdot \dfrac{4}{6} = \dfrac{1}{3}$ \qquad\qquad {\footnotesize (for $k \le 1$ there is no pair, and $\tau = 1$)}};
\end{tikzpicture}}
\caption{IRC on a small example, with interactions named by the actions $b_j$ of Figure~\ref{fig:pattern-example}. $\hat{S}$ misses $b_5$ and has an unexpected interaction $u$. (a)~$k = 4$ of the 6 interactions are shared, so $J = 4/6$. (b)~Turning $\hat{S}$ into $S$ takes $q = 3$ neighbour swaps. (c)~Of the 6 pairs of shared interactions, $c = 3$ keep their order, so $\tau = 3/6$.}
\Description{Three panels side by side and a formula below them. Panel a shows the groundtruth sequence S of five interactions b2, b3, b4, b5 and b7 above the recovered sequence of five interactions b2, b7, b4, b3 and u, with lines joining the four interactions that are in both. Three pairs of lines cross, and each crossing marks a pair in opposite order. The interaction b5 is only in S and is marked as missed, and u is only in the recovered sequence and is marked as unexpected, outside the groundtruth set. Four interactions are shared out of six different ones, so the Jaccard index J is 4 over 6. Panel b shows four rows of the four shared interactions. The first row is the recovered order b2, b7, b4, b3, and each of the next rows swaps one pair of neighbours, highlighted in red, until the last row is the groundtruth order b2, b3, b4, b7 after three swaps. Panel c is a triangle of the six pairs of shared interactions, with three green cells for pairs in the same order and three red cells for pairs in opposite order, so tau is 3 over 6. IRC is tau times J, which is 3 over 6 times 4 over 6, or one third.}
\label{fig:l3}
\end{figure*}

\FloatBarrier

\section{Recovery Output Format}
\label{app:output}

Figure~\ref{fig:role-mapping-json} shows the output of our pipeline for the login
handler of Figure~\ref{fig:code-example}: the role mappings $\hat{M}$ and presence
statuses $\hat{\mathcal{D}}$ that the LLM emits (Section~\ref{sec:method-roles}),
and the interaction sequence $\hat{S}$ that the CPG analysis derives from them
(Section~\ref{sec:interaction-sequence-recovery}).

\begin{figure*}[tb]
\centering
\begin{minipage}[t]{0.485\textwidth}
\vspace{0pt}
\begin{lstlisting}[style=json,basicstyle=\scriptsize\ttfamily,escapeinside={(*@}{@*)}]
{ "role_mappings": [
    {"patternRole": "(*@\jrole{Enforcer}@*)",
     "status": "(*@\jpres{present}@*)", "entity": "(*@\jpres{login}@*)",
     "file": "app.py", "lines": "16"},
    {"patternRole": "(*@\jrole{Verification manager}@*)",
     "status": "(*@\jpres{present}@*)", "entity": "(*@\jpres{authenticate}@*)",
     "file": "app.py", "lines": "9"},
    {"patternRole": "(*@\jrole{Password store}@*)",
     "status": "(*@\jpres{present}@*)", "entity": "(*@\jpres{get\_user}@*)",
     "file": "app.py", "lines": "6"},
    {"patternRole": "(*@\jrole{Comparator}@*)",
     "status": "(*@\jext{external}@*)", "entity": "(*@\jext{verify}@*)",
     "file": "app.py", "lines": "11"} ] }
\end{lstlisting}
\centering\small (a) Role mappings $\hat{M}$ and presence statuses $\hat{\mathcal{D}}$
\end{minipage}\hfill
\begin{minipage}[t]{0.485\textwidth}
\vspace{0pt}
\begin{lstlisting}[style=json,basicstyle=\scriptsize\ttfamily,escapeinside={(*@}{@*)}]
{ "interaction_order": {
    "backend": "joern",
    "entryRole": "(*@\jrole{Enforcer}@*)",
    "interactions": [
      {"rank": (*@\jrank{1}@*), "from": "(*@\jrole{Enforcer}@*)",
       "to": "(*@\jrole{Verification manager}@*)",
       "file": "app.py", "line": 17},
      {"rank": (*@\jrank{2}@*), "from": "(*@\jrole{Verification manager}@*)",
       "to": "(*@\jrole{Password store}@*)",
       "file": "app.py", "line": 10},
      {"rank": (*@\jrank{3}@*), "from": "(*@\jrole{Verification manager}@*)",
       "to": "(*@\jrole{Comparator}@*)",
       "file": "app.py", "line": 11} ] } }
\end{lstlisting}
\centering\small (b) Interaction sequence $\hat{S}$
\end{minipage}
\caption{The expected recovery output for the login handler of Figure~\ref{fig:code-example}, as JSON: (a) the role mappings $\hat{M}$ and presence statuses $\hat{\mathcal{D}}$ emitted by the LLM, and (b) the interaction sequence $\hat{S}$ derived from the CPG. }
\Description{Two JSON listings side by side. The left one lists four role mappings, Enforcer, Verification manager, Password store and Comparator, with a presence status, entity, file and line each; the first three are present and highlighted in green, the Comparator is external and highlighted in violet. The right one gives the interaction order starting from the entry role Enforcer: three ranked interactions, Enforcer to Verification manager, Verification manager to Password store, and Verification manager to Comparator, each with the file and line of the call.}
\label{fig:role-mapping-json}
\end{figure*}

\FloatBarrier

\section{Using Code Property Graphs for Interaction Recovery}
\label{app:order}

\lstdefinestyle{walkcode}{
  basicstyle=\scriptsize\ttfamily, columns=fullflexible, keepspaces=true,
  showstringspaces=false, keywordstyle=\bfseries,
  commentstyle=\color{gray!60!black}, frame=none,
  aboveskip=0pt, belowskip=0pt}
\lstdefinestyle{walklines}{style=walkcode, numbers=left, numbersep=5pt,
  numberstyle=\tiny\color{gray!70!black}, xleftmargin=1.4em}
\newcommand{\codefile}[1]{{\sffamily\scriptsize\color{gray!55!black}#1}\par\vspace{2pt}}
\newcommand{\codecase}[1]{{\sffamily\scriptsize\bfseries #1}\par\vspace{1pt}}

\newcommand{\jtype}[1]{{\sffamily\bfseries #1}}
\newcommand{\jprop}[1]{{\ttfamily\fontsize{6}{7.4}\selectfont #1}}
\tikzset{
  jbase/.style={draw, rounded corners=1.5pt, fill=white, align=center,
                inner sep=3pt, font=\scriptsize},
  jnode/.style={jbase, draw=gray!70},
  jext/.style={jbase, draw=gray!70, dashed},
  jread/.style={jbase, draw=gray!50, fill=gray!10},
  jrule/.style={jbase, draw=teal!60!black, dashed, fill=teal!4},
  jflow/.style={-{Stealth[length=1.8mm]}, gray!60!black},
  jcall/.style={-{Stealth[length=2mm]}, thick, dashed, orange!85!black},
  jadded/.style={-{Stealth[length=2.4mm]}, line width=1.6pt, teal!65!black},
  jelab/.style={font=\sffamily\tiny, text=gray!40!black, fill=white, inner sep=1pt,
               outer sep=1.5pt},
  jaddlab/.style={font=\sffamily\tiny\bfseries, text=white, fill=teal!65!black, solid,
                  rounded corners=1pt, inner sep=1.6pt},
  jproblem/.style={font=\sffamily\tiny\bfseries, text=red!70!black, solid,
                   draw=red!60!black, fill=red!4, rounded corners=1pt, inner sep=1.6pt},
  jpanel/.style={font=\sffamily\scriptsize\bfseries, text=gray!55!black,
                 anchor=south west, inner sep=0pt},
  jmiss/.style={-{Stealth[length=2mm]}, thick, dashed, red!65!black},
  jfaded/.style={-{Stealth[length=2mm]}, thick, dashed, orange!85!black, opacity=0.3},
  jfrom/.style={densely dotted, line width=0.8pt, teal!60!black},
  jelide/.style={font=\scriptsize\itshape, text=gray!50!black, inner sep=3pt},
}

\setlength{\emergencystretch}{2.5em}

To recover the interaction sequence $\hat{S}$, we traverse the code property graph (CPG)
of the codebase $C$ from the code symbol $\hat{e}$ that one role is mapped to, to the code symbols of the other roles, following the mappings $\hat{m} = (r, \hat{e}) \in \hat{M}$
(Section~\ref{sec:interaction-sequence-recovery}). We build the CPG with Joern, which
implements the CPG of Yamaguchi et al.~\cite{yamaguchi_modeling_2014}. This appendix first
explains why the CPG that Joern builds is not enough for this task on its own
(Appendix~\ref{app:walk-challenges}), and then how we added the missing links and defined
our own traversal rules to move between the roles (Appendix~\ref{app:walk-traversal}).

\subsection{Challenges in Building a Repository-Level CPG}
\label{app:walk-challenges}
\label{app:walk-joern}

In Joern's CPG, every node is one piece of code, a method or a statement or expression
inside it, with a type such as \texttt{METHOD} or \texttt{CALL} and properties such as
\texttt{name} or \texttt{fullName}. Every edge also has a type, and our traversal follows
the four kinds in Table~\ref{tab:app-walk-edges}. In the figures of this appendix, the code
is on the left and the graph on the right, where each box is a node (its type in bold, its
properties below) and each arrow an edge, labelled with its type.

\begin{table*}[t]
\caption{The edges of Joern's graph that our traversal follows. \emph{Joern step}
is the name of the traversal in Joern's query language; \emph{Label} is what
our traversal records on an interaction reached through the edge.}
\label{tab:app-walk-edges}
\footnotesize
\setlength{\tabcolsep}{4pt}
\begin{tabular}{@{}>{\raggedright\arraybackslash}p{0.16\linewidth} >{\raggedright\arraybackslash}p{0.28\linewidth} >{\raggedright\arraybackslash}p{0.28\linewidth} >{\raggedright\arraybackslash}p{0.20\linewidth}@{}}
\toprule
Edge & Joern step & Connects & Label \\
\midrule
Control flow & \texttt{cfgFirst}, \texttt{\_cfgOut} & a node to the node of
  the same method that runs next & --- \\
Call & \texttt{call.callee} & a call to the method it runs & \texttt{call} \\
Reference & \texttt{identifier.\_refOut} & a name to the variable, parameter
  or field it refers to & \texttt{data} \\
Method reference & \texttt{methodRef.referencedMethod} & a function passed as
  a value (a callback, a lambda, a nested function) to its method &
  \texttt{route-registration} if passed to a call, else
  \texttt{nested-definition} \\
\bottomrule
\end{tabular}
\end{table*}

\begin{figure*}[!htbp]
\centering
\begin{minipage}[t]{0.37\linewidth}
\vspace{0pt}
\begin{lstlisting}[style=walkcode, language=Python]
def login(user):
    check(user)
    log(user)
\end{lstlisting}
\end{minipage}\hfill
\begin{minipage}[t]{0.6\linewidth}
\vspace{0pt}\centering
\begin{tikzpicture}[jnode/.append style={minimum width=2.8cm}, node distance=5mm]
  \node[jnode] (n1) {\jtype{METHOD}\\\jprop{name = login}};
  \node[jnode, below=of n1] (n2) {\jtype{CALL}\\\jprop{code = check(user)}};
  \node[jnode, below=of n2] (n3) {\jtype{CALL}\\\jprop{code = log(user)}};
  \node[jnode, below=of n3] (n4) {\jtype{METHOD\_RETURN}\\\jprop{code = RET}};
  \foreach \a/\b in {n1/n2, n2/n3, n3/n4}
    \draw[jflow] (\a) -- node[jelab, right] {CFG} (\b);
\end{tikzpicture}
\end{minipage}
\caption{\texttt{CFG} edges: the chain runs from the \texttt{METHOD} node,
through each call in the order it runs, to the \texttt{METHOD\_RETURN} node.
The argument nodes, which come before each call, are left out.}
\Description{A three-line method login that calls check(user) and then
log(user). On the right, four Joern nodes in a column joined by CFG edges:
METHOD login, CALL check(user), CALL log(user) and METHOD_RETURN.}
\label{fig:app-walk-j-cfg}
\end{figure*}
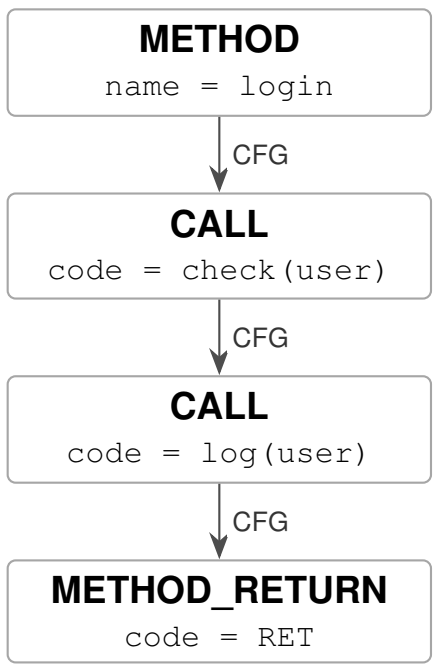

\textbf{The CPG is built per method.} The control-flow edges, which give the order in which
the code runs, never leave a method (Figure~\ref{fig:app-walk-j-cfg}). The graph of a
codebase is therefore a set of islands, one per method, and the only way from one island to
another is a call edge. The code symbols $\hat{e}$ that the roles $r \in R$ are mapped to, however, are rarely in the
same method, and often not in the same file, so recovering their interactions depends almost
entirely on the call edges between methods across the codebase.

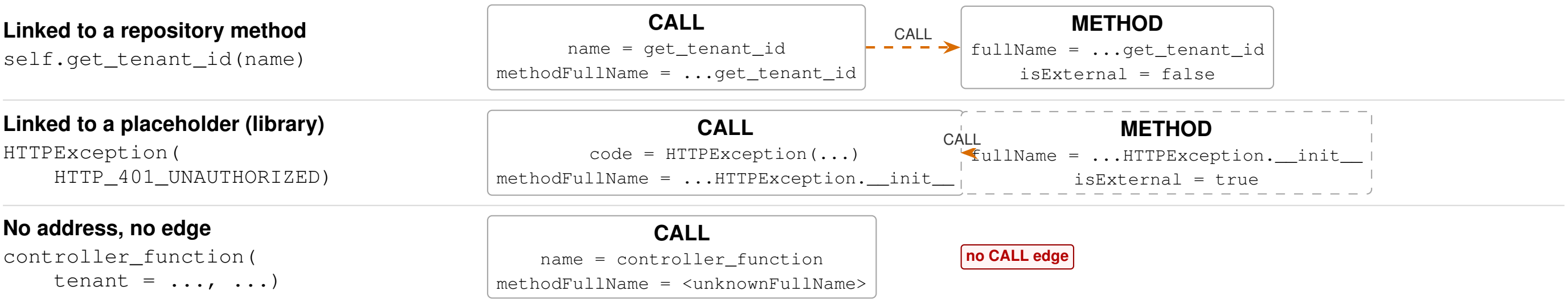
\begin{figure*}[!htbp]
\centering
\begin{minipage}[c]{0.29\linewidth}
\codecase{Linked to a repository method}
\begin{lstlisting}[style=walkcode, language=Python]
self.get_tenant_id(name)
\end{lstlisting}
\end{minipage}\hfill
\begin{minipage}[c]{0.69\linewidth}
\raggedright
\begin{tikzpicture}
  \node[jnode, anchor=west] (k) at (0,0) {\jtype{CALL}\\\jprop{name = get\_tenant\_id}\\\jprop{methodFullName = ...get\_tenant\_id}};
  \node[jnode, anchor=west] (m) at (5.5,0) {\jtype{METHOD}\\\jprop{fullName = ...get\_tenant\_id}\\\jprop{isExternal = false}};
  \draw[jcall] (k) -- node[jelab, above] {CALL} (m);
\end{tikzpicture}
\end{minipage}

\par\smallskip\noindent{\color{gray!30}\rule{\linewidth}{0.4pt}}\par\smallskip
\begin{minipage}[c]{0.29\linewidth}
\codecase{Linked to a placeholder (library)}
\begin{lstlisting}[style=walkcode, language=Python]
HTTPException(
    HTTP_401_UNAUTHORIZED)
\end{lstlisting}
\end{minipage}\hfill
\begin{minipage}[c]{0.69\linewidth}
\raggedright
\begin{tikzpicture}
  \node[jnode, anchor=west] (k) at (0,0) {\jtype{CALL}\\\jprop{code = HTTPException(...)}\\\jprop{methodFullName = ...HTTPException.\_\_init\_\_}};
  \node[jext, anchor=west] (m) at (5.5,0) {\jtype{METHOD}\\\jprop{fullName = ...HTTPException.\_\_init\_\_}\\\jprop{isExternal = true}};
  \draw[jcall] (k) -- node[jelab, above] {CALL} (m);
\end{tikzpicture}
\end{minipage}

\par\smallskip\noindent{\color{gray!30}\rule{\linewidth}{0.4pt}}\par\smallskip
\begin{minipage}[c]{0.29\linewidth}
\codecase{No address, no edge}
\begin{lstlisting}[style=walkcode, language=Python]
controller_function(
    tenant = ..., ...)
\end{lstlisting}
\end{minipage}\hfill
\begin{minipage}[c]{0.69\linewidth}
\raggedright
\begin{tikzpicture}
  \node[jnode, anchor=west] (k) at (0,0) {\jtype{CALL}\\\jprop{name = controller\_function}\\\jprop{methodFullName = <unknownFullName>}};
  \node[jproblem, anchor=west] at (5.5,0) {no \jtype{CALL} edge};
\end{tikzpicture}
\end{minipage}
\caption{The three outcomes of linking a call. Dashed orange arrows are \texttt{CALL} edges; the dashed
box is a placeholder method, which has no code.}
\Description{Three real calls, each with its code on the left and its Joern
nodes on the right. get_tenant_id: a CALL whose methodFullName equals the
fullName of a repository METHOD, joined by a CALL edge. HTTPException: a CALL
edge to a placeholder METHOD with isExternal true. controller_function:
methodFullName unknownFullName and no CALL edge.}
\label{fig:app-walk-j-call}
\end{figure*}

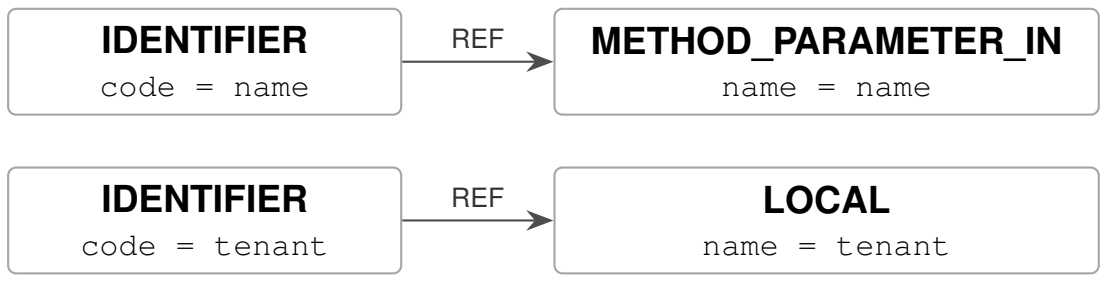
\begin{figure*}[!htbp]
\centering
\begin{minipage}[t]{0.4\linewidth}
\vspace{0pt}
\begin{lstlisting}[style=walkcode, language=Python, deletekeywords={id}]
def show_tenant(self, name):
  return self.proxy_controller_function(
    tenant.get_tenant,
    id=self.get_tenant_id(name))
\end{lstlisting}
\end{minipage}\hfill
\begin{minipage}[t]{0.57\linewidth}
\vspace{0pt}\centering
\begin{tikzpicture}[jnode/.append style={minimum height=7mm}]
  \node[jnode, minimum width=2.6cm] (r1) at (0,0) {\jtype{IDENTIFIER}\\\jprop{code = name}};
  \node[jnode, minimum width=3.6cm, right=10mm of r1] (r2) {\jtype{METHOD\_PARAMETER\_IN}\\\jprop{name = name}};
  \node[jnode, minimum width=2.6cm] (r3) at (0,-1.05) {\jtype{IDENTIFIER}\\\jprop{code = tenant}};
  \node[jnode, minimum width=3.6cm] (r4) at (r2 |- r3) {\jtype{LOCAL}\\\jprop{name = tenant}};
  \draw[jflow] (r1) -- node[jelab, above] {REF} (r2);
  \draw[jflow] (r3) -- node[jelab, above] {REF} (r4);
\end{tikzpicture}
\end{minipage}
\caption{\texttt{REF} edges: the name
\texttt{name} refers to the method's parameter, and the name \texttt{tenant} to a
local variable.}
\Description{Source code of show_tenant on the left. On the right, two
IDENTIFIER nodes with REF edges: name to a METHOD_PARAMETER_IN node, tenant to
a LOCAL node.}
\label{fig:app-walk-j-ref}
\end{figure*}

\begin{figure*}[!htbp]
\centering
\begin{minipage}[t]{0.34\linewidth}
\vspace{0pt}
\begin{lstlisting}[style=walkcode, language=Python]
self.proxy_controller_function(
    tenant.get_tenant, ...)
\end{lstlisting}
\end{minipage}\hfill
\begin{minipage}[t]{0.63\linewidth}
\vspace{0pt}\centering
\begin{tikzpicture}
  \node[jnode] (m1) at (0,0) {\jtype{CALL}\\\jprop{name = proxy\_controller\_function}};
  \node[jnode, below=6mm of m1] (m2) {\jtype{METHOD\_REF}\\\jprop{code = tenant.get\_tenant}\\\jprop{methodFullName = ...tenant.py:<module>.get\_tenant}};
  \node[jnode, below=6mm of m2] (m3) {\jtype{METHOD}\\\jprop{fullName = ...tenant.py:<module>.get\_tenant}};
  \draw[jflow] (m1) -- node[jelab, right] {ARGUMENT} (m2);
  \draw[jflow] (m2) -- node[jelab, right] {REF} (m3);
\end{tikzpicture}
\end{minipage}
\caption{A method reference: the function \texttt{tenant.get\_tenant}, passed as an argument, is a
\texttt{METHOD\_REF} node with a \texttt{REF} edge to its \texttt{METHOD}.}
\Description{A CALL proxy_controller_function with an ARGUMENT edge to a
METHOD_REF tenant.get_tenant, which has a REF edge to the METHOD get_tenant.}
\label{fig:app-walk-j-mref}
\end{figure*}
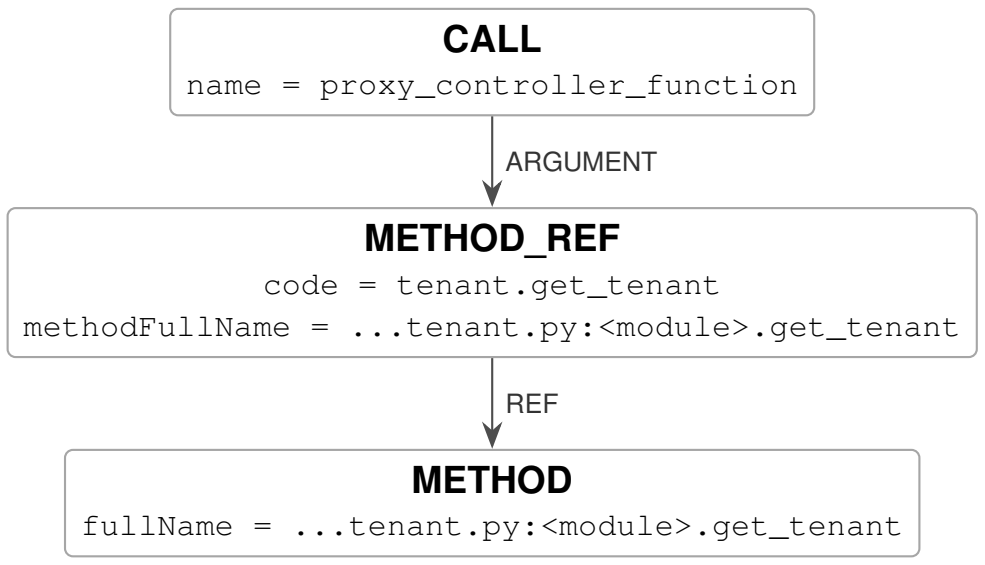

\textbf{Call edges are often missing.} Joern links a call to the method it runs through the
address it infers for the call, \texttt{methodFullName}, with three possible outcomes
(Figure~\ref{fig:app-walk-j-call}): the call is linked to a method of the codebase, which we
can traverse into; to a placeholder for a library method, which has no code; or to nothing at
all. A call between two files of the same codebase often ends up in the second or third case,
for example when the called function is re-exported by a Python package, or when Python
gives no type for the object a method is called on. Joern also links a name to the variable
it refers to (Figure~\ref{fig:app-walk-j-ref}), and a function passed as a value, such as a
callback, to its method (Figure~\ref{fig:app-walk-j-mref}).

\textbf{The way a security control is implemented can hide its calls.} Whether a call
between two roles appears in the graph also depends on the libraries and frameworks the
developers built the security control with. In Spring, the framework passes a login request to
the codebase's own provider inside library code, which is not in the graph, and builds the
objects that roles call through \texttt{@Bean} methods; in FastAPI, a dependency declared as
\texttt{Depends(f)} is called by the framework, and Joern's Python frontend drops it from the
graph altogether. In all these cases, two roles that do interact when the code runs have no
path between them in Joern's CPG, and a traversal of that graph alone misses their
interaction. We therefore add the missing edges ourselves, as described next.

\subsection{Traversing between Security Pattern Roles}
\label{app:walk-traversal}

\subsubsection{Traversal rules}
\label{app:walk-overview}

\begin{figure*}[t]
\centering
\begin{tikzpicture}[
  role/.style={draw=gray!70, rounded corners=2pt, font=\scriptsize, align=center,
               minimum width=2.1cm, minimum height=0.7cm, fill=blue!6},
  missing/.style={role, dashed, draw=gray!60, fill=white, text=gray!60!black},
  entry/.style={role, draw=teal!65!black, line width=1.4pt, fill=teal!10},
  wstep/.style={-{Stealth[length=1.8mm]}, gray!60!black},
  skipped/.style={wstep, dashed, gray!45},
  num/.style={circle, fill=orange!85!black, text=white, font=\scriptsize\bfseries,
              inner sep=1pt, minimum size=4mm},
  numoff/.style={num, fill=gray!45},
  tag/.style={font=\sffamily\tiny\bfseries, rounded corners=1pt, inner sep=1.6pt}
]
  \node[missing] (sub) at (0,0)      {Subject};
  \node[entry]   (enf) at (3.4,0)    {Enforcer};
  \node[role]    (ver) at (6.8,0)    {Verifier};
  \node[role]    (key) at (10.2,0.75)  {Key manager};
  \node[role]    (cry) at (10.2,-0.75) {Cryptographer};
  \node[role]    (sys) at (3.4,-1.9) {System};

  \draw[skipped] (sub) -- node[numoff] {1} (enf);
  \draw[wstep] (enf) -- node[num] {2} (ver);
  \draw[wstep] (ver.east) -- node[num] {3} (key.west);
  \draw[wstep] (ver.east) -- node[num] {4} (cry.west);
  \draw[wstep] (enf) -- node[num] {5} (sys);
  \draw[wstep] (sys.west) -| node[num, pos=0.25] {6} (sub.south);

  \node[tag, draw=red!60!black, text=red!70!black, fill=red!4, anchor=south]
    at ([yshift=3pt]sub.north) {\textsc{missing}: no code};
  \node[tag, text=white, fill=teal!65!black, anchor=south] (et)
    at ([yshift=12pt]enf.north) {entry role: the traversal starts here};
  \draw[-{Stealth[length=1.8mm]}, line width=1.2pt, teal!65!black] (et) -- (enf);
\end{tikzpicture}
\caption{Choosing the entry role from the actions $B$ of the
\emph{Verifiable token-based authentication} pattern. Boxes are the pattern's
roles $r \in R$ and numbered arrows its actions $b_1, \ldots, b_6$, in order. The sender of $b_1$, the
Subject, is \textsc{missing} (dashed), so $b_1$ is skipped (grey) and the
sender of $b_2$, the Enforcer, becomes the entry role.}
\Description{Six role boxes joined by six numbered arrows: Subject to Enforcer
(1), Enforcer to Verifier (2), Verifier to Key manager (3), Verifier to
Cryptographer (4), Enforcer to System (5) and System to Subject (6). The
Subject box is dashed and tagged missing, no code, and arrow 1 is grey. The
Enforcer box is highlighted and tagged as the entry role, where the walk
starts.}
\label{fig:app-walk-entry}
\end{figure*}
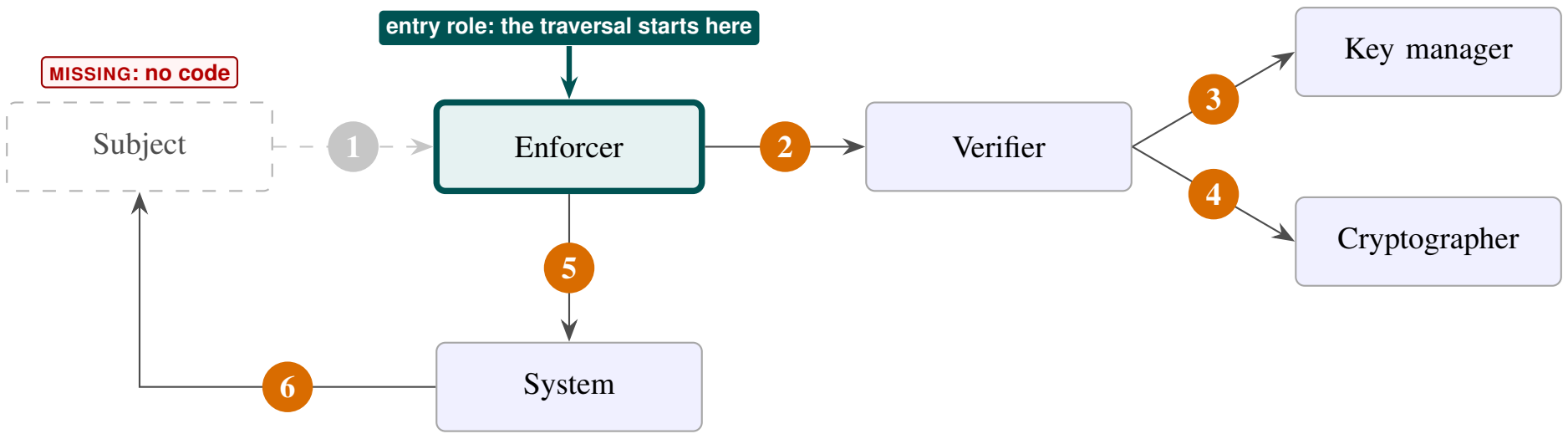

\textbf{Where the traversal starts.} The traversal starts at the code symbol $\hat{e}$ of the
\emph{entry role}: the role that sends the first action $b_1$ of the pattern $P = (R, B)$. If
that role is not mapped to any code ($\hat{\delta}(r) = \textsc{missing}$), the sender of the next
action $b_2$ is taken instead, and so on (Figure~\ref{fig:app-walk-entry}).

\textbf{Placing the roles on the graph.} Each mapping $\hat{m} = (r, \hat{e}) \in \hat{M}$ gives
the code symbol $\hat{e}$ (a method, class, variable or field) that role $r$ is mapped to, with
its file and line. We look for a node declared under that name in that file, preferring one on
the mapped line, and place $r$ on it. A role mapped to a class also covers the class's methods,
and a role mapped to a library function ($\hat{\delta}(r) = \textsc{external}$), which has no code
in $C$, is placed on the calls that $C$ makes to that function. One node may carry several
roles.

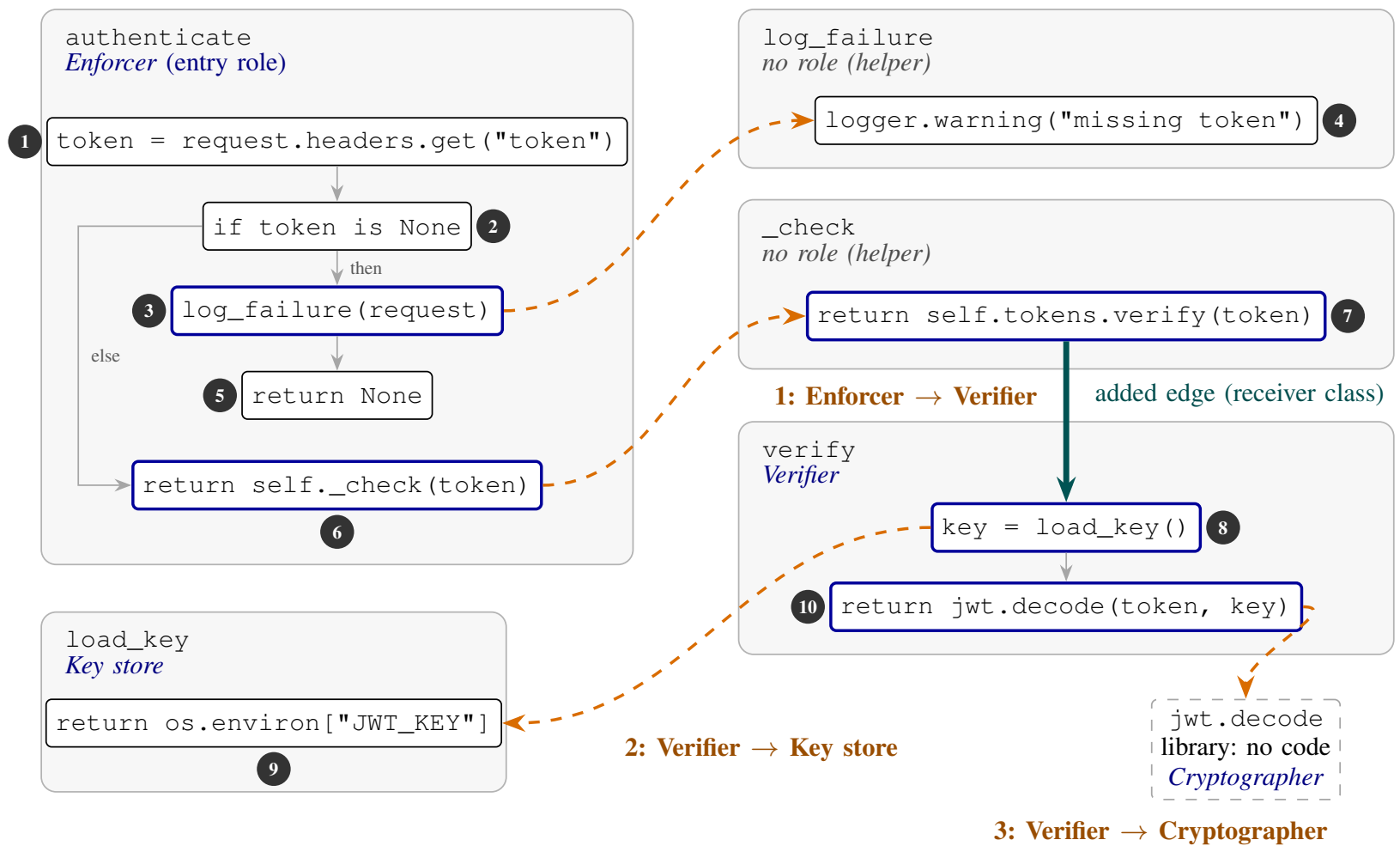
\begin{figure*}[t]
\centering
\settowidth{\dimen0}{\scriptsize\ttfamily xxxxxxxxxxxxxxxxxxxxxxxxxxxxxxxxxxxxxxxxxxxxxxxx}%
\edef\walkcodewd{\the\dimen0}%
\begin{minipage}[t]{\walkcodewd}
\vspace{0pt}
\begin{lstlisting}[basicstyle=\scriptsize\ttfamily, columns=fullflexible,
  keepspaces=true, frame=none, aboveskip=0pt, belowskip=0pt]
# app/middleware.py
 1| class AuthMiddleware:
 2|     def __init__(self):
 3|         self.tokens = TokenVerifier()
 4|
 5|     def authenticate(self, request):
 6|         token = request.headers.get("token")
 7|         if token is None:
 8|             log_failure(request)
 9|             return None
10|         return self._check(token)
11|
12|     def _check(self, token):
13|         return self.tokens.verify(token)
14|
15| def log_failure(request):
16|     logger.warning("missing token")
\end{lstlisting}

\medskip
\begin{lstlisting}[basicstyle=\scriptsize\ttfamily, columns=fullflexible,
  keepspaces=true, frame=none, aboveskip=0pt, belowskip=0pt]
# app/tokens.py
1| class TokenVerifier:
2|     def verify(self, token):
3|         key = load_key()
4|         return jwt.decode(token, key)
5|
6| def load_key():
7|     return os.environ["JWT_KEY"]
\end{lstlisting}

\medskip
{\scriptsize\setlength{\tabcolsep}{4pt}%
\begin{tabular}{@{}lll@{}}
\toprule
Role $r$ & Symbol $\hat{e}$ & Mapped at \\
\midrule
Enforcer & \texttt{authenticate} & \texttt{middleware.py:5} \\
Verifier & \texttt{verify} & \texttt{tokens.py:2} \\
Key store & \texttt{load\_key} & \texttt{tokens.py:6} \\
Cryptographer & \texttt{decode} (library) & \texttt{tokens.py:4} \\
\bottomrule
\end{tabular}}

\medskip
{\small (a) The code (illustrative; imports omitted) and its role mapping.}
\end{minipage}\hfill
\begin{minipage}[t]{\dimexpr\linewidth-\walkcodewd-1em\relax}
\vspace{0pt}\centering
\resizebox{\linewidth}{!}{%
\begin{tikzpicture}[
  cfg/.style={draw, rounded corners=1.5pt, font=\ttfamily\scriptsize,
              inner sep=2.5pt, minimum height=4.5mm, fill=white},
  callnode/.style={cfg, draw=blue!60!black, thick},
  lib/.style={draw=gray!70, dashed, rounded corners=1.5pt, font=\scriptsize,
              align=center, inner sep=2.5pt, fill=white},
  island/.style={draw=gray!60, rounded corners=4pt, fill=gray!6},
  islandlab/.style={font=\scriptsize, anchor=north west, align=left},
  flow/.style={-{Stealth[length=1.6mm]}, gray!70},
  joern/.style={-{Stealth[length=2.2mm]}, thick, dashed, orange!85!black},
  added/.style={-{Stealth[length=2.4mm]}, line width=1.6pt, teal!65!black},
  rec/.style={font=\scriptsize\bfseries, text=orange!60!black},
  addlab/.style={font=\scriptsize, text=teal!60!black},
  branch/.style={font=\tiny, text=gray!70!black},
  badge/.style={circle, fill=black!80, text=white, font=\tiny\bfseries,
                inner sep=0pt, minimum size=3.2mm}
]
  \draw[island] (0,0) rectangle (5.6,-5.25);
  \node[islandlab] at (0.1,-0.05) {\texttt{authenticate}\\[-1pt]\textcolor{blue!50!black}{\textit{Enforcer} (entry role)}};
  \node[cfg]      (a6)  at (2.8,-1.25) {token = request.headers.get("token")};
  \node[cfg]      (a7)  at (2.8,-2.05) {if token is None};
  \node[callnode] (a8)  at (2.8,-2.85) {log\_failure(request)};
  \node[cfg]      (a9)  at (2.8,-3.65) {return None};
  \node[callnode] (a10) at (2.8,-4.5)  {return self.\_check(token)};
  \draw[flow] (a6) -- (a7);
  \draw[flow] (a7) -- node[branch, right] {then} (a8);
  \draw[flow] (a8) -- (a9);
  \draw[flow] (a7.west) -- (0.35,-2.05) -- node[branch, right] {else} (0.35,-4.5) -- (a10.west);

  \draw[island] (6.6,0) rectangle (12.8,-1.5);
  \node[islandlab] at (6.7,-0.05) {\texttt{log\_failure}\\[-1pt]\textcolor{gray!60!black}{\textit{no role (helper)}}};
  \node[cfg] (l1) at (9.7,-1.05) {logger.warning("missing token")};

  \draw[island] (6.6,-1.8) rectangle (12.8,-3.4);
  \node[islandlab] at (6.7,-1.85) {\texttt{\_check}\\[-1pt]\textcolor{gray!60!black}{\textit{no role (helper)}}};
  \node[callnode] (k1) at (9.7,-2.9) {return self.tokens.verify(token)};

  \draw[island] (6.6,-3.9) rectangle (12.8,-6.1);
  \node[islandlab] at (6.7,-3.95) {\texttt{verify}\\[-1pt]\textcolor{blue!50!black}{\textit{Verifier}}};
  \node[callnode] (v3) at (9.7,-4.9)  {key = load\_key()};
  \node[callnode] (v4) at (9.7,-5.65) {return jwt.decode(token, key)};
  \draw[flow] (v3) -- (v4);

  \draw[island] (0,-5.7) rectangle (4.4,-7.4);
  \node[islandlab] at (0.1,-5.75) {\texttt{load\_key}\\[-1pt]\textcolor{blue!50!black}{\textit{Key store}}};
  \node[cfg] (lk) at (2.2,-6.75) {return os.environ["JWT\_KEY"]};

  \node[lib] (jwt) at (11.4,-7.0) {\texttt{jwt.decode}\\library: no code\\\textcolor{blue!50!black}{\textit{Cryptographer}}};

  \draw[joern] (a8.east)  to[out=0, in=180] (l1.west);
  \draw[joern] (a10.east) to[out=0, in=180] (k1.west);
  \draw[joern] (v3.west)  to[out=180, in=0] (lk.east);
  \draw[joern] (v4.east)  to[out=0, in=90]  (jwt.north);

  \draw[added] (k1.south) -- (v3.north);
  \node[addlab, anchor=west] at (9.85,-3.65) {added edge (receiver class)};

  \node[rec, anchor=east] at (9.55,-3.65) {1: Enforcer $\to$ Verifier};
  \node[rec, anchor=west] at (5.4,-7.0)   {2: Verifier $\to$ Key store};
  \node[rec, anchor=north east] at ([yshift=-2pt]jwt.south east) {3: Verifier $\to$ Cryptographer};

  \node[badge, left=1pt of a6]   {1};
  \node[badge, right=1pt of a7]  {2};
  \node[badge, left=1pt of a8]   {3};
  \node[badge, right=1pt of l1]  {4};
  \node[badge, left=1pt of a9]   {5};
  \node[badge, below=1pt of a10] {6};
  \node[badge, right=1pt of k1]  {7};
  \node[badge, right=1pt of v3]  {8};
  \node[badge, below=1pt of lk]  {9};
  \node[badge, left=1pt of v4]   {10};
\end{tikzpicture}}

\medskip
{\small (b) The graph the traversal sees.}
\end{minipage}
\caption{The worked example. In (b), each box is one method. Grey arrows are
control-flow edges; dashed orange arrows are call edges from Joern; the thick
teal arrow is an edge we add. The dashed box is Joern's placeholder
for a library function, which has no code. Black circles give the order in
which the traversal visits the statements, and the three interactions it records
are numbered on the edges that produce them.}
\Description{Part (a) shows two short Python files and a four-row role
mapping. Part (b) shows five boxes, one per method: authenticate (Enforcer,
the entry role) with a branch, the helpers log\_failure and \_check, verify
(Verifier) and load\_key (Key store), plus a dashed box for the library
function jwt.decode (Cryptographer). Dashed orange arrows mark Joern's call
edges, a thick teal arrow from \_check to verify marks the added edge, and
numbered circles mark the ten steps of the traversal. The edges are labelled with
the three recorded interactions: Enforcer to Verifier, Verifier to Key store,
and Verifier to Cryptographer.}
\label{fig:app-walk-example}
\end{figure*}

\textbf{How the traversal moves.} The traversal follows the graph in the order the code would
run: from one statement to the next, into a called method and back to the statement after the
call, and through both branches of a condition, the first-written one first. It never visits
the same method or statement twice, so loops and recursion end. Throughout, it acts as one
role, starting with the entry role. When it reaches a node while acting as role $r_i$:
\begin{itemize}
\item if the node carries another role $r_j$, we record the interaction $r_i \to r_j$ and
  continue into the code of $r_j$, now acting as $r_j$;
\item if the node is a method with no role, such as a helper, we record nothing and continue
  into it, still acting as $r_i$, since roles are often linked only through such helpers;
\item if the node carries the same role $r_i$, we record nothing and do not go in.
\end{itemize}
A node with several roles gives one interaction for each, in the order of the actions $B$.
The recorded interactions, in order and without repeats, form $\hat{S}$. Since both branches
are followed, $r_i \to r_j$ is recorded whenever some path leads from the code of $r_i$ to that
of $r_j$, and a role that cannot be reached from the entry role is left out of $\hat{S}$.
Figure~\ref{fig:app-walk-code} outlines these rules, and Figure~\ref{fig:app-walk-example}
shows the order in which the traversal visits the statements of a small example.

\begin{figure*}[t]
\begin{lstlisting}[basicstyle=\footnotesize\ttfamily, columns=fullflexible,
  keepspaces=true, breaklines=true, frame=tb, aboveskip=4pt, belowskip=4pt]
walk(m, start, ctx, via):                  -- ctx: the roles the walk acts as
  if (m, start, ctx) was walked before: return
  if start is m's first statement:
    for t in edges added at the start of m: reach(t, ctx, via)
  stack <- [start];  seen <- {}
  while stack is not empty:
    n <- stack.pop()
    if n in seen: continue
    seen.add(n)
    for t in edges(n): reach(t, ctx, via)          -- Joern's edges and added ones
    push the control-flow successors of n in reverse order

reach(t, ctx, via):
  pairs <- { (a, b) : a in ctx, b in roles(t), a != b }
  if pairs is not empty:
    record each pair, ordered by the pattern's actions B
    if t has a body: walk(t, first(t), targets of pairs, [])  -- act as the new role
  else if roles(t) is empty and t has a body:
    walk(t, first(t), ctx, via + [t])                        -- helper: same role
\end{lstlisting}
\caption{Our traversal in outline. \texttt{walk} steps one method; \texttt{reach}
decides what reaching a node means.}
\Description{Pseudocode for two procedures, walk and reach. Walk steps a
method's control-flow graph with a stack and calls reach for every node an edge
leads to; reach records an interaction for each new role and continues into the
reached method.}
\label{fig:app-walk-code}
\end{figure*}

\subsubsection{Edges we add}
\label{app:walk-added}

Where Joern's graph lacks a link between two roles that interact, we add an edge of our own
(Table~\ref{tab:app-walk-added}). At a call, we always try Joern's edge first and use ours
only when Joern's leads to no code of the codebase. We never add an edge by searching the
codebase for a method with a matching name: an earlier version of our traversal did so, and on
one codebase it recovered six interactions, five of them wrong, because one wrong link sends
the traversal into unrelated code and every interaction found there is credited to the wrong
role. Each case below comes with a short example: the code on the left, (a)~Joern's graph,
where a red dashed arrow marks the missing link, and (b)~the same graph with the edge we add
(solid teal), where the shaded boxes are what we read to build it.

\begin{table*}[t]
\caption{Edges we add to Joern's graph. \emph{Label} is recorded on every
interaction the edge produces, so interactions that rest on an added edge can be
told apart from ordinary calls.}
\label{tab:app-walk-added}
\footnotesize
\setlength{\tabcolsep}{4pt}
\begin{tabular}{@{}>{\raggedright\arraybackslash}p{0.17\linewidth} >{\raggedright\arraybackslash}p{0.31\linewidth} >{\raggedright\arraybackslash}p{0.24\linewidth} >{\raggedright\arraybackslash}p{0.20\linewidth}@{}}
\toprule
Added edge & Leads to & Built from & Label \\
\midrule
Import binding & the repository method that the calling file's import names &
  Joern's import nodes & \texttt{call} \\
Receiver class (Python) & the method on the class the code assigns to the
  object a call is made on & Joern's assignment nodes & \texttt{call} \\
Field access & the field that a read or write refers to & Joern's
  field-access node and the object's class & \texttt{data} \\
Override dispatch (Java) & the repository's implementations of a library
  interface method & Joern's class hierarchy, method names and parameter counts
  & \texttt{dispatch} \\
Framework route (Java) & the repository code that Spring Security passes a
  login request to & a rule written for that framework & \texttt{dispatch} \\
Injection provider (Java) & the \texttt{@Bean} method that builds the object a
  call is made on & Joern's annotation nodes and return types &
  \texttt{dependency-injection} \\
Parameter default (Python) & the functions named in \texttt{Depends(f)}
  parameter defaults & the method's signature in the source text &
  \texttt{dependency-injection} \\
\bottomrule
\end{tabular}
\end{table*}

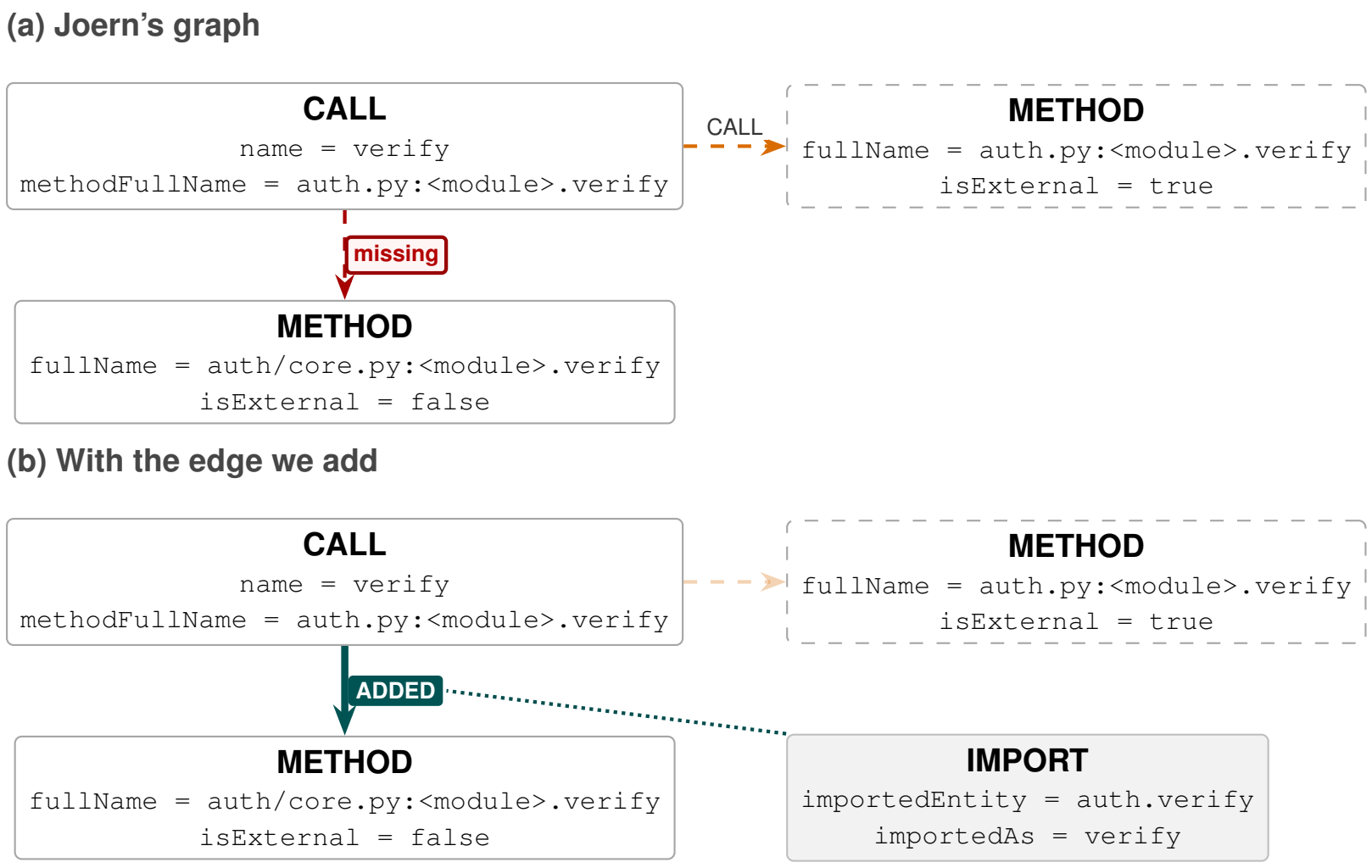
\begin{figure*}[!htbp]
\centering
\begin{minipage}[t]{0.27\linewidth}
\vspace{0pt}
\codefile{app.py}
\begin{lstlisting}[style=walkcode, language=Python]
from auth import verify
verify(token)
\end{lstlisting}
\medskip
\codefile{auth/\_\_init\_\_.py}
\begin{lstlisting}[style=walkcode, language=Python]
from .core import verify
\end{lstlisting}
\medskip
\codefile{auth/core.py}
\begin{lstlisting}[style=walkcode, language=Python]
def verify(token): ...
\end{lstlisting}
\end{minipage}\hfill
\begin{minipage}[t]{0.7\linewidth}
\vspace{0pt}\centering
\begin{tikzpicture}[wide/.style={minimum width=4.5cm}]
  \node[jpanel] at (0,0.3) {(a) Joern's graph};
  \node[jnode, wide, anchor=north west] (c) at (0,0) {\jtype{CALL}\\\jprop{name = verify}\\\jprop{methodFullName = auth.py:<module>.verify}};
  \node[jext, right=8mm of c] (p) {\jtype{METHOD}\\\jprop{fullName = auth.py:<module>.verify}\\\jprop{isExternal = true}};
  \node[jnode, wide, below=7mm of c] (m) {\jtype{METHOD}\\\jprop{fullName = auth/core.py:<module>.verify}\\\jprop{isExternal = false}};
  \draw[jcall] (c) -- node[jelab, above] {CALL} (p);
  \draw[jmiss] (c) -- node[jproblem, right] {missing} (m);
  \begin{scope}[yshift=-3.4cm]
  \node[jpanel] at (0,0.3) {(b) With the edge we add};
  \node[jnode, wide, anchor=north west] (c) at (0,0) {\jtype{CALL}\\\jprop{name = verify}\\\jprop{methodFullName = auth.py:<module>.verify}};
  \node[jext, right=8mm of c] (p) {\jtype{METHOD}\\\jprop{fullName = auth.py:<module>.verify}\\\jprop{isExternal = true}};
  \node[jnode, wide, below=7mm of c] (m) {\jtype{METHOD}\\\jprop{fullName = auth/core.py:<module>.verify}\\\jprop{isExternal = false}};
  \node[jread, anchor=west] (r) at (p.west |- m) {\jtype{IMPORT}\\\jprop{importedEntity = auth.verify}\\\jprop{importedAs = verify}};
  \draw[jfaded] (c) -- (p);
  \draw[jadded] (c) -- node[jaddlab, right] (add) {ADDED} (m);
  \draw[jfrom] (r.north west) -- (add.east);
  \end{scope}
\end{tikzpicture}
\end{minipage}
\caption{Import binding: Joern's graph (a) and the edge we add (b).}
\Description{Code on the left. Panel a: the CALL verify has a CALL edge to a placeholder METHOD at auth.py and no link to the METHOD verify in auth/core.py, marked missing. Panel b: we read the IMPORT node and adds an edge from the CALL to the METHOD in auth/core.py.}
\label{fig:app-walk-a-import}
\end{figure*}

\paragraph{Import binding}
\label{app:walk-added-import}
Joern sometimes gives a call to an imported function the wrong address. In Python,
\texttt{from auth import verify} is read as the file \texttt{auth.py}; when \texttt{auth} is a
package that re-exports \texttt{verify} from \texttt{auth/core.py}, no method has that
address, and the call is linked to a placeholder, as if it were a library call
(Figure~\ref{fig:app-walk-a-import}a). Java statically imported methods have the same problem.
The import statement of the calling file names the right method, so we follow it, through the
re-export, to the file that defines it, and link the call there
(Figure~\ref{fig:app-walk-a-import}b).

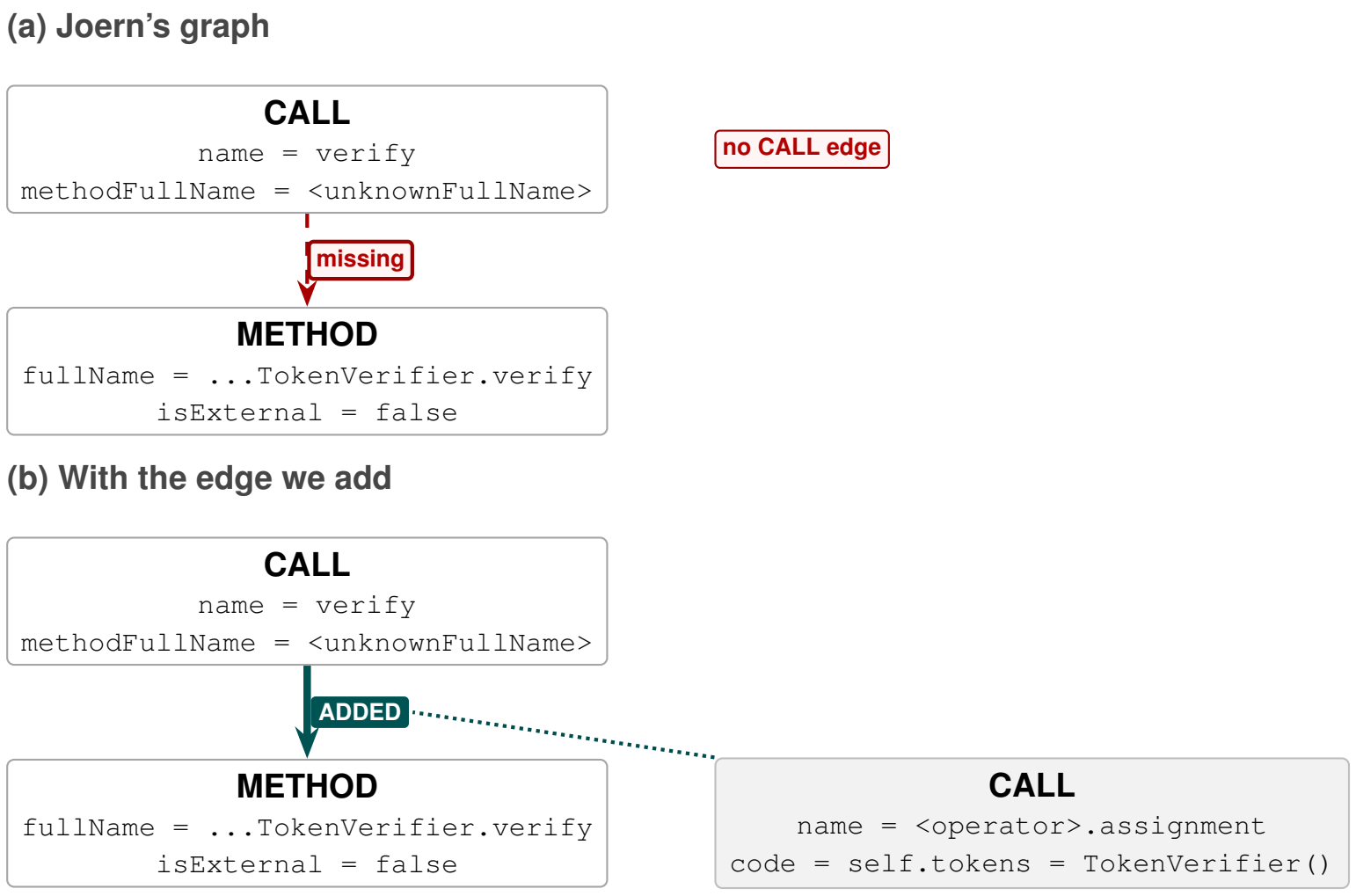
\begin{figure*}[!htbp]
\centering
\begin{minipage}[t]{0.31\linewidth}
\vspace{0pt}
\begin{lstlisting}[style=walkcode, language=Python]
def __init__(self):
    self.tokens = TokenVerifier()

def check(self, token):
    self.tokens.verify(token)
\end{lstlisting}
\end{minipage}\hfill
\begin{minipage}[t]{0.66\linewidth}
\vspace{0pt}\centering
\begin{tikzpicture}[wide/.style={minimum width=3.9cm}]
  \node[jpanel] at (0,0.3) {(a) Joern's graph};
  \node[jnode, wide, anchor=north west] (c) at (0,0) {\jtype{CALL}\\\jprop{name = verify}\\\jprop{methodFullName = <unknownFullName>}};
  \node[jproblem, anchor=west] at ([xshift=8mm]c.east) {no \jtype{CALL} edge};
  \node[jnode, wide, below=7mm of c] (m) {\jtype{METHOD}\\\jprop{fullName = ...TokenVerifier.verify}\\\jprop{isExternal = false}};
  \draw[jmiss] (c) -- node[jproblem, right] {missing} (m);
  \begin{scope}[yshift=-3.4cm]
  \node[jpanel] at (0,0.3) {(b) With the edge we add};
  \node[jnode, wide, anchor=north west] (c) at (0,0) {\jtype{CALL}\\\jprop{name = verify}\\\jprop{methodFullName = <unknownFullName>}};
  \node[jnode, wide, below=7mm of c] (m) {\jtype{METHOD}\\\jprop{fullName = ...TokenVerifier.verify}\\\jprop{isExternal = false}};
  \node[jread, anchor=west] (r) at ([xshift=8mm]m.east) {\jtype{CALL}\\\jprop{name = <operator>.assignment}\\\jprop{code = self.tokens = TokenVerifier()}};
  \draw[jadded] (c) -- node[jaddlab, right] (add) {ADDED} (m);
  \draw[jfrom] (r.north west) -- (add.east);
  \end{scope}
\end{tikzpicture}
\end{minipage}
\caption{Receiver class: Joern's graph (a) and the edge we add (b).}
\Description{Code on the left. Panel a: the CALL verify has address unknownFullName and no CALL edge; the link to METHOD TokenVerifier.verify is marked missing. Panel b: we read the assignment node and adds that edge.}
\label{fig:app-walk-a-receiver}
\end{figure*}

\paragraph{Receiver class (Python)}
\label{app:walk-added-receiver}
To link a call \texttt{x.m()}, Joern needs the type of \texttt{x}, which Python does not
declare: \texttt{self.tokens.verify(token)} gets no address and no call edge, although
\texttt{TokenVerifier.verify} is in the codebase (Figure~\ref{fig:app-walk-a-receiver}a). We
look at what the code assigns to the object, here \texttt{self.tokens = TokenVerifier()} in
\texttt{\_\_init\_\_}, and link the call to \texttt{verify} on that class
(Figure~\ref{fig:app-walk-a-receiver}b). This is the edge used at step~7 of
Figure~\ref{fig:app-walk-example}b.

\begin{figure*}[!htbp]
\centering
\begin{minipage}[t]{0.28\linewidth}
\vspace{0pt}
\begin{lstlisting}[style=walkcode, language=Java]
class Signer {
  private Key key;
  byte[] sign(byte[] m) {
    return mac(this.key, m);
  }
}
\end{lstlisting}
\end{minipage}\hfill
\begin{minipage}[t]{0.69\linewidth}
\vspace{0pt}\centering
\begin{tikzpicture}
  \node[jpanel] at (0,0.3) {(a) Joern's graph};
  \node[jnode, anchor=north west] (i) at (0,-1.45) {\jtype{IDENTIFIER}\\\jprop{code = this}\\\jprop{typeFullName = Signer}};
  \node[jnode, anchor=north west] (f) at ([xshift=3mm]i.north east) {\jtype{FIELD\_IDENTIFIER}\\\jprop{canonicalName = key}};
  \path (i.north west) -- (f.north east) coordinate[midway] (mid);
  \node[jnode, anchor=north] (c) at (mid |- 0,0) {\jtype{CALL}\\\jprop{name = <operator>.fieldAccess}\\\jprop{code = this.key}};
  \node[jnode, anchor=north] (t) at ([xshift=16mm]f.north east) {\jtype{TYPE\_DECL}\\\jprop{fullName = Signer}};
  \node[jnode] (m) at (t.center |- c.center) {\jtype{MEMBER}\\\jprop{name = key}};
  \draw[jflow] (c) -- (i);
  \draw[jflow] (c) -- (f);
  \draw[jflow] (t) -- node[jelab, right] {AST} (m);
  \draw[jmiss] (c) -- node[jproblem, above] {missing} (m);
  \begin{scope}[yshift=-3.25cm]
  \node[jpanel] at (0,0.3) {(b) With the edge we add};
  \node[jread, anchor=north west] (i) at (0,-1.45) {\jtype{IDENTIFIER}\\\jprop{code = this}\\\jprop{typeFullName = Signer}};
  \node[jread, anchor=north west] (f) at ([xshift=3mm]i.north east) {\jtype{FIELD\_IDENTIFIER}\\\jprop{canonicalName = key}};
  \path (i.north west) -- (f.north east) coordinate[midway] (mid);
  \node[jnode, anchor=north] (c) at (mid |- 0,0) {\jtype{CALL}\\\jprop{name = <operator>.fieldAccess}\\\jprop{code = this.key}};
  \node[jread, anchor=north] (t) at ([xshift=16mm]f.north east) {\jtype{TYPE\_DECL}\\\jprop{fullName = Signer}};
  \node[jnode] (m) at (t.center |- c.center) {\jtype{MEMBER}\\\jprop{name = key}};
  \draw[jflow] (c) -- (i);
  \draw[jflow] (c) -- (f);
  \draw[jflow] (t) -- node[jelab, right] {AST} (m);
  \draw[jadded] (c) -- node[jaddlab, above] {ADDED} (m);
  \end{scope}
\end{tikzpicture}
\end{minipage}
\caption{Field access: Joern's graph (a) and the edge we add (b). Unlabelled
grey arrows are \texttt{ARGUMENT} edges.}
\Description{Code on the left. Panel a: a fieldAccess CALL with IDENTIFIER this and FIELD_IDENTIFIER key as arguments; the MEMBER key of TYPE_DECL Signer is not linked, marked missing. Panel b: we read the arguments and the class and adds the edge from the access to the MEMBER.}
\label{fig:app-walk-a-field}
\end{figure*}
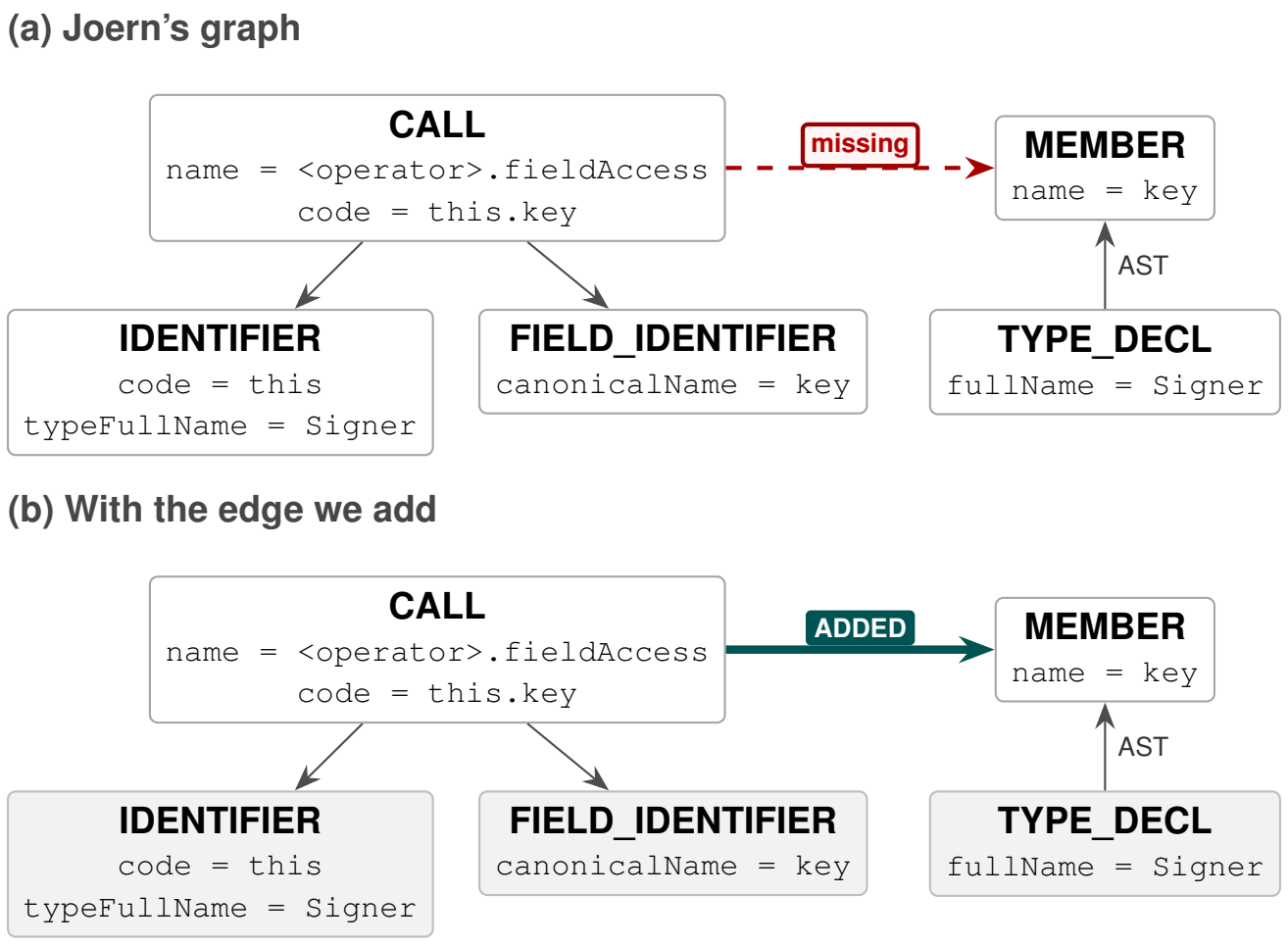

\paragraph{Field access}
\label{app:walk-added-field}
Joern stores a read such as \texttt{this.key} as a field-access call on the object and the
field's name, without an edge to the field itself (Figure~\ref{fig:app-walk-a-field}a), so a
role mapped to the field is never reached from the code that reads it. We take the object's
class, find the field with that name in it or in its parent classes, and link the read to it
(Figure~\ref{fig:app-walk-a-field}b).

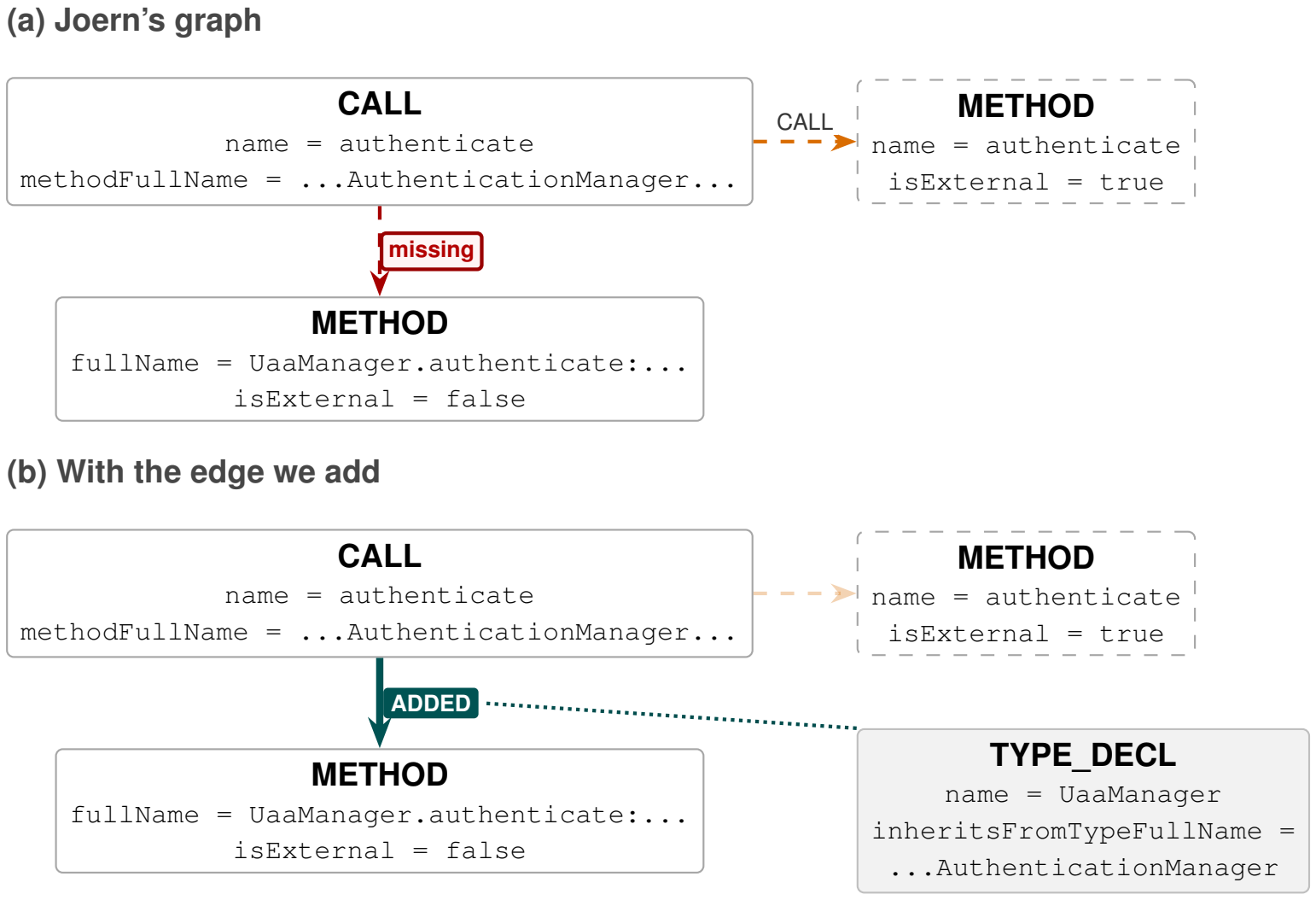
\begin{figure*}[!htbp]
\centering
\begin{minipage}[t]{0.31\linewidth}
\vspace{0pt}
\begin{lstlisting}[style=walkcode, language=Java]
authManager.authenticate(auth);

class UaaManager implements
    AuthenticationManager {
  Authentication authenticate(...)
  { ... }
}
\end{lstlisting}
\end{minipage}\hfill
\begin{minipage}[t]{0.66\linewidth}
\vspace{0pt}\centering
\begin{tikzpicture}[wide/.style={minimum width=4.9cm}]
  \node[jpanel] at (0,0.3) {(a) Joern's graph};
  \node[jnode, wide, anchor=north west] (c) at (0,0) {\jtype{CALL}\\\jprop{name = authenticate}\\\jprop{methodFullName = ...AuthenticationManager...}};
  \node[jext, right=8mm of c] (p) {\jtype{METHOD}\\\jprop{name = authenticate}\\\jprop{isExternal = true}};
  \node[jnode, wide, below=7mm of c] (m) {\jtype{METHOD}\\\jprop{fullName = UaaManager.authenticate:...}\\\jprop{isExternal = false}};
  \draw[jcall] (c) -- node[jelab, above] {CALL} (p);
  \draw[jmiss] (c) -- node[jproblem, right] {missing} (m);
  \begin{scope}[yshift=-3.5cm]
  \node[jpanel] at (0,0.3) {(b) With the edge we add};
  \node[jnode, wide, anchor=north west] (c) at (0,0) {\jtype{CALL}\\\jprop{name = authenticate}\\\jprop{methodFullName = ...AuthenticationManager...}};
  \node[jext, right=8mm of c] (p) {\jtype{METHOD}\\\jprop{name = authenticate}\\\jprop{isExternal = true}};
  \node[jnode, wide, below=7mm of c] (m) {\jtype{METHOD}\\\jprop{fullName = UaaManager.authenticate:...}\\\jprop{isExternal = false}};
  \node[jread, anchor=west] (r) at (p.west |- m) {\jtype{TYPE\_DECL}\\\jprop{name = UaaManager}\\\jprop{inheritsFromTypeFullName =}\\\jprop{...AuthenticationManager}};
  \draw[jfaded] (c) -- (p);
  \draw[jadded] (c) -- node[jaddlab, right] (add) {ADDED} (m);
  \draw[jfrom] (r.north west) -- (add.east);
  \end{scope}
\end{tikzpicture}
\end{minipage}
\caption{Override dispatch: Joern's graph (a) and the edge we add (b).}
\Description{Code on the left. Panel a: the CALL authenticate on AuthenticationManager has a CALL edge to a placeholder METHOD and no link to UaaManager.authenticate, marked missing. Panel b: we read the TYPE_DECL UaaManager, which inherits from AuthenticationManager, and adds the edge.}
\label{fig:app-walk-a-dispatch}
\end{figure*}

\paragraph{Override dispatch (Java)}
\label{app:walk-added-dispatch}
Spring code often calls a method on a library interface, such as
\texttt{authManager.\allowbreak authenticate(auth)}, while at run time Spring supplies a class
of the codebase that implements it. The interface's code is not in the graph, so the call is
linked to a placeholder (Figure~\ref{fig:app-walk-a-dispatch}a). We find the classes of the
codebase that implement the interface, directly or through a parent, and link the call to
their method with the same name and number of parameters
(Figure~\ref{fig:app-walk-a-dispatch}b). We take at most one such step along a chain of
helper calls, so that one guess is never built on another.

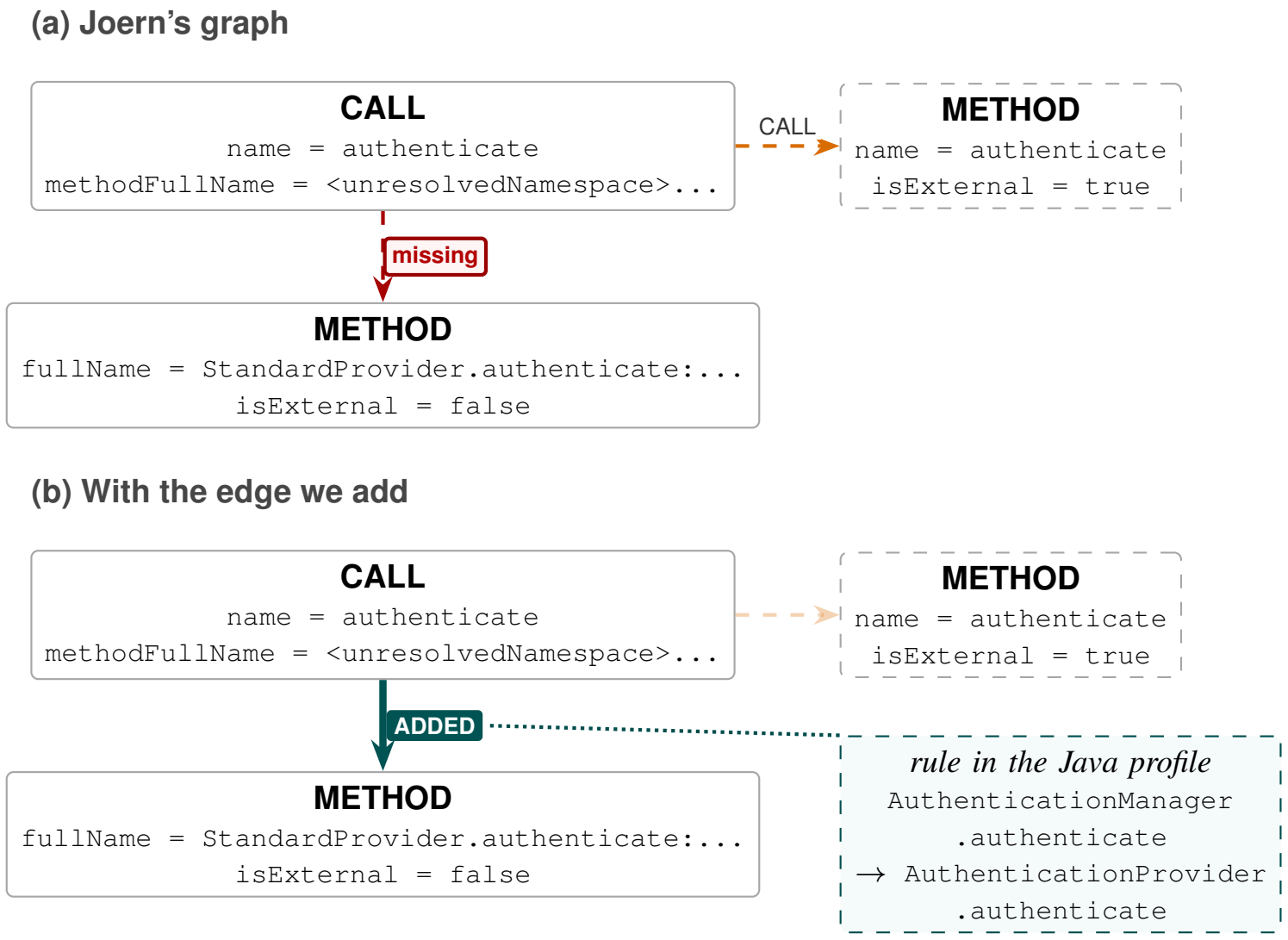
\begin{figure*}[!htbp]
\centering
\begin{minipage}[t]{0.31\linewidth}
\vspace{0pt}
\begin{lstlisting}[style=walkcode, language=Java]
// login filter
getAuthenticationManager()
    .authenticate(req);

class StandardProvider implements
    AuthenticationProvider {
  Authentication authenticate(...)
  { ... }
}
\end{lstlisting}
\end{minipage}\hfill
\begin{minipage}[t]{0.66\linewidth}
\vspace{0pt}\centering
\begin{tikzpicture}[wide/.style={minimum width=4.9cm}]
  \node[jpanel] at (0,0.3) {(a) Joern's graph};
  \node[jnode, wide, anchor=north west] (c) at (0,0) {\jtype{CALL}\\\jprop{name = authenticate}\\\jprop{methodFullName = <unresolvedNamespace>...}};
  \node[jext, right=8mm of c] (p) {\jtype{METHOD}\\\jprop{name = authenticate}\\\jprop{isExternal = true}};
  \node[jnode, wide, below=7mm of c] (m) {\jtype{METHOD}\\\jprop{fullName = StandardProvider.authenticate:...}\\\jprop{isExternal = false}};
  \draw[jcall] (c) -- node[jelab, above] {CALL} (p);
  \draw[jmiss] (c) -- node[jproblem, right] {missing} (m);
  \begin{scope}[yshift=-3.6cm]
  \node[jpanel] at (0,0.3) {(b) With the edge we add};
  \node[jnode, wide, anchor=north west] (c) at (0,0) {\jtype{CALL}\\\jprop{name = authenticate}\\\jprop{methodFullName = <unresolvedNamespace>...}};
  \node[jext, right=8mm of c] (p) {\jtype{METHOD}\\\jprop{name = authenticate}\\\jprop{isExternal = true}};
  \node[jnode, wide, below=7mm of c] (m) {\jtype{METHOD}\\\jprop{fullName = StandardProvider.authenticate:...}\\\jprop{isExternal = false}};
  \node[jrule, anchor=west] (r) at (p.west |- m) {\textit{rule in the Java profile}\\\jprop{AuthenticationManager}\\\jprop{.authenticate}\\\jprop{$\to$ AuthenticationProvider}\\\jprop{.authenticate}};
  \draw[jfaded] (c) -- (p);
  \draw[jadded] (c) -- node[jaddlab, right] (add) {ADDED} (m);
  \draw[jfrom] (r.north west) -- (add.east);
  \end{scope}
\end{tikzpicture}
\end{minipage}
\caption{Framework route: Joern's graph (a) and the edge we add (b). The
dashed teal box is a rule in the Java profile, not a node of Joern's graph.}
\Description{Code on the left. Panel a: the CALL authenticate with an unresolved address has a CALL edge to a placeholder and no link to StandardProvider.authenticate, marked missing. Panel b: a rule in the Java profile leads the added edge to that method.}
\label{fig:app-walk-a-route}
\end{figure*}

\paragraph{Framework route (Java)}
\label{app:walk-added-route}
Some hand-offs happen entirely inside library code. A Spring Security login filter calls
\texttt{getAuthenticationManager().authenticate(req)}, and Spring's \texttt{ProviderManager}
then passes the request to the codebase's own \texttt{AuthenticationProvider}. The codebase
does not implement the interface that the call names, so even override dispatch finds nothing
(Figure~\ref{fig:app-walk-a-route}a). We write this hand-off once, as a rule for Java: a call
to \texttt{AuthenticationManager.authenticate} runs \texttt{AuthenticationProvider.authenticate}.
We apply it only when nothing else links the call, and link the call to the codebase's
implementation (Figure~\ref{fig:app-walk-a-route}b).

\begin{figure*}[!htbp]
\centering
\begin{minipage}[t]{0.27\linewidth}
\vspace{0pt}
\begin{lstlisting}[style=walkcode, language=Java]
// class SecurityConfig
@Bean
PasswordEncoder encoder() {
  return new BCrypt...();
}

passwordEncoder.matches(
    raw, hash);
\end{lstlisting}
\end{minipage}\hfill
\begin{minipage}[t]{0.7\linewidth}
\vspace{0pt}\centering
\begin{tikzpicture}[wide/.style={minimum width=4.4cm}]
  \node[jpanel] at (0,0.3) {(a) Joern's graph};
  \node[jnode, wide, anchor=north west] (c) at (0,0) {\jtype{CALL}\\\jprop{name = matches}\\\jprop{methodFullName = ...PasswordEncoder...}};
  \node[jext, right=8mm of c] (p) {\jtype{METHOD}\\\jprop{name = matches}\\\jprop{isExternal = true}};
  \node[jnode, wide, below=7mm of c] (m) {\jtype{METHOD}\\\jprop{fullName = SecurityConfig.encoder:...}\\\jprop{isExternal = false}};
  \draw[jcall] (c) -- node[jelab, above] {CALL} (p);
  \draw[jmiss] (c) -- node[jproblem, right] {missing} (m);
  \begin{scope}[yshift=-3.4cm]
  \node[jpanel] at (0,0.3) {(b) With the edge we add};
  \node[jnode, wide, anchor=north west] (c) at (0,0) {\jtype{CALL}\\\jprop{name = matches}\\\jprop{methodFullName = ...PasswordEncoder...}};
  \node[jext, right=8mm of c] (p) {\jtype{METHOD}\\\jprop{name = matches}\\\jprop{isExternal = true}};
  \node[jnode, wide, below=7mm of c] (m) {\jtype{METHOD}\\\jprop{fullName = SecurityConfig.encoder:...}\\\jprop{isExternal = false}};
  \node[jread, anchor=west] (r2) at (p.west |- m) {\jtype{METHOD\_RETURN}\\\jprop{typeFullName =}\\\jprop{...PasswordEncoder}};
  \node[jread, anchor=west] (r1) at ([xshift=3pt]r2.east) {\jtype{ANNOTATION}\\\jprop{name = Bean}};
  \draw[jfaded] (c) -- (p);
  \draw[jadded] (c) -- node[jaddlab, right] (add) {ADDED} (m);
  \draw[jfrom] (r2.north west) -- (add.east);
  \end{scope}
\end{tikzpicture}
\end{minipage}
\caption{Injection provider: Joern's graph (a) and the edge we add (b).}
\Description{Code on the left. Panel a: the CALL matches has a CALL edge to a placeholder and no link to the METHOD encoder, marked missing. Panel b: we read the METHOD_RETURN type of encoder and its Bean ANNOTATION and adds the edge.}
\label{fig:app-walk-a-provider}
\end{figure*}
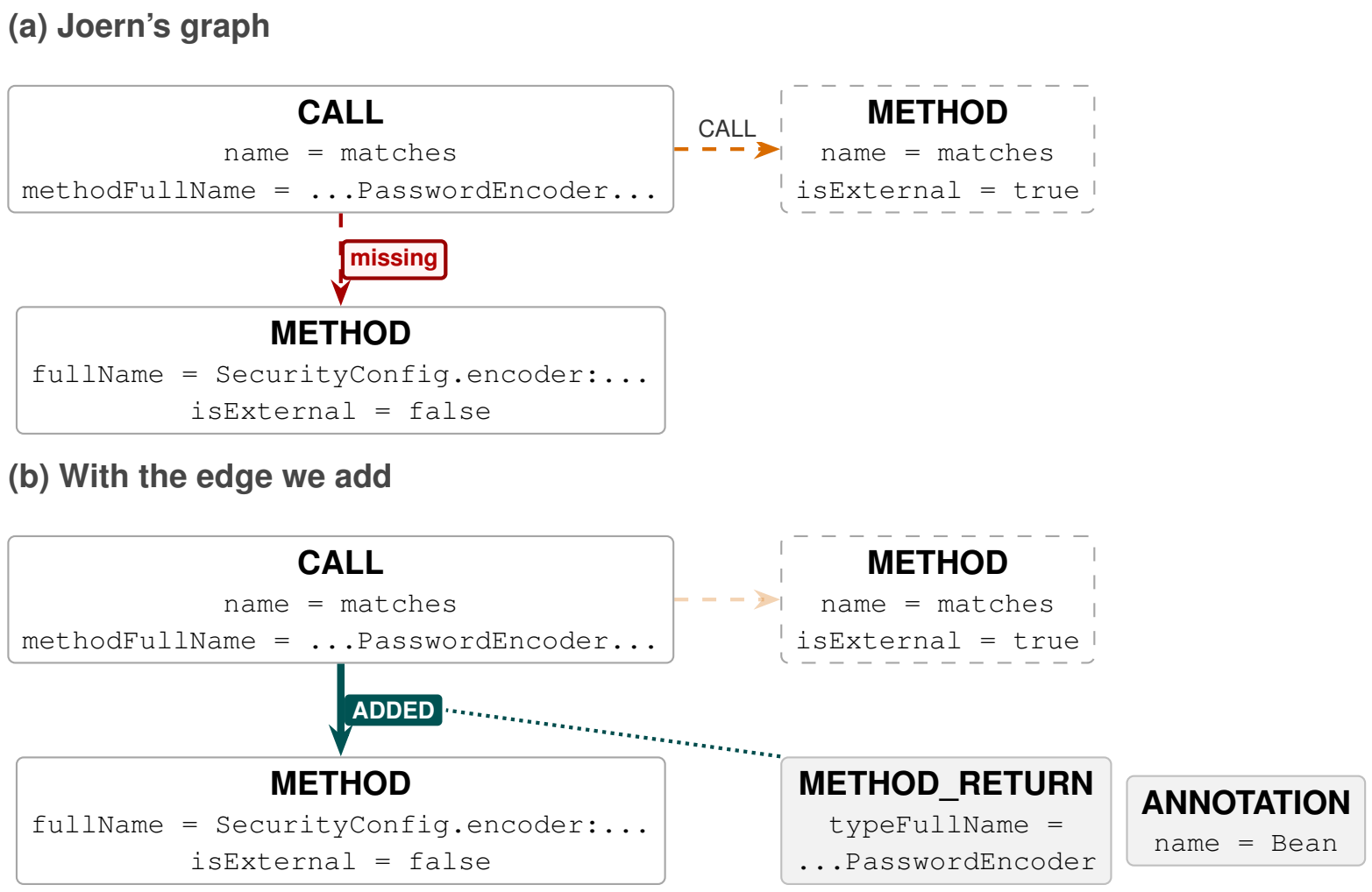

\paragraph{Injection provider (Java)}
\label{app:walk-added-provider}
A call such as \texttt{passwordEncoder.\allowbreak matches(raw, hash)} runs a library object
that Spring built by calling a method of the codebase marked \texttt{@Bean}. Joern links the
call to a placeholder (Figure~\ref{fig:app-walk-a-provider}a), so a role mapped to that
method, such as the Hasher, is never reached. We find the \texttt{@Bean} methods whose return
type is the type of the object the call is made on, and record their role at the call
(Figure~\ref{fig:app-walk-a-provider}b). We do not enter the method, which runs when the
application starts, not during the request.

\begin{figure*}[!htbp]
\centering
\begin{minipage}[t]{0.3\linewidth}
\vspace{0pt}
\begin{lstlisting}[style=walklines, language=Python]
def read_me(
    user = Depends(get_user)):
    return user.name

def get_user(): ...
\end{lstlisting}
\end{minipage}\hfill
\begin{minipage}[t]{0.67\linewidth}
\vspace{0pt}\centering
\begin{tikzpicture}[jnode/.append style={minimum width=2.2cm}]
  \node[jpanel] at (0,0.3) {(a) Joern's graph};
  \node[jnode, anchor=north west] (c) at (0,0) {\jtype{METHOD}\\\jprop{name = read\_me}};
  \node[jnode, right=24mm of c] (m) {\jtype{METHOD}\\\jprop{name = get\_user}};
  \node[jnode, anchor=north west] (q) at (0,-1.3) {\jtype{METHOD\_PARAMETER\_IN}\\\jprop{name = user}\\\jprop{code = user}};
  \draw[jflow] (c.south) -- node[jelab, right] {AST} (c.south |- q.north);
  \draw[jmiss] (c) -- node[jproblem, above] {missing} (m);
  \node[jproblem, anchor=west] at ([xshift=8mm]q.east) {no child for the default};
  \begin{scope}[yshift=-3.3cm]
  \node[jpanel] at (0,0.3) {(b) With the edge we add};
  \node[jnode, anchor=north west] (c) at (0,0) {\jtype{METHOD}\\\jprop{name = read\_me}};
  \node[jnode, right=24mm of c] (m) {\jtype{METHOD}\\\jprop{name = get\_user}};
  \node[jnode, anchor=north west] (q) at (0,-1.3) {\jtype{METHOD\_PARAMETER\_IN}\\\jprop{name = user}\\\jprop{code = user}};
  \draw[jflow] (c.south) -- node[jelab, right] {AST} (c.south |- q.north);
  \draw[jadded] (c) -- node[jaddlab, above] (add) {ADDED} (m);
  \node[jrule, anchor=west] (r) at ([xshift=8mm]q.east) {\textit{source text, line 2}\\\jprop{user = Depends(get\_user)}};
  \draw[jfrom] (r.north) -- (add.south);
  \end{scope}
\end{tikzpicture}
\end{minipage}
\caption{Parameter default: Joern's graph (a) and the edge we add (b). The
dashed teal box is read from the source file, not from Joern's graph.}
\Description{Code on the left. Panel a: METHOD read_me with its METHOD_PARAMETER_IN user, which has no child for the default; the link to METHOD get_user is marked missing. Panel b: we read the signature from the source text and adds the edge.}
\label{fig:app-walk-a-default}
\end{figure*}
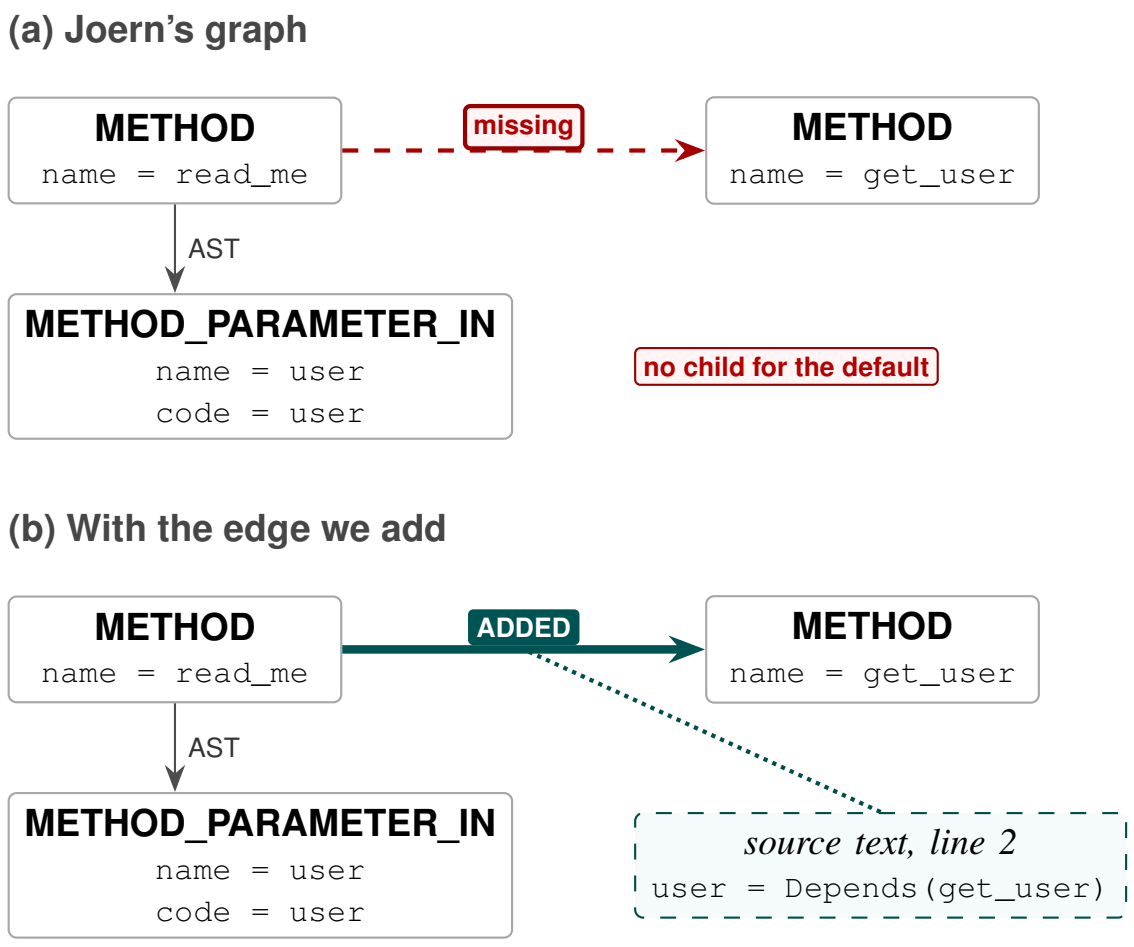

\paragraph{Parameter default (Python)}
\label{app:walk-added-default}
In FastAPI, a parameter written as \texttt{user = Depends(get\_user)} makes the framework call
\texttt{get\_user} before the function body runs. Joern's Python frontend drops parameter
default values, so nothing in the graph links \texttt{read\_me} to \texttt{get\_user}
(Figure~\ref{fig:app-walk-a-default}a). We read the signature from the source file and link
\texttt{read\_me} to \texttt{get\_user} before its first statement
(Figure~\ref{fig:app-walk-a-default}b).

\subsubsection{Lessons learned}
\label{app:walk-added-use}

\textbf{Choosing among several methods.} Some calls lead to several possible methods: a Java
interface call has one Joern edge per implementing class, and override dispatch may find
several implementations. Only one of them runs, and the graph does not say which, so we try
each one separately and keep the one that reaches the most roles.

\textbf{One traversal, one rule set per language.} The traversal is the same for every
language. What differs between Python and Java, such as how an import is mapped to a file,
which framework rules apply and how parameter defaults are read, is kept in a small rule set
per language. A language without a rule set is not analysed.

\textbf{Assumptions behind the added edges.} Override dispatch and framework routes assume
that the framework behaves as documented, for instance that Spring passes the login request to
the codebase's provider; the codebase's configuration is not checked. Each added edge labels
the interactions it produces (Table~\ref{tab:app-walk-added}), so those that rest on such an
assumption can be set apart.

\FloatBarrier

\section{Groundtruth Annotation}
\label{app:annotation}

\subsection{Generating Candidate Mappings with Claude Code}
\label{app:annotation-claude}
\textbf{Chat transcript from Claude Code.} We publish the chat transcript that produced the preminilary mappings for the 100 recovery instances of our evaluation among 5 patterns. It is included in our customized Huggingface bucket for this paper, and can be downloaded from \url{https://huggingface.co/buckets/AnonymousSpottedToad/securityPatternsRecovery}.

\subsection{Annotation Interface}
\label{app:annotation-ui}

\begin{figure*}[tb]
  \centering
  \includegraphics[width=0.78\textwidth]{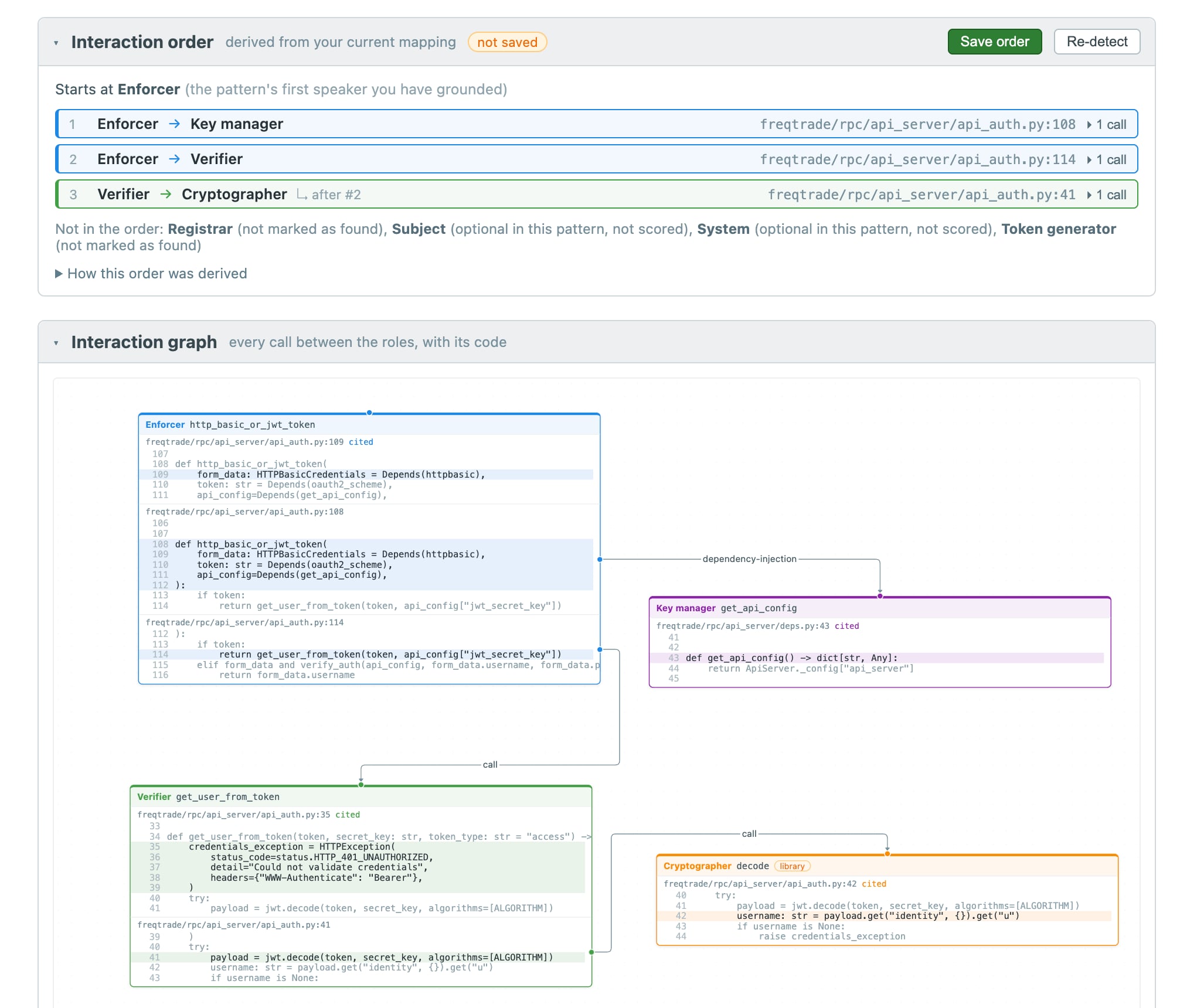}
  \caption{Annotation interface for P2 Verifiable-token authentication in \texttt{freqtrade}. Top: the interaction order derived from the current mappings. Bottom: every call between the mapped roles, with its code.}
  \label{fig:annotation-interface}
\end{figure*}

Figure~\ref{fig:annotation-interface} shows the part of our annotation interface that annotators use to verify the interaction sequence, here for P2 Verifiable-token authentication in \texttt{freqtrade}. The \emph{Interaction order} panel lists the candidate interactions that the CPG traversal derives from the current mappings, each with the file and line of its call, together with the roles that are left out of the order and why. Whenever annotators change a mapping, they can re-detect the order and save it once it reflects the pattern. The \emph{Interaction graph} panel shows every call between the mapped roles with its code, so that annotators can check each interaction against the source code before approving it.

\subsection{Annotation Process}
\label{app:annotation-process}
Of the 500 roles in our 100 groundtruth records (Section~\ref{app:annotation-stats}), annotators changed 114 (22.8\%) from the candidate roles recovered by Claude Code and kept the other 386 (77.2\%) unchanged. The two annotators first annotated 114 roles independently, with an inter-annotator agreement of 0.17 (Cohen's $\kappa$). They then resolved their disagreements on these roles together, and annotated the rest of the roles together.

\subsection{Dataset Statistics}
\label{app:annotation-stats}
\begin{figure}[tb]
  \centering
  \includegraphics[width=\linewidth,trim=0 22 0 0,clip]{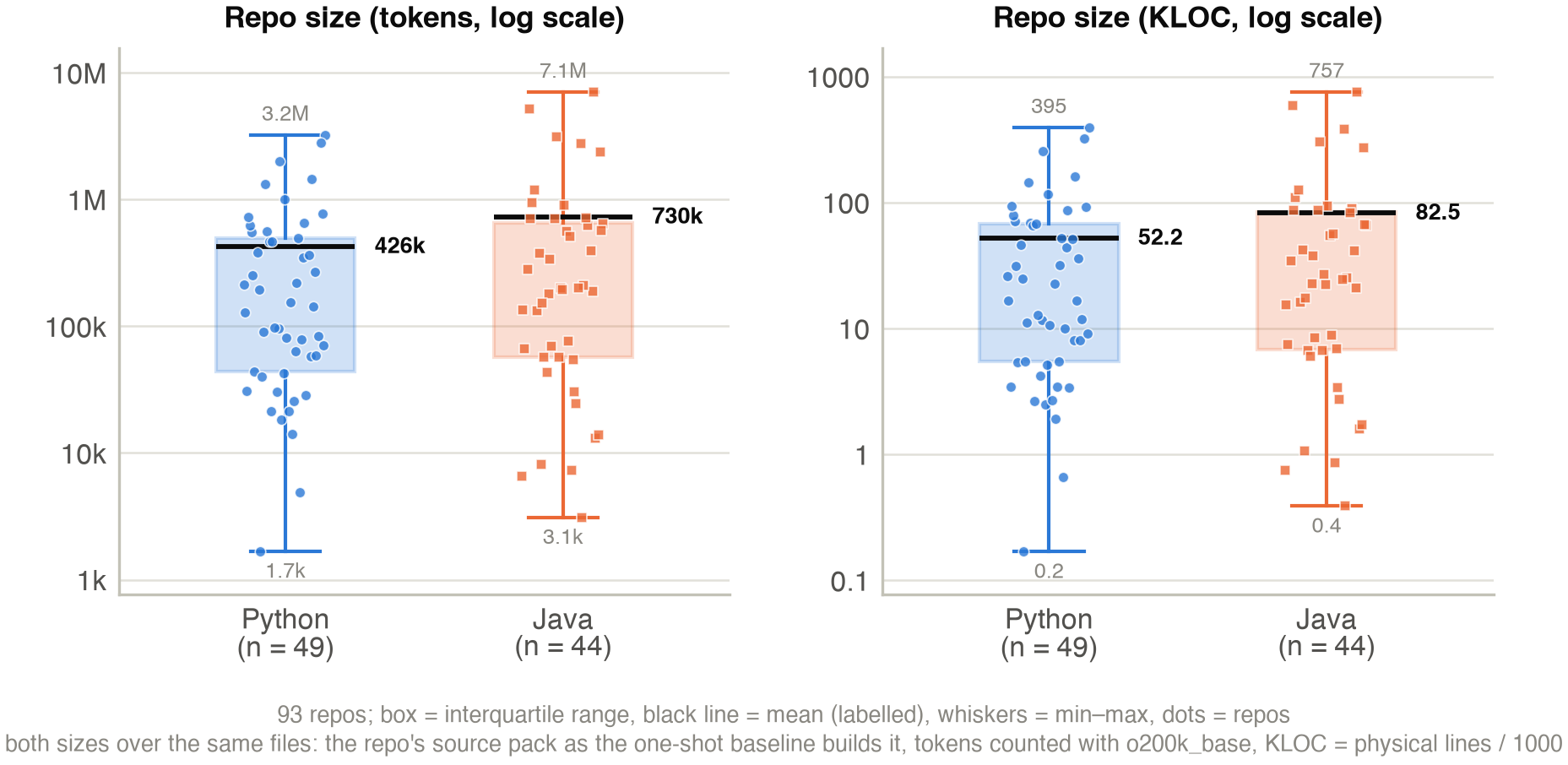}
  \caption{Size of the 93 codebases (49 Python, 44 Java) in \texttt{o200k\_base} tokens and in KLOC.}
  \label{fig:boxplotdatasize}
\end{figure}

Figure~\ref{fig:boxplotdatasize} shows the size of the 93 codebases, in tokens and in KLOC. Tables~\ref{tab:ground-truth-java} and~\ref{tab:ground-truth-python} give the roles by presence status and the interactions of our groundtruth, per pattern, for Java and Python.

\begin{table}[t]
\caption{Groundtruth per pattern for Java (10 codebases each): roles by presence status, and interactions.}
\label{tab:ground-truth-java}
\small
\setlength{\tabcolsep}{3pt}
\renewcommand{\arraystretch}{1.15}
\resizebox{\ifdim\width>\columnwidth\columnwidth\else\width\fi}{!}{%
\begin{tabular}{>{\raggedright\arraybackslash}m{2.9cm}rrrrr}
\toprule
& & \multicolumn{3}{c}{\#Roles} & \\
\cmidrule(lr){3-5}
Pattern & \#Repos & Present & External & Missing & \#Interactions \\
\midrule
P1 Password-based authentication & 10 & 33 & 14 & 23 & 39 \\
P2 Verifiable-token authentication & 10 & 28 & 12 & 0 & 27 \\
P3 Obscure-token access control & 10 & 30 & 6 & 4 & 28 \\
P4 Session-based access control & 10 & 31 & 11 & 18 & 25 \\
P5 Rate limiting & 10 & 34 & 6 & 0 & 34 \\
\midrule
Total & 50 & 156 & 49 & 45 & 153 \\
\bottomrule
\end{tabular}}
\end{table}

\begin{table}[t]
\caption{Groundtruth per pattern for Python (10 codebases each): roles by presence status, and interactions.}
\label{tab:ground-truth-python}
\small
\setlength{\tabcolsep}{3pt}
\renewcommand{\arraystretch}{1.15}
\resizebox{\ifdim\width>\columnwidth\columnwidth\else\width\fi}{!}{%
\begin{tabular}{>{\raggedright\arraybackslash}m{2.9cm}rrrrr}
\toprule
& & \multicolumn{3}{c}{\#Roles} & \\
\cmidrule(lr){3-5}
Pattern & \#Repos & Present & External & Missing & \#Interactions \\
\midrule
P1 Password-based authentication & 10 & 32 & 17 & 21 & 31 \\
P2 Verifiable-token authentication & 10 & 31 & 9 & 0 & 24 \\
P3 Obscure-token access control & 10 & 33 & 0 & 7 & 19 \\
P4 Session-based access control & 10 & 42 & 0 & 18 & 19 \\
P5 Rate limiting & 10 & 40 & 0 & 0 & 34 \\
\midrule
Total & 50 & 178 & 26 & 46 & 127 \\
\bottomrule
\end{tabular}}
\end{table}

\FloatBarrier

\section{Additional Results}
\label{app:results}

\begin{figure*}[!tbp]
  \centering
  \includegraphics[width=\textwidth]{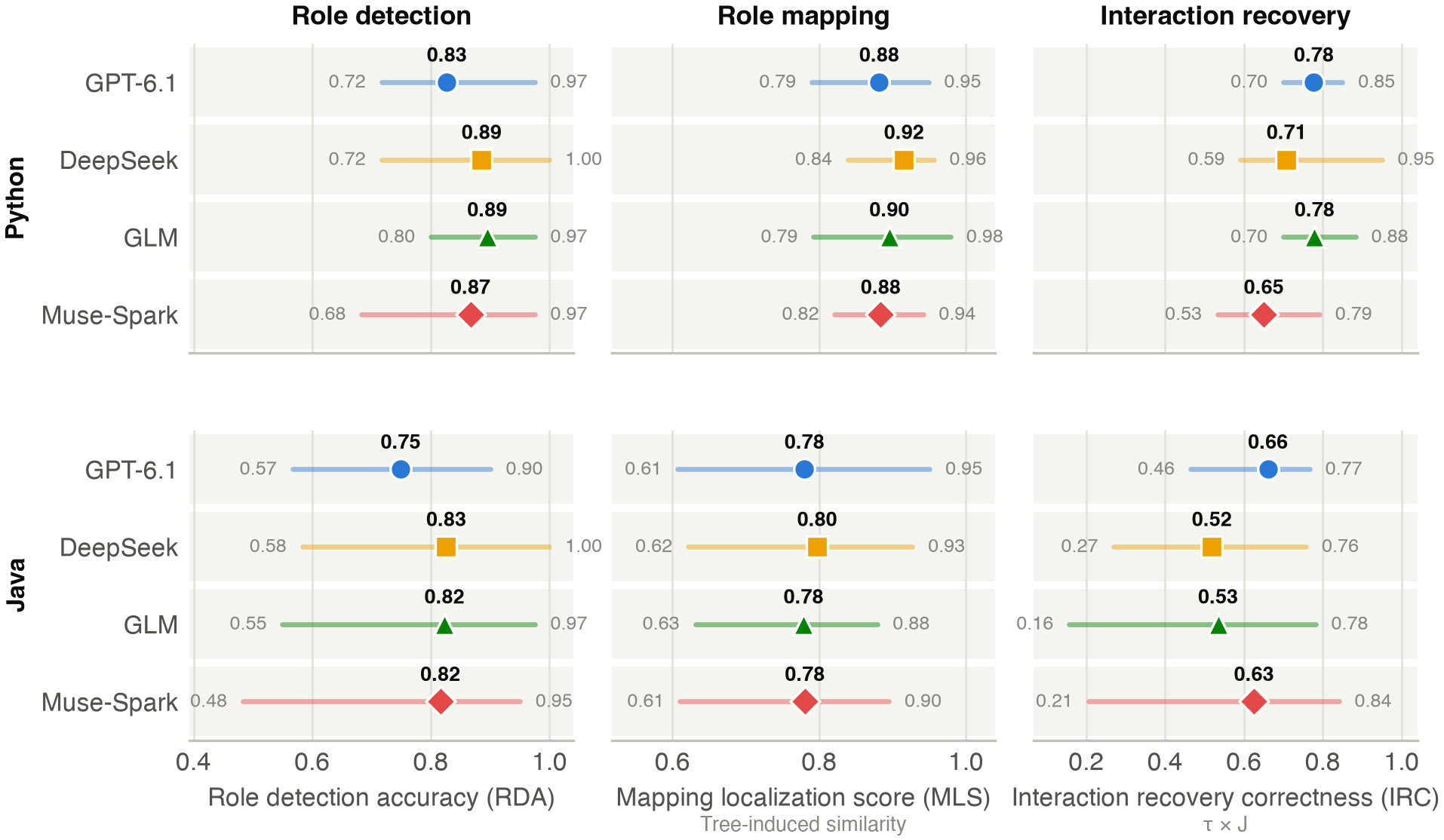}
  \caption{Mean RDA, MLS and IRC of our pipeline per LLM and language. Bold: mean over the 50 instances; whiskers: lowest and highest pattern mean.}
  \label{fig:mean-performance-all-models}
\end{figure*}

\begin{figure*}[!tbp]
  \centering
  \includegraphics[width=\textwidth]{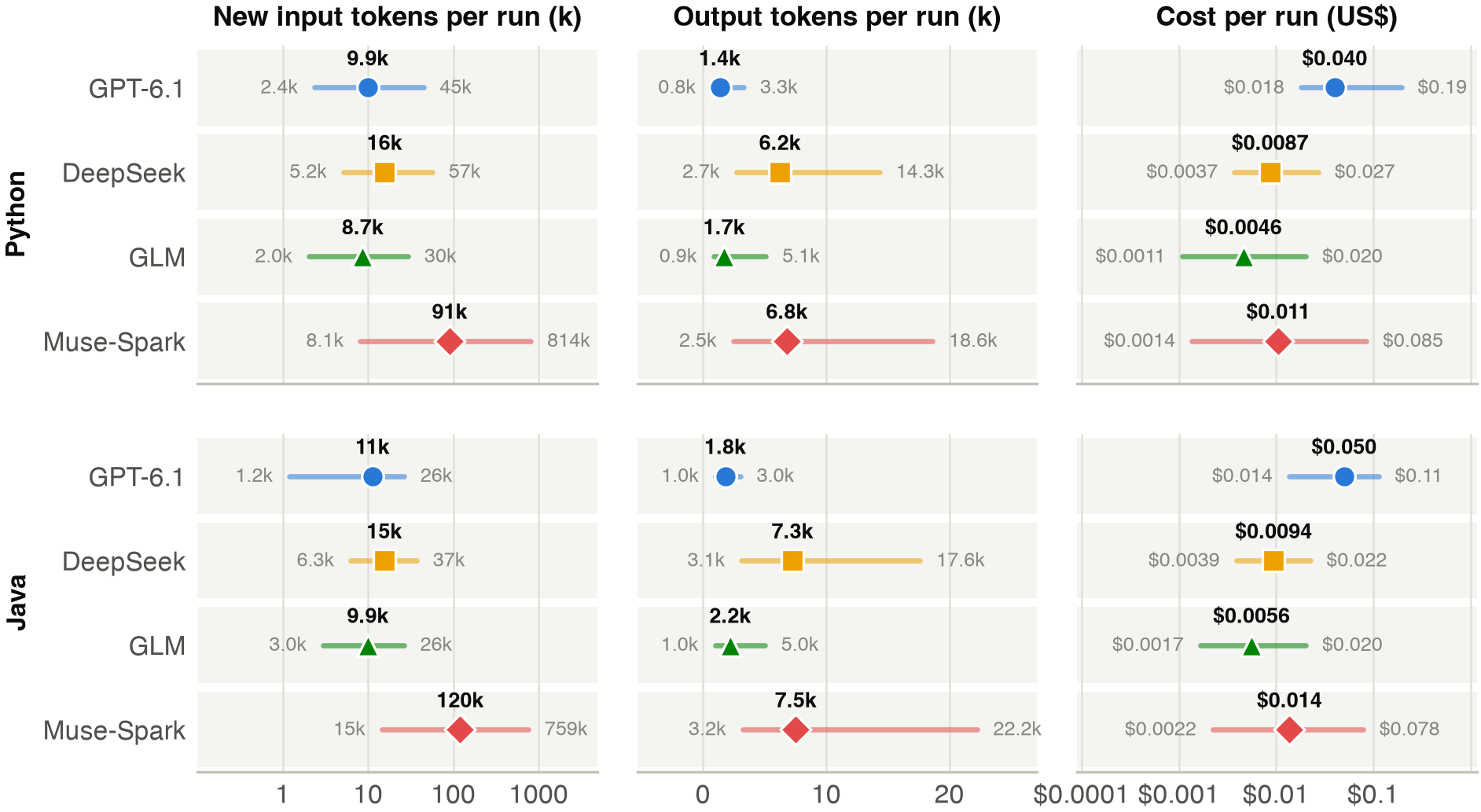}
  \caption{Tokens and cost per run of our pipeline, per LLM and language. Bold: mean over the 50 runs; line: lowest to highest run. New input excludes prompt-cache hits; cost uses public API prices, also for the self-hosted DeepSeek and GLM.}
  \label{fig:mean-tokens-all-models-pipeline}
\end{figure*}

\begin{figure*}[p]
  \centering
  \includegraphics[width=\textwidth,height=0.45\textheight,keepaspectratio]{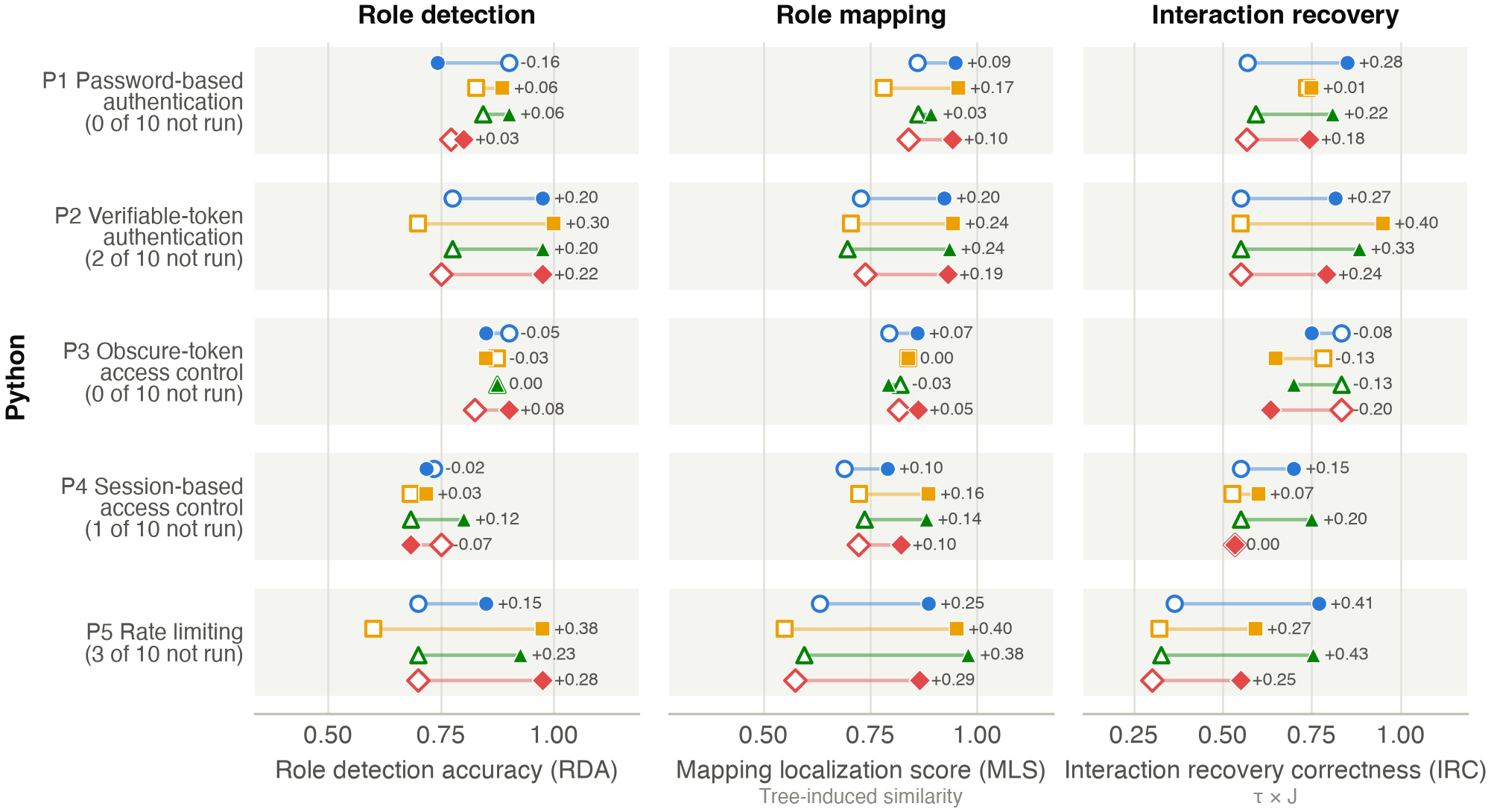}\\[-1pt]
  {\small (a) Python}\\[6pt]
  \includegraphics[width=\textwidth,height=0.45\textheight,keepaspectratio]{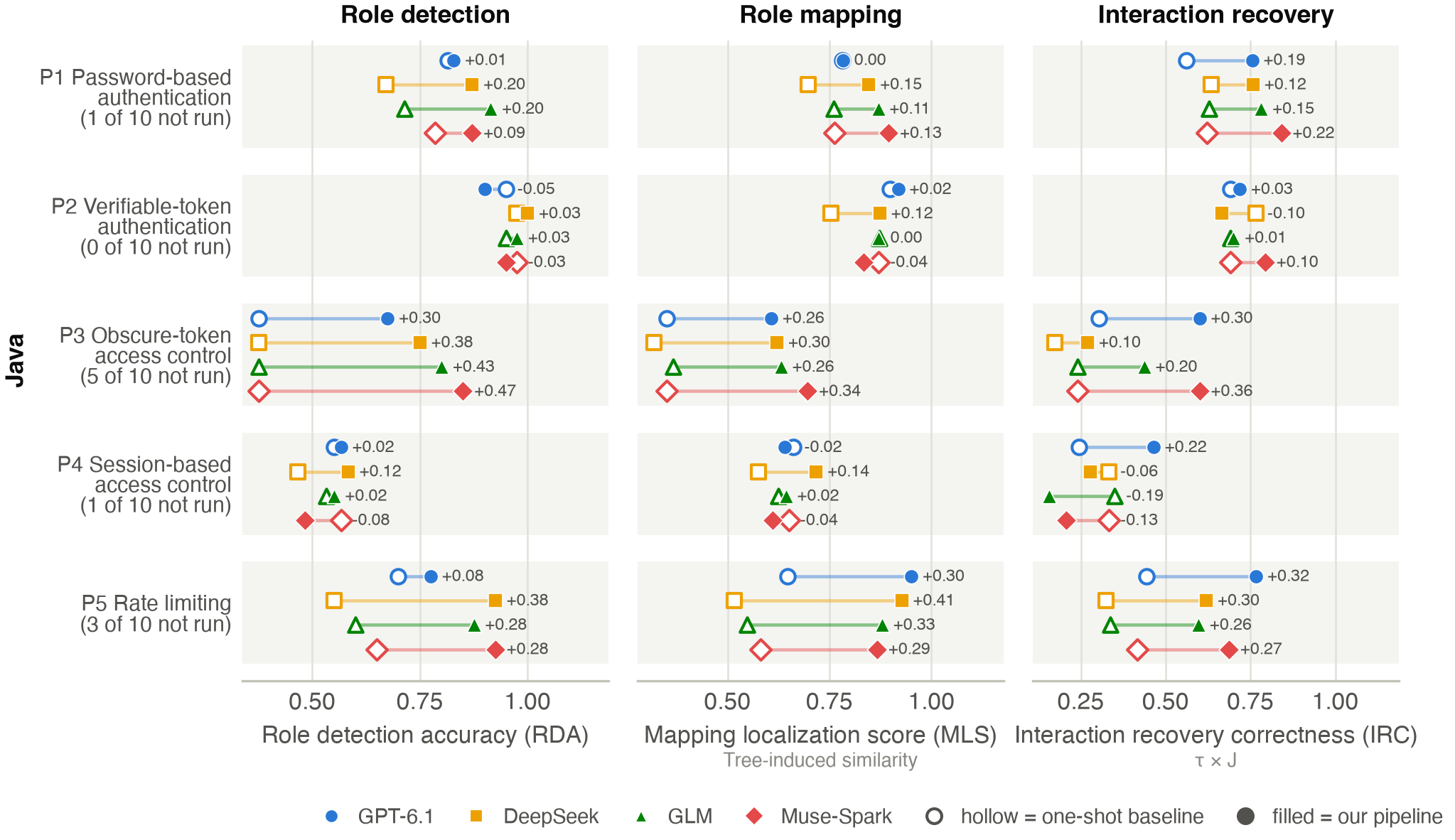}\\[-1pt]
  {\small (b) Java}
  \caption{Mean RDA, MLS and IRC per pattern for (a)~Python and (b)~Java: one-shot baseline (hollow) against our pipeline (filled), with the difference beside each pair. Instances over the baseline's 900K-token input limit score 0 for the baseline; their number is under each pattern.}
  \label{fig:mean-performance-all-pattern-all-model-baseline-pipeline}
\end{figure*}

\begin{figure*}[!tbp]
  \centering
  \includegraphics[width=\textwidth]{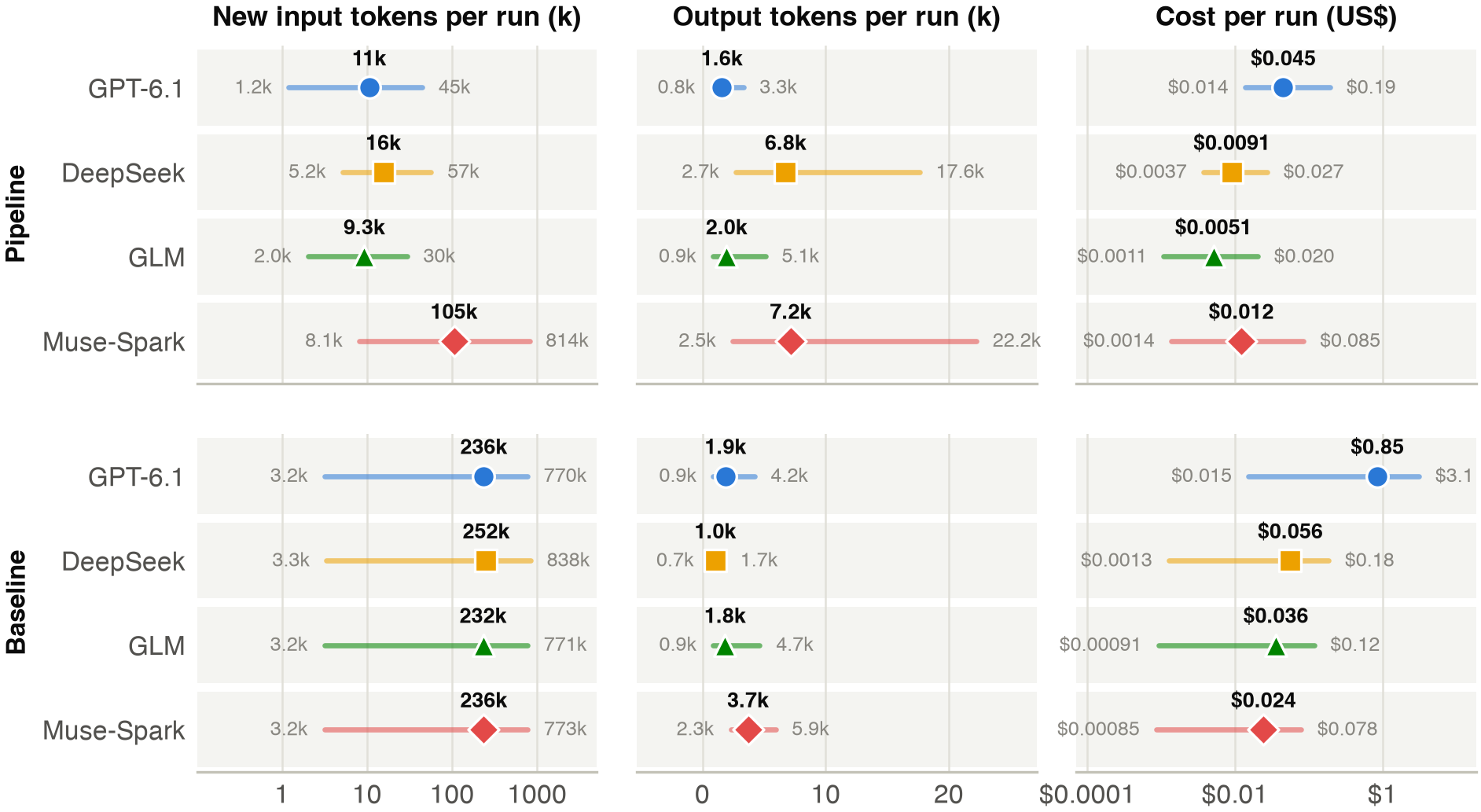}
  \caption{Tokens and cost per run of our pipeline (top) and the one-shot baseline (bottom), both languages, drawn as in Figure~\ref{fig:mean-tokens-all-models-pipeline}. The baseline covers the 84 instances under its 900K-token input limit.}
  \label{fig:mean-tokens-baseline-vs-pipeline}
\end{figure*}

\begin{figure*}[!tbp]
  \centering
  \includegraphics[width=\textwidth]{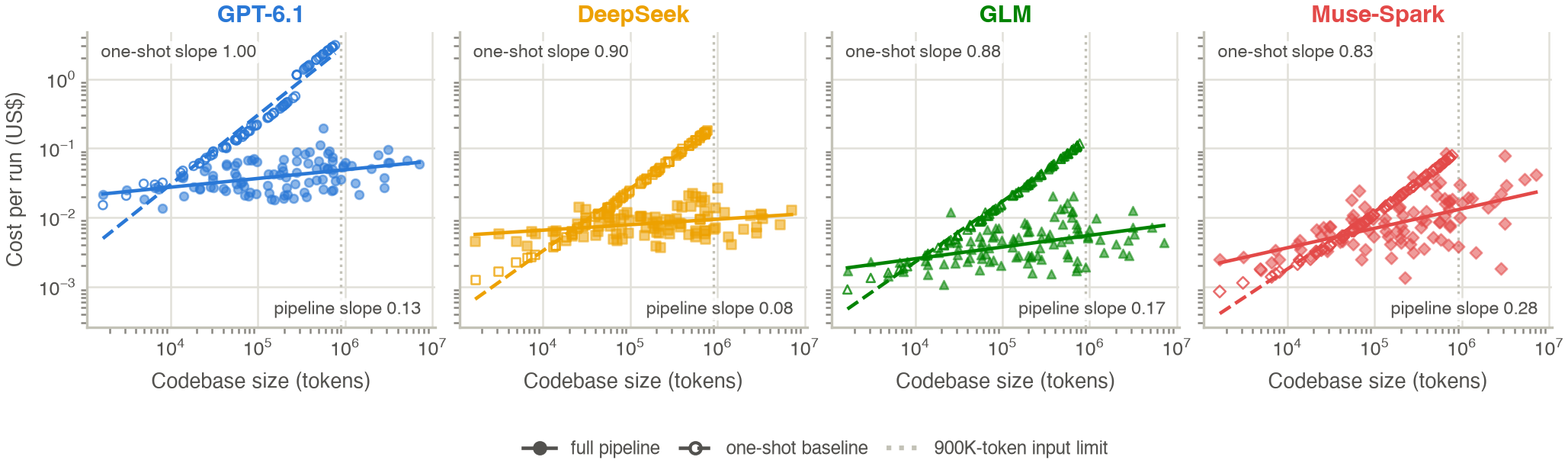}
  \caption{Cost per run against codebase size, log-log: our pipeline (filled, solid fit) against the one-shot baseline (hollow, dashed fit). A slope of 1 means the cost grows in proportion to the size. Dotted: the baseline's 900K-token input limit.}
  \label{fig:cost-vs-size}
\end{figure*}

\begin{figure*}[!tbp]
  \centering
  \includegraphics[width=0.6\textwidth]{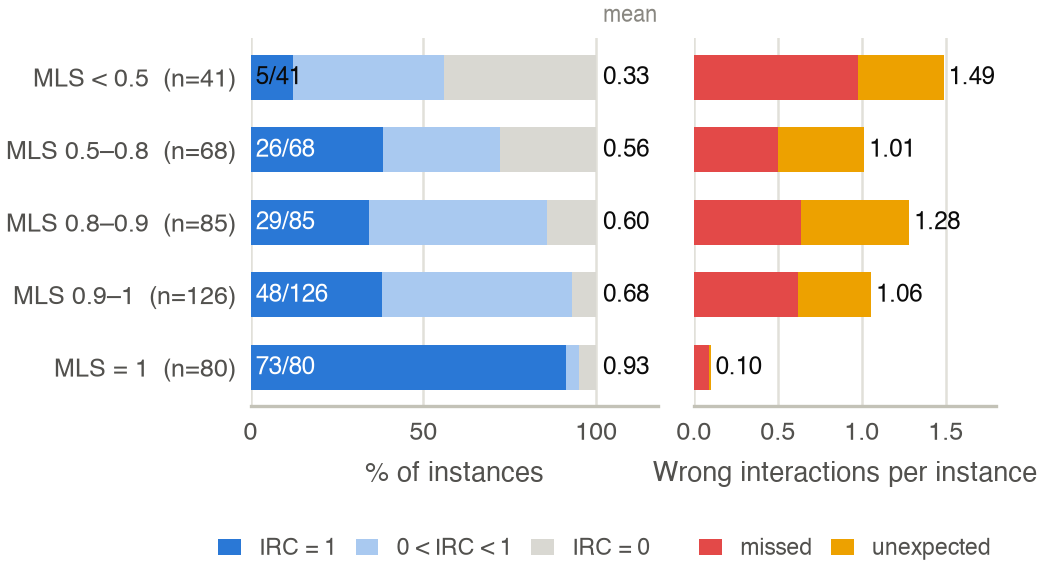}
  \caption{IRC of our pipeline by MLS band, over 400 instances (4 LLMs $\times$ 100). Left: share of instances with a perfect, partial or zero IRC (the number of perfect ones in the bar) and the band's mean IRC. Right: missed and unexpected interactions per instance.}
  \label{fig:irc-by-mls}
\end{figure*}

This appendix gives the figures and the table that Section~\ref{sec:evaluation} refers
to: the mean RDA, MLS and IRC of our pipeline per LLM and language
(Figure~\ref{fig:mean-performance-all-models}, RQ1) and the instances with the highest MLS (Table~\ref{tab:l2-records}, RQ1); the tokens and cost per run of our pipeline
(Figure~\ref{fig:mean-tokens-all-models-pipeline}, RQ2); for RQ3, RDA, MLS and IRC per
pattern against the one-shot baseline (Figure~\ref{fig:mean-performance-all-pattern-all-model-baseline-pipeline}),
the mean RDA, MLS and IRC against the baseline with and without the instances over its input limit (Table~\ref{tab:baseline-runnable}), the tokens and cost per run of the baseline (Figure~\ref{fig:mean-tokens-baseline-vs-pipeline}) how the
cost per run grows with codebase size (Figure~\ref{fig:cost-vs-size}) and IRC by MLS band (Figure~\ref{fig:irc-by-mls}); and the change
when one search tool is removed (Table~\ref{tab:delta-ablation-codesearch}, RQ4).

\begin{table}[!tbp]
\caption{Instances with the highest MLS (ties count for each LLM) and with MLS $=1$, out of 100.}
\label{tab:l2-records}
\centering
\small
\begin{tabular}{lrr}
\toprule
LLM & Highest MLS & MLS $=1$ \\
\midrule
GPT-6.1 & \textbf{54} & \textbf{27} \\
DeepSeek & 44 & 20 \\
GLM & 37 & 17 \\
Muse-Spark & 43 & 16 \\
\bottomrule
\end{tabular}
\end{table}

\begin{table}[!tbp]
\caption{Mean RDA, MLS and IRC of the one-shot baseline (Base) and our pipeline (Ours) per LLM, over all 100 instances (the 16 over the baseline's input limit score 0 for the baseline) and over the 84 instances both can run. Bold: the higher of the two.}
\label{tab:baseline-runnable}
\centering
\small
\setlength{\tabcolsep}{4pt}
\begin{tabular}{lrrrrrr}
\toprule
& \multicolumn{2}{c}{RDA} & \multicolumn{2}{c}{MLS} & \multicolumn{2}{c}{IRC} \\
\cmidrule(lr){2-3}\cmidrule(lr){4-5}\cmidrule(lr){6-7}
LLM & Base & Ours & Base & Ours & Base & Ours \\
\midrule
\multicolumn{7}{l}{\emph{All 100 instances}} \\
GPT-6.1 & 0.74 & \textbf{0.79} & 0.70 & \textbf{0.83} & 0.51 & \textbf{0.72} \\
DeepSeek & 0.67 & \textbf{0.86} & 0.65 & \textbf{0.86} & 0.51 & \textbf{0.61} \\
GLM & 0.70 & \textbf{0.86} & 0.69 & \textbf{0.84} & 0.51 & \textbf{0.66} \\
Muse-Spark & 0.71 & \textbf{0.84} & 0.69 & \textbf{0.83} & 0.51 & \textbf{0.64} \\
\midrule
\multicolumn{7}{l}{\emph{84 instances under the input limit}} \\
GPT-6.1 & \textbf{0.88} & 0.78 & \textbf{0.84} & 0.83 & 0.61 & \textbf{0.73} \\
DeepSeek & 0.80 & \textbf{0.86} & 0.77 & \textbf{0.87} & 0.61 & \textbf{0.64} \\
GLM & 0.84 & \textbf{0.85} & 0.82 & \textbf{0.84} & 0.61 & \textbf{0.68} \\
Muse-Spark & \textbf{0.85} & 0.83 & 0.82 & \textbf{0.84} & 0.61 & \textbf{0.65} \\
\bottomrule
\end{tabular}
\end{table}

\begin{table*}[!tbp]
\caption{Change from our pipeline when one search tool is removed: RDA, MLS and IRC over both
languages; new input and output tokens (\%), steps and memory (MB) per run and language. Without
semantic search, neither the vector DB nor the embedding model is loaded.}
\label{tab:delta-ablation-codesearch}
\centering
\small
\begin{tabular}{lrrrrrrrrrrr}
\toprule
& \multicolumn{3}{c}{Accuracy} & \multicolumn{2}{c}{New input (\%)} & \multicolumn{2}{c}{Output (\%)}
& \multicolumn{2}{c}{Steps} & \multicolumn{2}{c}{Memory (MB)} \\
\cmidrule(lr){2-4}\cmidrule(lr){5-6}\cmidrule(lr){7-8}\cmidrule(lr){9-10}\cmidrule(lr){11-12}
LLM & $\Delta$RDA & $\Delta$MLS & $\Delta$IRC & Py & Java & Py & Java & Py & Java & Py & Java \\
\midrule
\multicolumn{12}{l}{\emph{No keyword search}} \\
GPT-6.1 & -0.009 & -0.015 & -0.001 & 0 & -2 & -8 & -6 & -0.6 & -0.5 & 0 & 0 \\
DeepSeek & -0.002 & -0.002 & +0.088 & +16 & +11 & -11 & +4 & -0.6 & +0.8 & 0 & 0 \\
GLM & +0.008 & +0.001 & -0.032 & +40 & +12 & -6 & -3 & -0.7 & -0.5 & 0 & 0 \\
Muse-Spark & +0.023 & +0.009 & +0.032 & -33 & -30 & -7 & -5 & -0.4 & -1.0 & 0 & 0 \\
\midrule
\multicolumn{12}{l}{\emph{No semantic search}} \\
GPT-6.1 & -0.009 & +0.007 & -0.016 & -1 & -3 & 0 & -3 & +0.5 & -0.1 & -165 & -188 \\
DeepSeek & -0.024 & -0.027 & +0.037 & -3 & +5 & -9 & +9 & -1.0 & +0.7 & -165 & -188 \\
GLM & +0.019 & +0.003 & -0.038 & +4 & +8 & +2 & +7 & +0.2 & +2.1 & -165 & -188 \\
Muse-Spark & +0.023 & +0.006 & +0.029 & -41 & -29 & +2 & +8 & +0.1 & +0.7 & -165 & -188 \\
\bottomrule
\end{tabular}
\end{table*}

\section{Failure Cases}
\label{appendix:failure-cases}

This appendix shows instances where the CPG analysis cannot recover interactions of the
groundtruth although the LLM detected the roles correctly.

\subsection{Unconnected Enforcers in Session-Based Access Control}
\label{app:session-failure}

Figure~\ref{fig:app-sf} shows why P4 Session-based access control scores a low IRC in
Java, on the codebase \texttt{ncats/gsrs-spring-starter}. The Authentication enforcer is
the servlet filter's \texttt{doFilterInternal}. It checks the session and then hands the
request to Spring with \texttt{filterChain.doFilter}. The Authorisation enforcer is the
\texttt{@PreAuthorize} annotation that \texttt{UserController} carries through
\texttt{@canManageUsers}. Spring connects the two at run time, in framework code, and the
annotation's call to the Decider exists only inside a string. No call in the codebase
links the two enforcers, so the CPG traversal from the Authentication enforcer
(Section~\ref{sec:interaction-sequence-recovery}) never reaches the authorisation roles.
On this instance, GPT-6.1 detects all six roles correctly (RDA $= 1$) but scores
IRC $= 0$: its sequence shares only two authentication interactions with the groundtruth,
in the opposite order. DeepSeek and GLM score IRC $= 0$ on this instance for the same
reason.

\newcommand{\sfel}{{\color{black!45}\ldots}}
\newcommand{\sfin}{\phantom{xxxx}}
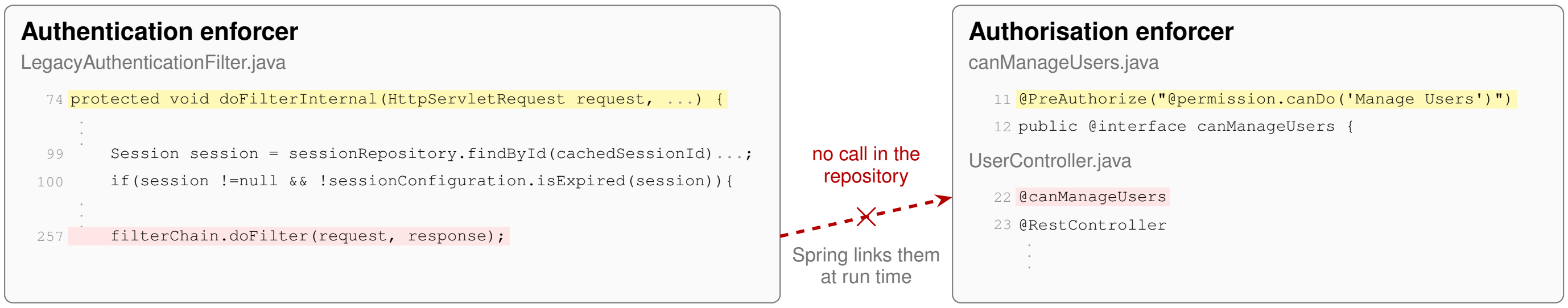
\begin{figure*}[t]
\centering
\resizebox{\textwidth}{!}{%
\begin{tikzpicture}[
  box/.style={draw=black!50, rounded corners=3pt, fill=black!2},
  head/.style={font=\sffamily\bfseries\small, anchor=north west, inner sep=0pt},
  file/.style={font=\sffamily\scriptsize, text=black!55, anchor=west, inner sep=0pt},
  ln/.style={font=\ttfamily\fontsize{5.5}{6.5}\selectfont, text=black!45,
             anchor=east, inner sep=1pt},
  code/.style={font=\ttfamily\fontsize{6}{7}\selectfont, anchor=west, inner sep=1pt},
  ent/.style={code, fill=yellow!35},
  hand/.style={code, fill=red!10},
  gap/.style={font=\fontsize{6}{6}\selectfont, text=black!40, anchor=west, inner sep=0pt}
]
\fill[box] (0,0) rectangle (9.9,-3.8);
\draw[box, fill=none] (0,0) rectangle (9.9,-3.8);
\node[head] at (0.2,-0.2) {Authentication enforcer};
\node[file] at (0.2,-0.75) {LegacyAuthenticationFilter.java};
\node[ln] at (0.8,-1.2)  {74};
\node[ent] at (0.8,-1.2) {protected void doFilterInternal(HttpServletRequest request, \sfel) \{};
\node[gap] at (0.95,-1.53) {$\vdots$};
\node[ln] at (0.8,-1.9)  {99};
\node[code] at (0.8,-1.9) {\sfin Session session = sessionRepository.findById(cachedSessionId)\sfel;};
\node[ln] at (0.8,-2.25) {100};
\node[code] at (0.8,-2.25) {\sfin if(session !=null \&\& !sessionConfiguration.isExpired(session))\{};
\node[gap] at (0.95,-2.58) {$\vdots$};
\node[ln] at (0.8,-2.95) {257};
\node[hand] (l257) at (0.8,-2.95) {\sfin filterChain.doFilter(request, response);};

\fill[box] (12.1,0) rectangle (19.9,-3.8);
\draw[box, fill=none] (12.1,0) rectangle (19.9,-3.8);
\node[head] at (12.3,-0.2) {Authorisation enforcer};
\node[file] at (12.3,-0.75) {canManageUsers.java};
\node[ln] at (12.9,-1.2)  {11};
\node[ent] at (12.9,-1.2) {@PreAuthorize("@permission.canDo(\textquotesingle Manage Users\textquotesingle)")};
\node[ln] at (12.9,-1.55) {12};
\node[code] at (12.9,-1.55) {public @interface canManageUsers \{};
\node[file] at (12.3,-2.0) {UserController.java};
\node[ln] at (12.9,-2.45) {22};
\node[hand] (l22) at (12.9,-2.45) {@canManageUsers};
\node[ln] at (12.9,-2.8) {23};
\node[code] at (12.9,-2.8) {@RestController};
\node[gap] at (13.05,-3.1) {$\vdots$};

\draw[-{Stealth[length=2.2mm]}, dashed, very thick, red!70!black]
  (9.9,-2.95) -- node[font=\Large\bfseries, text=red!70!black] {$\times$}
  node[above=2.5mm, font=\sffamily\scriptsize, text=red!70!black, align=center]
    {no call in the\\repository}
  node[below=2.5mm, font=\sffamily\scriptsize, text=black!55, align=center]
    {Spring links them\\at run time}
  (12.1,-2.45);
\end{tikzpicture}}
\caption{The two enforcers of session-based access control in
\texttt{ncats/gsrs-spring-starter}. The Authentication enforcer (left) ends by
passing the request to Spring (line 257). Spring then applies the Authorisation
enforcer (right) to \texttt{UserController}. No call in the repository joins the
two, so the code property graph has no edge between them. Yellow marks each
enforcer's entity, and red marks the two ends of the missing link.}
\Description{Two code boxes side by side. The left box, Authentication enforcer,
shows LegacyAuthenticationFilter.java: the method doFilterInternal at line 74,
highlighted, a session lookup at line 99, an expiry check at line 100, and
filterChain.doFilter at line 257. The right box, Authorisation enforcer, shows
canManageUsers.java, whose line 11, highlighted, is
PreAuthorize("@permission.canDo('Manage Users')"), and UserController.java,
which carries the canManageUsers annotation at line 22. A red dashed arrow with
a cross runs from line 257 on the left to line 22 on the right, labelled no call
in the repository; Spring links them at run time.}
\label{fig:app-sf}
\end{figure*}

\FloatBarrier

\section{Artifacts}
\label{app:artifacts}

We release the dataset and the code of this paper as five anonymized artifacts
(Table~\ref{tab:app-artifacts}). This appendix maps each of them to the parts of the
paper it supports, gives the names under which the paper's terms appear in them, and
shows how the numbers we report are recomputed from them.

\begin{table*}[!htbp]
\caption{The artifacts of this paper and the parts of the paper each one supports.}
\label{tab:app-artifacts}
\small
\setlength{\tabcolsep}{4pt}
\renewcommand{\arraystretch}{1.15}
\begin{tabular}{>{\raggedright\arraybackslash}p{2.5cm}>{\raggedright\arraybackslash}p{5.4cm}>{\raggedright\arraybackslash}p{5.5cm}>{\raggedright\arraybackslash}p{3.2cm}}
\toprule
Artifact & Link & Contents & Used in \\
\midrule
Dataset & \url{https://huggingface.co/buckets/AnonymousSpottedToad/securityPatternsRecovery}
  & The 100 groundtruth architectures; the 93 codebases and their CPGs; the outputs of
    every experiment; the five specifications $P$; the Claude Code session that drafted
    the candidate mappings
  & Sections~\ref{sec:setup-groundtruth} and~\ref{sec:evaluation};
    Appendices~\ref{app:adl} and~\ref{app:annotation} \\
\textsc{SecPatRec} & \url{https://github.com/anonymousspottedtoad/secpatrec}
  & The role detection and mapping agent, the one-shot baseline, the two ablations and
    the \texttt{secpat} command that runs them
  & Sections~\ref{sec:method-pipeline} and~\ref{sec:evaluation} \\
Interaction recovery & \url{https://github.com/anonymousspottedtoad/secpat_interaction_recovery}
  & The traversal of the Joern CPG: \texttt{walk.sc}, with the rules per language in
    \texttt{java.sc} and \texttt{python.sc}
  & Section~\ref{sec:interaction-sequence-recovery}; Appendix~\ref{app:order} \\
Evaluation metrics & \url{https://github.com/anonymousspottedtoad/secpat_evaluation_metrics}
  & RDA, MLS and IRC
  & Section~\ref{sec:setup-evaluation}; Appendix~\ref{app:irc-example} \\
Annotation interface & \url{https://github.com/anonymousspottedtoad/secpat_annotation_portal}
  & The web interface the annotators used to verify the candidate architectures
  & Appendix~\ref{app:annotation-ui} \\
\bottomrule
\end{tabular}
\end{table*}

\subsection{Names Used in the Artifacts}
\label{app:artifacts-names}
The code and the dataset name some of the paper's terms differently:
{\raggedright
\begin{itemize}
  \setlength{\itemsep}{0pt}
  \item \emph{Metrics}: RDA, MLS and IRC are Level~1, Level~2 and Level~3, and
        $1 - \text{score}$ is reported as \emph{drift}.
  \item \emph{Patterns}: P1 is \texttt{authentication\_pwd}, P2
        \texttt{verifiable\_token\_based\_authentication}, P3
        \texttt{obscure\_token\_based\_access\_control}, P4
        \texttt{session\_based\_access\_control} and P5 \texttt{limit\_request\_rate}.
  \item \emph{LLMs}: GPT-6.1 is \texttt{gpt-6.1-sol}, Muse Spark
        \texttt{muse-spark-1.3-contributor}, DeepSeek \texttt{deepseek-v4-flash-0731}
        and GLM \texttt{glm-5.3-flash}.
  \item \emph{Search methods}: lexical (keyword) search is the agent's \texttt{grep}
        tool and semantic search its \texttt{search} tool, so the two settings of RQ4
        are \texttt{ablation\_no\_grep} and \texttt{ablation\_no\_search}. The full
        pipeline is \texttt{pipeline} and the one-shot baseline \texttt{baseline}.
\end{itemize}}

\subsection{From the Dataset to the Reported Numbers}
\label{app:artifacts-numbers}
\textbf{Groundtruth.} Each file under \texttt{ground\_truth/} keeps every role the
annotators recorded, 758 rows in total, including Subject and System and some roles of
the lifecycle flows, such as the Registrar of P1. The 500 roles of
Table~\ref{tab:ground-truth} are the roles marked as required in the pattern's
\texttt{04\_pattern.json}, and a required role without a row is counted as missing,
which is the case for the Pepper store in three P1 records. Counted this way, the files
give the 334 present, 75 external and 91 missing roles and the 280 interactions of
Table~\ref{tab:ground-truth}, cell by cell.

\textbf{Scores.} Under \texttt{experiment\_outputs/}, one \texttt{evaluation.json}
per experiment, LLM and language holds the RDA, MLS and IRC of its 50 instances. Averaging them gives the scores
of Section~\ref{sec:evaluation}; for the full pipeline, the means over the four LLMs are
an RDA of 0.84, an MLS of 0.84 and an IRC of 0.66. The \texttt{secpat evaluate} command
of the \textsc{SecPatRec} repository recomputes these scores from the stored recoveries
and the groundtruth, without calling any LLM. For the pipeline and the ablations, IRC
derives the recovered interaction sequence $\hat{S}$ from the role mapping on the same
CPG and with the same traversal as the groundtruth; for the baseline, it scores the
sequence that the LLM states in its answer.

\textbf{Baseline input limit.} The one-shot baseline is called only when the packed
codebase fits in 900K tokens of the \texttt{o200k\_base} tokenizer, below the 1M-token
limit of Section~\ref{sec:setup-models}. The 16 codebases above it have no recovery in
the dataset and score 0 (Section~\ref{sec:setup-scalability}).

\clearpage

\end{document}